%% file: main.tex
\providecommand{\onlineappendixmode}{0}
\ifdefined\divideanddivergeclassloaded
\else
  \documentclass[12pt]{article}
\fi
\usepackage{graphicx} 
\usepackage{cmap} 
\usepackage[T1]{fontenc}
\usepackage[utf8]{inputenc} 
\usepackage{lmodern}
\input{glyphtounicode}
\usepackage{geometry}
\usepackage{floatrow}
\usepackage{amsfonts}
\usepackage{booktabs}
\usepackage{amssymb}
\usepackage{amsmath}
\usepackage{mathtools}
\usepackage{float}
\usepackage{caption}
\usepackage{subcaption}
\usepackage{graphicx}
\usepackage{lscape}
\usepackage{xcolor}
\definecolor{partyblue}{RGB}{46,82,128}
\definecolor{partyrust}{RGB}{174,78,32}
\definecolor{softgray}{RGB}{225,225,225}
\usepackage{url}
\usepackage{footmisc}
\newtheorem{theorem}{Theorem}

\newtheorem{corollary}{Corollary}

\newtheorem{definition}{Definition}

\newtheorem{lemma}{Lemma}

\newtheorem{proposition}{Proposition}
\newtheorem{remark}{Remark}

\newtheorem{assumption}{Assumption}
\usepackage{setspace}
\usepackage{footmisc}
\usepackage{needspace}
\usepackage[round]{natbib}
\usepackage{xr-hyper}
\usepackage{hyperref}
\definecolor{DJETBlue}{HTML}{0C769F}
\definecolor{LJETBlue}{HTML}{0080AC}
\hypersetup{
    colorlinks=true,
    linkcolor=LJETBlue,
    citecolor=LJETBlue,
    filecolor=LJETBlue,
    urlcolor=LJETBlue
}
\ifnum\onlineappendixmode=1
  \hypersetup{
    pdftitle={Online Appendix to Divide and Diverge},
    pdfauthor={Giampaolo Bonomi}
  }
\else
  \hypersetup{
    pdftitle={Divide and Diverge: Political Competition and the Organization of Conflict},
    pdfauthor={Giampaolo Bonomi}
  }
\fi
\ifnum\onlineappendixmode=1
  \IfFileExists{main.aux}{%
    \externaldocument[main:]{main}%
  }{%
    \externaldocument[main:]{main_xr}[main.pdf]%
  }
\else
  \IfFileExists{online_appendix.aux}{%
    \externaldocument[oa:]{online_appendix}%
  }{%
    \externaldocument[oa:]{online_appendix_xr}[online_appendix.pdf]%
  }
\fi
\newenvironment{proof}[1][Proof]{\noindent \textbf{#1.} }{\  \rule{0.5em}{0.5em}}
\usepackage{enumerate}   
\usepackage{tikz}
\usetikzlibrary{patterns}
\input{pictures.tex}

\newcommand{\ubar}[1]{\underline{#1}}
\usepackage{chngcntr}
\usepackage{apptools}
\AtAppendix{\counterwithin{lemma}{section}}
\AtAppendix{\counterwithin{proposition}{section}}
\AtAppendix{\counterwithin{theorem}{section}}
\AtAppendix{\counterwithin{definition}{section}}
\AtAppendix{\counterwithin{corollary}{section}}
\AtAppendix{\counterwithin{remark}{section}}
\AtAppendix{\counterwithin{equation}{section}}
\IfFileExists{kpfonts.sty}{%
  \usepackage{kpfonts}%
}{%
  \PackageWarningNoLine{divide-diverge}{%
    kpfonts.sty not found; using the Latin Modern fallback%
  }%
}

\title{Divide and Diverge: Political Competition and the Organization of Conflict\thanks{I am grateful to Renee Bowen, Peter Buisseret, Chris Bidner, Adam Brzezinski, Carlo Cusumano, Camilo García-Jimeno, Germán Gieczewski, Gleason Judd, Anton Kolotilin, John Londregan, Simone Galperti, Nicola Gennaioli, Aram Grigoryan, Adam Meirowitz, Marta Mojoli, Matias Iaryczower, Federica Izzo, Freddie Papazyan, Kris Ramsay, Joel Sobel, Guido Tabellini, Leeat Yariv, and participants at the Princeton RPPE seminar, the Stony Brook International Conference on Game Theory, the SIOE 2026 Conference, and the Econometric Society NASM for helpful comments and suggestions.}}
\author{Giampaolo Bonomi\thanks{Princeton University. Email: bonomi@princeton.edu}}
\date{August 2026}

\begin{document}

\ifnum\onlineappendixmode=1
\title{Online Appendix to\\[3pt] Divide and Diverge: Political Competition and the Organization of Conflict}
\author{Giampaolo Bonomi\thanks{Princeton University. Email: bonomi@princeton.edu}\\[0.75em]}
\maketitle
\newpage
\makeatletter
\renewcommand*\l@section{\@dottedtocline{1}{1.5em}{3.5em}}
\renewcommand*\l@subsection{\@dottedtocline{2}{5.0em}{4.7em}}
\renewcommand*\l@subsubsection{\@dottedtocline{3}{9.7em}{5.7em}}
\makeatother
\tableofcontents
\input{online_appendix_content.tex}
\newpage
\bibliographystyle{apalike}
\begin{spacing}{1.15}
\bibliography{biblio}
\end{spacing}
\else
\maketitle
\vspace{.5cm}
\begin{abstract}
\begin{spacing}{1.15}
Why do politicians sharpen divisions when elections reward broad appeal? I model how two identical, office-motivated parties turn voter disagreement into partisan conflict. Elections allocate political power. Each party empowers its most valuable members first, so power has diminishing returns. Platforms never converge: differentiation builds constituencies that survive adverse popularity shocks. Voter polarization pushes platforms apart and benefits both parties---even when a voter group grows more hostile to one party. Across issues, parties divide along the direction that generates more voter conflict---for elliptical electorates, a leading principal component of voter positions, consistent with ANES survey evidence.

\vspace{1cm}


\end{spacing}
\end{abstract}
\newpage

\section{Introduction}

\begin{flushright}
\begin{minipage}{0.74\textwidth}
\small\itshape
``The outcome of the game of politics depends on which of a multitude of
possible conflicts gains the dominant position.''
\par\smallskip
\raggedleft\upshape
--- E.~E. Schattschneider, \emph{The Semisovereign People}
(1960, p.~62)
\end{minipage}
\end{flushright}
\vspace{0.4em}

Political polarization increasingly structures democratic politics. In the United States and other democracies, economic and cultural views have aligned more closely with partisan identity, while party platforms and elite rhetoric have moved apart.\footnote{See \citealp{Gentzkow2016,Gentzkow2019,BertrandKamenica2023,BGT21}.} Political actors help organize this conflict: they emphasize divisive issues, direct communication and mobilization toward partisan constituencies, and portray political stakes in antagonistic or zero-sum terms.\footnote{For instance, \citealp{Gleaser2005,MartinYurukoglu2017,Eunji2022,Feddersen2022,ali2024}. \citet{BoeriNikiforovaTabellini2026} document pre-election divergence in both issue emphasis and within-issue rhetoric.} Yet electoral competition is supposed to reward broad appeal. Why does it instead reward efforts to sharpen social divisions?

An extensive literature studies how economic shocks, identity, media, and information create or reinforce divisions among voters.\footnote{See, among others, \citealp{Autor2020,Alesina2020,BGT21,GT23,Bowen2023,Stantcheva2021,MartinYurukoglu2017,DellaVigna2007,Gleaser2005,BonomiSpillovers}.} I ask a complementary supply-side question: why and how do two ex ante identical, office-motivated parties turn voter disagreement into partisan conflict? More specifically, does party polarization track voter polarization? When opinions span several issues, which pattern of disagreement organizes voters into opposing party constituencies? And why might rival parties both benefit as voter divisions sharpen? All three answers turn on the same mechanism: sharper divisions stabilize each party’s political power. Power is constant-sum—one party’s gain is the other’s loss—but the value each party derives from it is not. Both parties can therefore prefer a more stable split of power to alternating landslides.

The reason lies in how parties use political power. I model them as political organizations \citep{CaillaudTirole2002,SnyderTing2002}: each party allocates its power share to affiliates through positions and channels of influence within government and the broader political economy. Scarce influence opportunities go first to affiliates generating the highest political or economic returns, so a foothold preserves a party’s most valuable uses of power, whereas a wider landslide reaches progressively lower-return affiliates. When platforms coincide, changes in national conditions move all voter groups in lockstep and can sweep power to one side. Differentiation breaks the lockstep: each party cedes support mainly in good times and retains a constituency in bad---a hedge against adverse electoral swings. How effective that hedge is depends on voter disagreement. Along one policy dimension, greater voter polarization pushes equilibrium platforms apart and strengthens the hedge, so both parties gain. A party can even gain when voters already opposed to it move farther away: they pull its rival outward, and the wider platform gap makes its remaining constituency less vulnerable to an adverse swing. With several issues, the same logic determines what partisan conflict is about. Both sides prefer the cleavage that best sorts the electorate into two conflicting constituencies. When voter disagreements become more correlated across issues, the equilibrium partisan cleavage better separates the same voter blocs on multiple fronts, increasing equilibrium platform polarization and benefiting both parties.

To formalize this logic, I study a probabilistic-voting model with two ex ante identical, purely office-motivated parties and no inherited constituencies or electoral advantages. Voters choose between the parties based on their platforms and an aggregate popularity shock realized after platforms are chosen. An electoral rule translates vote shares into constant-sum allocations of political power. The rule is strictly increasing, but otherwise unrestricted: it may have jumps at majority and other thresholds, thereby accommodating different electoral systems. Each party's payoff is strictly increasing and strictly concave in its political power, reflecting the diminishing organizational returns described above. The induced payoff need not be concave---or even continuous---in vote share.

Despite the parties’ ex ante symmetry, policy convergence is impossible in equilibrium, reversing Downsian and standard probabilistic-voting logic
\citep{Downs1957,LW87,Lindbeck,PT2000}. Classic spatial models obtain divergence
from candidates' policy preferences,
a literature surveyed by
\citet{Duggan2006}, and when candidates face different effective electorates, for instance because of primaries \citep[e.g.][]{Brady2007,Hirano2009} or asymmetric audiences \citep{Gleaser2005, GT23}. Platform polarization has also been explained by relaxing the concavity of voter utility \citep{KamadaKojima2014} or combining electoral sorting with endogenous caucus composition \citep{PolbornSnyder2017}. Here, instead, organizational incentives push parties to differentiate to build stable constituencies that stay loyal in adverse times.\footnote{The non-convergence result is related to \citet{ZakharovSorokin2014} and
\citet{SorokinZakharov2018}, who study platform convergence and divergence with nonlinear objectives, and \citet{Buisseret}, discussed in more detail below.} 
This channel is consistent with empirical evidence that polarized and sorted voters respond less to economic news
and candidate quality and cross-party lines less often
\citep{Moreira2025,ButticeStone2012,DavisMason2016}.

With a single policy dimension, the more left-wing platform places greater weight on voters to
the left of the median, while the more right-wing platform does the same on the other side.
Over-weighted voters anchor the constituency that protects the corresponding party after
unfavorable aggregate shocks. With quadratic policy loss and uniform shocks, the
pure-strategy equilibrium is essentially unique up to party labels: each platform is a weighted average of voter positions, with voter group weights proportional to the party's payoff gain when the group joins its coalition. Both parties' payoffs rise with equilibrium platform distance, as more differentiated platforms better stabilize political power.

This equilibrium characterization reveals a distinctive implication of the model: voter polarization benefits both parties. When voters to the left of the median move farther left and those to the right move farther right, equilibrium platforms move
farther apart and both parties gain, even if the two sides move by different amounts in an
asymmetric electorate. Economic change, identity, or political communication that
separates opposing groups of voters can therefore become a jointly valuable input into
party competition.

Strikingly, a party can benefit when voters who already oppose it become still more hostile. If a group lies to the left of the median and both platforms, pushing it farther
left draws the more left-wing platform after it. This wider platform gap helps the party
offering the more right-wing platform preserve the support of right wing voters in adverse states. That party loses
the distant group only in strongly favorable states, when additional votes are least
valuable, yet becomes better protected when conditions turn against it. \citet{GT23} show that parties may engage in polarizing campaigns provided that they radicalize their supporters more than the opponents. My results explain why a party may want to strategically alienate its opponents.

Changes in party organization translate into changes in platform polarization.
Different affiliates---officeholders, economic actors, and allied organizations---generate
different returns through governing, fundraising, patronage, or rent
extraction. If a few generate most of the party's returns, preserving enough influence to
empower this core becomes especially valuable. Holding total affiliate value fixed, such concentration pushes the
two platforms towards opposite ideological poles, increasing party
polarization.

With several policy dimensions, parties' platform choices define a direction in policy space along which they separate. I call the resulting axis the \textit{partisan cleavage}; it may correspond to a single issue or to a weighted combination of issues \citep{LipsetRokkan1967,Riker1986}. Its value to the parties depends not only on voter positions on each issue but also on how positions align across issues. Two electorates with identical marginal distributions of tax and immigration preferences may thus sustain different partisan cleavages: if views are cross-cutting, the blocs that form on one issue dissolve on the other. The model endogenizes the axis along which parties organize political competition.\footnote{In multidimensional electoral competition models, \citet{Mack2022} studies how parties rotate platforms to manage learning across repeated elections, whereas \citet{Vaeth2026} studies how voter learning produces issue alignment. Both feature policy-motivated parties whose primitive ideals anchor the equilibrium axis. Here, identical parties have no such ideals; equilibrium locations reflect voter disagreement.}

Comparing possible cleavages requires comparing the constituencies they organize. For centrally symmetric electorates, I show that the game admits an exact potential---a function whose change equals a deviating party's payoff change under every unilateral deviation. The potential establishes pure-strategy Nash equilibrium existence and reduces the comparison of cleavages to one of conflict values. To construct the conflict value of a cleavage, voters are ordered along that axis and equal-mass groups are paired from opposite ends of the ranking. The distance between the two groups in a pair is then weighted by how much more a party gains from adding a group of that mass to its coalition when it is weak rather than strong. In the equilibria both parties prefer, the partisan cleavage maximizes this conflict value; its maximum determines platform distance and the parties' common payoff. Under regularity conditions, the potential's maximizers are exactly the conflict-value-maximizing equilibria, hence the party-preferred equilibria.\footnote{On each finite platform grid, a standard noisy adjustment process selects the grid's potential maximizers as mistakes vanish. Online Appendix \ref{oa:app:logitselection} provides the grid-refinement qualification and proof.}

For voter distributions with elliptical contours, this criterion reduces to a familiar empirical object: the selected cleavage is the combination of issues on which voter positions vary most---the first principal component when that direction is unique. In the two-issue case, the alignment result takes a sharp form: holding the marginal distribution of views on each issue fixed, platform separation and both parties’ payoffs rise with the absolute correlation between views. Positive correlation bundles views in the same direction and negative correlation in opposing directions, but either alignment organizes the same citizens into opposing camps across both issues, increasing the parties’ ability to secure different segments of the electorate. Consistently, I show that, in the 2024 ANES, the first principal component of voter issue positions sharply sorts partisan identification and lies only \(3.0^\circ\)
from the direction of Trump--Harris position difference.\footnote{Voters are asked to locate both presidential candidates on the same issue scales.}

The closest political-economy comparison is \emph{The Race to the Base} \citep[henceforth BBH]{Buisseret}. In reduced form, the papers share a basic force: an increment of electoral support can be more valuable when it protects a weak party’s foothold than when it widens a strong party’s victory. BBH use this force to study one-dimensional competition within a fixed conflict structure: partisan attachments are inherited, and the parties face an exogenous expected popularity imbalance. Under their main assumptions, the force governs targeting along the inherited partisan ordering---the parties converge when the imbalance is small, the underdog retreats toward its base at intermediate imbalances, and a sufficiently advantaged party chases its rival into its ideological territory.\footnote{In BBH, electorally relevant voters follow a fixed uniform distribution and each party's payoff is piecewise linear in seat share, with weakly greater marginal value below the majority threshold than above it. Their Online Appendix also obtains divergence (toward the distinct primitive bases) at low popularity imbalance, when majority control is sufficiently unimportant. In their model, a platform-independent partisan-affinity term fixes voters' ordering by relative party appeal: platforms move the coalition boundary along that ordering but cannot change it. Switching thresholds therefore remain heterogeneous even at identical platforms, inducing convergence at some interior majority premia. Here, by contrast, the voter ordering is fully endogenous. Identical platforms give every group the same switching threshold, ruling out equilibrium convergence (Proposition \ref{divergence}). Separation creates heterogeneous thresholds and, with several issues, its direction determines the endogenous ordering that organizes voters into opposing constituencies.}

The central difference is the object of study. BBH characterize how competition moves along an inherited partisan map as the parties become increasingly asymmetric ex ante; here, the parties begin identical and competition draws the map itself from preexisting voter disagreements. Because voters carry no inherited party attachments, a protected constituency exists only if platform separation creates it. With several issues, the direction of separation determines which disagreement becomes the partisan divide; the exact potential makes alternative directions globally comparable, and the conflict value identifies the divide both parties prefer. This difference shifts the relevant comparative statics from ex ante popularity imbalances to the distribution of voter disagreement and the parties’ internal organization. Both parties can gain when opposing voter groups move apart, even when a group grows more hostile to one of them; top-heavy organizations widen the divide; and the party-preferred divide depends not only on disagreement within individual issues but also on how disagreements align across issues.

Formal applications in Online Appendix \ref{oa:appendixapplications} show how parties may engineer sharper divisions through issue emphasis and divisive rhetoric, political information, and partisan identity. Emphasizing an issue on which groups have opposing interests, revealing those opposing interests, or strengthening identity around an existing cleavage
makes the same voters more likely to remain on opposite sides of party competition,
benefiting both parties. Information about common-interest policies moves
voters together rather than organizing stable blocs and creates no comparable gain.

The platform polarization that benefits both parties can harm voters. Suppose implemented policy
averages the two platforms according to the parties' election-determined shares of
political influence. Under proportional power, parties that value each additional unit
equally converge to the utilitarian optimum. Diminishing organizational returns induce
platform differentiation, so electoral swings move policy between the two positions and
impose volatility on an unchanged electorate. Electoral rules can mitigate this cost: as
majority control determines nearly all such influence, platforms converge toward the
median voter and policy variance disappears. A maximal majority premium is first-best
when that voter's ideal point is utilitarian. Otherwise, reduced variance
accompanies residual policy bias, and some degree of proportionality may therefore improve welfare.

Taken together, the paper offers a portable framework for studying the supply side of
political polarization: rival parties compete over political power while sharing an interest in the
divisions that structure the fight. It delivers closed-form equilibrium
platforms with intuitive voter weights; squared platform distance is a sufficient
statistic for the parties' common gain from differentiation. The same structure maps
voter disagreement, party organization, political communication, electoral institutions,
and electoral uncertainty into polarization's extent, direction, and welfare consequences.

Section \ref{model} is the model setup. Section \ref{Conflict1D} studies single-issue platform competition. Section \ref{ConflictKD} studies the multidimensional problem,
endogenizing the partisan cleavage. Section \ref{applications} considers engineered divisions and electoral institutions. Section \ref{conclusion} concludes.

\section{Political Power and Party Organizations}\label{model}

The model treats each party as an organization. Elections determine the share of contestable political power that accrues to each party. Power is then optimally allocated among rent-generating affiliates. Separating votes, power, and its organizational uses allows me to isolate the role of voter disagreement, electoral rules, and organizational characteristics in shaping the equilibrium partisan conflict.

\subsection{Electoral environment}

Two parties, $A$ and $B$, choose platforms in a $K$-dimensional policy space. Voters are
grouped into finitely many types \(I=\{1,\ldots,N\}\). Type \(i\) has population share
\(s_i>0\), with \(\sum_{i\in I}s_i=1\), and policy bliss point
\(x_i\in\mathbb R^K\). Types sharing a bliss point are aggregated, so the indexed bliss
points are pairwise distinct; at least two types remain.\footnote{If the electorate were perfectly homogeneous, platform convergence would
obtain. The present paper, however, shows that parties gain from voter disagreements, and highlights supply-side strategies that sharpen such disagreements (Online Appendix \ref{oa:appendixapplications}). Through the lens of this model, electoral homogeneity is unlikely to persist.}
The utility of a type-\(i\) voter from policy $x\in\mathbb{R}^K$ is
\[
v(x,x_i)=\sum_{k\le K}h^k(|x^k-x^k_i|),
\]
where each $h^k:\mathbb{R}_+\to\mathbb{R}$ is differentiable, strictly concave,
and strictly decreasing in $|x^k-x^k_i|$, with \(h^{k\prime}(0)=0\). Thus policy
preferences are smooth and single-peaked: on every issue, utility falls with distance
from the bliss point.

The parties announce platforms $x_A,x_B\in\mathbb{R}^K$ simultaneously. Before voting sincerely, all voters observe a single popularity shock \(\epsilon\in\mathbb R\)
in favor of party \(B\).\footnote{With atomistic voters, sincere voting rules out arbitrary weakly
dominated behavior while keeping the focus on platform competition. The model has a
finite number of voter types, not a finite number of voters.} The shock captures common
electoral developments---economic news, political scandals, or shifts in candidate
appeal---and has a continuous cdf \(F\) with density \(f\), continuous at zero, symmetric
around zero, and satisfying \(f(0)>0\). Because every voter observes the same
realization, it exposes both parties to aggregate electoral movement.

For any pair of platforms $(x_A,x_B)$, define
\begin{equation*}
\Delta_i(x_A,x_B)=v(x_A,x_i)-v(x_B,x_i).
\end{equation*}
Thus $\Delta_i$ is party $A$'s policy advantage among type \(i\), whose voters support
\(A\) exactly when \(\Delta_i(x_A,x_B)\ge\epsilon\), an event with probability
\(\pi_i(x_A,x_B)=F(\Delta_i(x_A,x_B))\). Platform choice therefore determines
both expected support and its responsiveness to common news. As
standard in probabilistic-voting models, I assume that the support of \(F\) is wide enough
that no group is mechanically committed to a party for any pair of undominated
platforms.\footnote{Let
\(D=\prod_{k\le K}[\min_i x_i^k,\max_i x_i^k]\) and
\(M=\sup_{a,b\in D,\,i\in I}|v(a,x_i)-v(b,x_i)|\).
Formally, I assume that \([-M,M]\) lies in the interior of the support of \(F\).
A coordinatewise projection of a platform outside \(D\) onto \(D\) strictly improves its appeal to
every voter.}
For a realized shock, party \(A\)'s vote share is
\[
S_A(x_A,x_B,\epsilon)
=
\sum_{i\in I}s_i\mathbf 1\{\Delta_i(x_A,x_B)\ge\epsilon\},
\qquad
S_B(x_A,x_B,\epsilon)=1-S_A(x_A,x_B,\epsilon).
\]
The baseline omits voter-specific shocks so that this aggregate-risk channel remains
transparent. Online Appendix \ref{oa:Robustness} follows the mechanism as idiosyncratic noise
increases.

\subsection{Votes and political power}

Parties are office motivated: they value the capacity of their coalition and its
affiliates to occupy positions of influence in the political and economic system. Let
$\rho:[0,1]\to\mathbb{R}$ translate electoral support into a scalar index of contestable
political power. Separating votes from power matters because electoral institutions may
generate proportional allocations, seat bonuses, or discrete jumps when a party acquires
a majority, supermajority, or executive control.

\begin{assumption}\label{officerents}
The power allocation function $\rho$ satisfies two properties: (i) $\rho$ is strictly increasing on
$[0,1]$; (ii) there exists a constant $\bar \rho$ such that $\rho(s)+\rho(1-s)=\bar \rho$ for all
$s\in[0,1]$.
\end{assumption}

Assumption \ref{officerents} isolates the contestable component of political power:
electoral gains for one coalition are losses for the other. This component can include
legislative representation, committee and leadership positions, staff and agenda access,
executive and administrative appointments, policy implementation, and influence through
allied organizations. The map \(\rho\) need not be proportional or continuous; it can
combine a strictly increasing baseline with discrete majority, supermajority, or executive
premia. Presidential systems fit the model whenever the strategic actor values the executive
prize together with the wider coalition's legislative, administrative, and organizational
influence. 

\subsection{The organizational value of power}\label{organization}

Assumption \ref{officerents} specifies how elections divide contestable power between
the parties; its use within each party explains why the marginal organizational return
to power may decline. A party is a coalition of affiliated politicians, personnel,
activists, economic actors, and allied organizations. Political power allows these
affiliates to occupy, retain, or gain access to positions and channels of influence within
government and the broader political economy. Because affiliates differ in the political
or economic returns they can generate for the coalition, scarce influence is rationally allocated
first toward the affiliates generating the highest returns. Larger victories therefore extend influence to
progressively lower-return members of the coalition.

Let \(a=\rho(0)\), \(b=\rho(1)\), and
\[
r(s)=\frac{\rho(s)-a}{b-a},
\qquad
r(s)+r(1-s)=1,
\]
so \(r(s)\) is the fraction of contestable power controlled with vote share \(s\).
Normalize the party's affiliates to unit mass and index them by descending
organizational value, with rank \(k\in[0,1]\). Let \(q(k)>0\) denote the strictly
decreasing return generated for the party when the rank-\(k\) affiliate is empowered
through a position of influence. If the party controls power \(p\in[a,b]\),
it can empower the highest-value fraction \((p-a)/(b-a)\) of its affiliates. Its
normalized organizational value is therefore
\begin{equation}\label{assignmentvalue}
U_q(p)
=
\frac{\displaystyle
      \int_0^{(p-a)/(b-a)}q(k)\,dk}
     {\displaystyle
      \int_0^1 q(k)\,dk}.
\end{equation}
The numerator is the value generated by the affiliates empowered at power level \(p\);
the denominator is the value the coalition would generate if all its affiliates were
empowered. Thus \(U_q(p)\) is the share of total affiliate value activated by political
power \(p\). Because \(q\) is positive and strictly decreasing, \(U_q\) is increasing and
strictly concave. Diminishing returns arise naturally from empowering the most valuable
affiliates first.

This interpretation encompasses both party insiders placed in formal positions and
external allies whose access or influence expands with the party's power. Legislative
strength brings representation, committee and leadership positions, staff, and agenda
access, while executive strength brings appointments, administrative authority, and
influence through a wider partisan network. Closed lists make ranked placement
especially literal but are only one application
\citep{BuisseretFolkePratoRickne2022,SvitakovaSoltes2024,
ColonnelliPremTeso2020,Lewis2008PoliticsPresidentialAppointments}. The return \(q(k)\) may reflect an affiliate's contribution
to effective governing, policy implementation, fundraising, mobilization, or coalition
maintenance, as well as political or economic rents enabled by control of an office,
access channel, or network.\footnote{These returns are measured from the party's perspective
and need not coincide with candidate quality or social welfare.}

The main results of the paper require only an increasing, strictly concave value
function \(U\) over political power.

\begin{assumption}\label{concavity}
The value function
\(U:[\rho(0),\rho(1)]\to\mathbb{R}\) is continuous, strictly increasing, and strictly
concave.
\end{assumption}

Let \(\nu=U\circ\rho\) denote the reduced-form payoff from electoral support, normalized
so that \(\nu(1)-\nu(0)=1\). Electoral support \(s\) first becomes political power
\(\rho(s)\), and that power then produces organizational value \(U(\rho(s))\).
Electoral institutions determine \(\rho\), which may contain majority premia or other
threshold discontinuities, while party organization determines \(U\). Concavity is
imposed on the value of political power, so the induced payoff from vote share need not
be concave or continuous. Online Appendix \ref{oa:AppConc} develops alternative foundations
based on rent sharing between party members and organizational costs. There, I show that the concavity of \(U\) can also be obtained from party affiliates' decreasing marginal utility from consumption or convex organizational costs. 

Party \(p\)'s expected payoff at platform profile \((x_A,x_B)\) is therefore
\begin{equation}\label{expectedpayoff}
V_p(x_A,x_B)
=
\mathbb{E}_{\epsilon}\!\left[
\nu\!\left(S_p(x_A,x_B,\epsilon)\right)
\right],
\qquad
p\in\{A,B\}.
\end{equation}
A pure-strategy Nash equilibrium is a platform pair at which neither party can raise
this expected payoff by changing its own platform.

The key strategic implication follows by comparing a constituency that raises a
minority party's support from \(s\) to \(s'\) with the corresponding constituency that
raises a majority party's support from \(1-s'\) to \(1-s\). Constant-sum power makes the
two increments in political power equal, but strict concavity makes the increment
received by the weaker party more valuable.

\begin{remark}\label{remarknu}
For any \(s,s'\in[0,\tfrac{1}{2}]\) with \(s'>s\), it holds that
\[
\nu(s')-\nu(s)>\nu(1-s)-\nu(1-s').
\]
\end{remark}

Under ranked allocation, the same-sized increment in power empowers higher-ranked
affiliates when it preserves a party's foothold and lower-ranked affiliates when it
enlarges a landslide. Thus the same electoral bloc creates more organizational value by
protecting a weak party, even though political power itself is constant-sum.


\subsection{Differentiation as constituency insurance}

Convergence cannot be sustained in pure strategies. When platforms coincide, every voter
group switches at the same national shock, so either party can gain by creating a
constituency that remains with it in adverse electoral states.

\begin{proposition}[Party Divergence]\label{divergence}
If a pure-strategy equilibrium exists, the two parties offer different policy platforms.
\end{proposition}

Pure office motivation therefore fails to deliver Downsian convergence: the gain from
differentiation comes from constituency insurance, not necessarily from higher expected
votes. At a common platform, every voter faces the same comparison between the parties,
so all groups switch at \(\epsilon=0\) and party \(A\)'s political power forms a
cliff---\(\rho(1)\) when the shock favors it, \(\rho(0)\) when it does not
(Figure~\ref{fig:staircase}(a)). A small unilateral move by \(A\) turns the cliff into a
staircase of switching thresholds: voters closer to its new position remain through less
favorable shocks, while those on the opposite side leave under more favorable ones
(Figure~\ref{fig:staircase}(b)). The party gives up power where it is already strong to
retain a constituency where it would otherwise hold little, and diminishing
organizational returns make the retained power the more valuable of the
two.\footnote{The exchange is exact at a common platform that is locally optimal under
constant returns to power: a small move leaves expected political power unchanged to
first order, relinquishing as much in favorable states as it preserves in adverse ones,
so differentiation is profitable. At other common platforms the two quantities need not
balance, but a move in at least one direction along a coordinate with voter disagreement
is profitable; hence convergence is never a pure-strategy equilibrium.}


\StaircaseFigure

The force is more general than the foundation: divergence requires neither ranked
allocation nor concavity of the reduced-form payoff in vote share. The following
inequality, implied by Remark~\ref{remarknu}, is sufficient---dividing power between two
viable coalitions must create more combined organizational value than concentrating it
in one:
\[
\nu(s)+\nu(1-s)>\nu(1)+\nu(0)
\qquad\text{for every }s\in(0,1).
\]
The mechanism vanishes as \(U\) becomes linear; when \(\rho\) is the proportional-with-majority-premium
family described in Online Appendix~\ref{oa:appendixA4}, platform separation likewise diminishes as the majority prize approaches the entire contestable stock \(\bar{\rho}\).\footnote{Online Appendix~\ref{oa:app:convexconverse} also
gives a converse boundary result. If convergent platforms form an equilibrium when
parties value political power linearly, applying an increasing convex payoff
transformation to power preserves that convergent equilibrium.}

\section{How Much Conflict? The Unidimensional Case}\label{Conflict1D}

I first study competition along a given policy divide, such as redistribution or an
established left--right policy bundle. With \(K=1\), voters have a natural ordering (e.g., from left to right), so
the question is how sharply the parties differentiate and which groups anchor their
platforms along that ordering. To obtain a closed-form equilibrium, I specialize to quadratic policy loss and a uniform
popularity shock,
\[
h(|x-x_i|)=-\alpha(x-x_i)^2,
\qquad
\epsilon\sim U(-\phi,\phi),
\]
with $\alpha>0$. Rescaling the shock by \(\alpha\) replaces \(\phi\) by
\(\phi/\alpha\), so I set \(\alpha=1\) without loss of generality. 

Without affecting the equilibrium set, I restrict attention to platforms in
$D=[x_1,x_N]$.\footnote{Outside platforms are weakly dominated by their
projections onto $D$; if one were used in equilibrium, either the party could
profitably switch to its projection or its opponent could profitably match it.} Order voter types so that $x_1\le x_2\le\cdots\le x_N$. Let the median set contain the
points with at least one half of the electorate weakly on either side, and let \(x_m\)
denote its midpoint.\footnote{If the
cumulative mass equals one half in \([x_i,x_j]\), then
\(x_m=(x_i+x_j)/2\). Otherwise, \(x_m\) is the usual median type. } Before stating the equilibrium, it is useful to define the common
platform the parties would choose if organizational value were linear in political power:
\[
x^{RN}=\sum_{i=1}^N\frac{\rho(s_i+\sum_{j<i}s_j)-\rho(\sum_{j<i}s_j)}{\rho(1)-\rho(0)}x_i,
\]
with $s_0=0$. Its weights are the increments in political power obtained as voter groups
are added in ideological order; under proportional power it equals the population mean bliss point,
although with a general power rule it need not coincide with it. Linear
valuation sends both parties to this common electoral center, whereas organizational
concavity assigns additional policy influence to groups whose support protects a weak
party and pulls the platforms toward opposite sides of the electorate.

\begin{proposition}[Equilibrium]\label{NE1D}
A pure-strategy equilibrium exists. The following properties characterize the pure-strategy equilibria of the game.
\begin{enumerate}[(i)]
\item There exist two platforms $\ubar{x},\bar{x}\in\mathbb{R}$ with
$x_1<\ubar{x}<x^{RN}<\bar{x}<x_N$ such that, in every pure-strategy equilibrium, one party offers $\ubar{x}$ and the other offers $\bar{x}$. Each platform is a convex combination of voter bliss points,
\[
\ubar{x}=\sum_{i=1}^N\ubar{w}_ix_i,\qquad
\bar{x}=\sum_{i=1}^N\bar{w}_ix_i,
\]
where $\ubar{w}_i,\bar{w}_i\in(0,1)$ and weights sum to one.

\item The equilibrium weights reflect the marginal payoff from obtaining the support of
each voter group:
\[
\ubar{w}_i=\frac{\nu(\ubar{s}_i)-\nu(\ubar{s}_{i-1})}{\nu(1)-\nu(0)},\qquad
\bar{w}_i=\frac{\nu(\bar{s}_{i})-\nu(\bar{s}_{i+1})}{\nu(1)-\nu(0)},
\]
where $\ubar{s}_i=\sum_{j\le i}s_j$ and $\bar{s}_i=\sum_{j\ge i}s_j$ for each \(i=1,..., N\), and \(\ubar{s}_0=\bar{s}_{N+1}=0\).

\item One party assigns greater weight to voters to the left of the median, while the
other assigns greater weight to voters to the right. In particular,
\(\bar w_i>\ubar w_i\) when \(x_i>x_m\), and
\(\bar w_i<\ubar w_i\) when \(x_i<x_m\).

\item Both parties obtain the same equilibrium expected payoff, which is increasing in the squared distance between the two equilibrium platforms $(\bar{x}-\ubar{x})^2$.
\end{enumerate}
\end{proposition}

The weights describe how each party builds a constituency. The left-positioned platform
is constructed from left to right, so \(\ubar w_i\) is the organizational gain from adding
group \(i\) after securing every group farther left; the right-positioned platform is
constructed analogously from right to left. Each weight combines group size, the political
power generated by its votes, and the organizational return to that increment. Because an
equal increment matters more when the party is weak, mirrored voter groups receive
asymmetric weights, tilting the two platforms toward opposite sides of the electorate.

Equilibrium payoffs make the common return to this differentiation explicit:
\[
\hat V_A=\hat V_B
=
\underbrace{\frac{\nu(1)+\nu(0)}{2}}_{\substack{\text{convergence}\\\text{lottery payoff}}}
+\underbrace{\frac{(\bar x-\ubar x)^2}{2\phi}}_{\substack{\text{differentiation}\\\text{insurance gain}}}.
\]
The first term values the endpoint lottery over contestable power at convergence, while
the second measures the insurance gain. Platform distance is therefore a sufficient
statistic for such a gain.

This electoral protection has familiar empirical counterparts. Candidate-quality
differences matter more when U.S. House candidates are ideologically close
\citep{ButticeStone2012}, while voters punish presidential incumbents less for adverse
economic performance in more polarized environments \citep{Moreira2025}.

Online Appendix \ref{oa:Robustness} studies the equilibrium as we add idiosyncratic voter noise. The
original platform gap is
initially unchanged, then contracts, and finally disappears for large idiosyncratic noise. Polarization is therefore stronger when electoral movement is synchronized by common forces, suggesting a connection to the nationalization of electoral behavior \citep{Hopkins2018}.

\subsection{Voter polarization}\label{voterseparation}

I now study how equilibrium payoffs change with voter conflict. Compare two electorates after centering each at its median. In the second, the
voters to the left of the median have shifted farther left and those to its right farther
right, so that mass has shifted from the (relative) center to the tails of the electorate, potentially in asymmetric ways. I call the second electorate ``more polarized" as voters in opposite halves of the spectrum now are, on average, farther apart than in the first electorate. The stochastic order below captures this outward separation.

For a voter distribution $(\mathbf{x},\mathbf{s})$, let
$\hat V_p(\mathbf{x},\mathbf{s})$ denote party $p$'s equilibrium expected payoff.

\begin{definition}\label{defspread}
For a given voter distribution $(\mathbf{x},\mathbf{s})$, split any mass at the median
between two half-electorates so that each has mass one half. After normalizing each
half-electorate to have unit mass, let
$(\mathbf{x}_L,\mathbf{s}_L)$ and $(\mathbf{x}_R,\mathbf{s}_R)$ denote the resulting left
and right distributions. A distribution $(\mathbf{x}',\mathbf{s}')$ is a spread of
$(\mathbf{x},\mathbf{s})$ if
$(\mathbf{x}_L,\mathbf{s}_L)$ first-order stochastically dominates
$(\mathbf{x}'_L,\mathbf{s}'_L)$ and $(\mathbf{x}'_R,\mathbf{s}'_R)$ first-order
stochastically dominates $(\mathbf{x}_R,\mathbf{s}_R)$, with at least one relation strict.
\end{definition}

A spread thus increases disagreement between voters on opposite sides of the median. For
any distribution \((\textbf{x},\textbf{s})\), let
\(\textbf{x}_m\in\mathbb{R}^{I}\) be the vector with all entries equal to
\(x_m\).

\begin{proposition}[Voter polarization]\label{polarization}
Fix \(\nu\) and \(\phi\), and suppose the wide-support condition holds for both electorates.
Let $(\mathbf{x}',\mathbf{s}')$ and $(\mathbf{x},\mathbf{s})$ be two voter distributions. If
$(\mathbf{x}'-\mathbf{x}'_m,\mathbf{s}')$ is a spread of
$(\mathbf{x}-\mathbf{x}_m,\mathbf{s})$, then
\[
\bar x'-\ubar x'>\bar x-\ubar x
\quad\text{and}\quad
\hat V_p(\mathbf{x}',\mathbf{s}')>\hat V_p(\mathbf{x},\mathbf{s}),
\qquad p=A,B.
\]
\end{proposition}

The two platforms overweight opposite halves of the electorate, so each responds more strongly to movement in its own half: the same outward movement widens their distance. Greater voter polarization therefore generates greater platform differentiation without any change in party objectives or electoral institutions. Both parties gain through better constituency insurance: the wider platform gap further stabilizes each party's foothold, making sharper conflict a jointly valuable input into party competition.

\SpreadAlienationFigure

\subsection{Voter alienation}\label{alienpolar1D}

A party can profit from alienating those who oppose it. When a group already hostile to
party \(A\) radicalizes, moving farther toward the rival's side, \(A\) loses votes it
would have collected only in landslides. In equilibrium, however, the rival chases the
radicalizing group, abandoning the contested ground next to \(A\)'s equilibrium constituency; the
insurance this creates can be worth more than the lost votes. Let party \(A\) offer the
right platform \(\bar{x}\) and party \(B\) the left platform \(\ubar{x}\). When a small
movement of a group raises a party's equilibrium payoff, I call the gain
\emph{attraction} if the group becomes more likely to support that party, and
\emph{alienation} if it becomes less likely to do so.

\begin{proposition}[Alienation]\label{alienation}
Let \(x_m\) denote the median voter type. Hold voter shares fixed and follow the equilibrium branch
on which \(A\) remains the right-hand party. Party $A$ benefits from attracting group
$i$ if
\(
x_i>\max\{\ubar{x}(\mathbf{x},\mathbf{s}),x_m\},
\) and from alienating it if
\(
x_i<\min\{\ubar{x}(\mathbf{x},\mathbf{s}),x_m\}.
\)
Symmetrically, party $B$ benefits from attracting group $i$ if
\(
x_i<\min\{\bar{x}(\mathbf{x},\mathbf{s}),x_m\},
\)
and from alienating it if
\(
x_i>\max\{\bar{x}(\mathbf{x},\mathbf{s}),x_m\}.
\)
\end{proposition}

Figure~\ref{fig:spreadalienation}(b) illustrates. As the most left-wing group moves
farther left, both platforms follow it, but the group carries more weight in \(B\)'s
constituency, so \(B\) moves farther and the platform gap widens. Symmetrically, \(B\)
gains when the most right-wing group moves farther right even as its support for \(B\)
declines. The calculation for this case makes the decomposition explicit. Consider a
marginal increase in $x_N$, the bliss point of the most right-wing voter group.\footnote{With a small abuse of notation, I
suppress the electorate arguments of \(V_B\) in the derivatives below.} Since in equilibrium
$x_A=\bar{x}(\mathbf{x},\mathbf{s})$ and
$x_B=\ubar{x}(\mathbf{x},\mathbf{s})$, the total equilibrium derivative is
\begin{equation}\label{derivative}
\frac{d\hat V_B(\mathbf{x},\mathbf{s})}{d x_N}
=
\underbrace{\frac{\partial V_B(\bar{x},\ubar{x})}{\partial x_N}}_{\substack{\text{support lost in}\\\text{favorable states}<0}}
+
\underbrace{\frac{\partial V_B(\bar{x},\ubar{x})}{\partial \bar{x}}
\frac{\partial \bar{x}(\mathbf{x},\mathbf{s})}{\partial x_N}}_{\substack{\text{protection created}\\\text{by the rival}>0}}.
\end{equation}
Party $B$'s own platform response drops out by the envelope theorem.

The first term is ordinary vote accounting: as group \(N\) moves right, its support for
\(B\) falls in states in which \(B\) is already strong. The second term is the
equilibrium response: party \(A\) moves right by
\[
\frac{\partial \bar{x}(\mathbf{x},\mathbf{s})}{\partial x_N}=\bar w_N>0,
\]
away from the groups most disposed to support \(B\). Proposition \ref{alienation}
identifies when the protection created by this response exceeds the value of the lost
votes.

The spread result ranks entire electorates; the alienation result isolates a single
group's movement, from which both parties can gain even as the group becomes less
likely to support one of them. Once the rival's platform response is included, a
party's payoff and a group's support for it need not move together.


\paragraph{Organizational concentration.}
The preceding comparative statics hold party organization fixed, but the assignment
foundation also links internal organization to platforms. Recall that affiliate value
measures the returns an affiliate generates for the party
(Section~\ref{organization}); ranking affiliates by value thus defines the party's internal
hierarchy---who is served first when power is scarce. Hold total value fixed and make
the distribution more top-heavy, so that every top fraction of affiliates accounts for
a weakly larger share: the hierarchy steepens, as when a leadership circle, a few major
donors, or an allied machine generates most of the party's rents. The first units of
political power then secure especially large returns, while influence added after a
landslide yields little, so the party values constituency insurance more.

Proposition~\ref{orgconcentration} in the appendix formalizes the comparison: holding
voter views and electoral rules fixed, greater concentration widens the platform
gap and raises both payoffs. Voter polarization, group alienation, and
organizational concentration thus alter different primitives---the distribution of
disagreement, a group's political position, and the value of power---while
strengthening the same demand for protected constituencies.

\section{Choosing the Political Divide}\label{ConflictKD}
I now turn to multidimensional politics, \(K\ge 2\). With one policy issue, voters have a natural ordering before the parties choose their
platforms: they can be ranked from left to right. Several issues remove that ordering:
a voter who lies near one party on redistribution may lie near the other on
immigration, so the direction of the platform contrast determines how positions bundle
into rival camps. Platform choices thus determine not only how far apart the parties
stand but also which social disagreement they stand apart on. Which of society's many
disagreements becomes the partisan divide?

For this section, voters have quadratic Euclidean policy loss,
\[
v(x,x_i)=-\|x-x_i\|^2,
\]
the popularity shock is uniform on $[-\phi,\phi]$, and the electorate is centrally
symmetric around $\mu\in\mathbb R^K$.\footnote{For a finite electorate, this means that
for every type $i$ there is a type $\bar i$ such that $x_{\bar i}=2\mu-x_i$ and
$s_{\bar i}=s_i$.} Central symmetry is what disciplines comparisons across directions:
it lets a single function---an exact potential, constructed below---record both
parties' incentives, so that platform contrasts that reorder the electorate can be
compared globally.\footnote{The role of symmetry is related in spirit to \citet{Plott1967}, although the strategic
environment and the potential argument developed in this section are different.}

\subsection{Comparing political divides: exact potential and equilibrium}\label{NEKD}
Call the axis through the two platforms the \emph{partisan cleavage}. For any proposed
cleavage, projecting voters onto it reproduces the unidimensional characterization of
Proposition~\ref{NE1D}. The difficulty arises when a party rotates the platform
contrast: a rotation can reorder the entire electorate and assemble different
coalitions, so a platform pair optimal against one ordering may invite a profitable
deviation toward a direction that organizes a more valuable division. Ruling out every
such deviation requires a global representation of the parties' incentives.

Equilibrium existence is part of the same problem. Payoffs are not quasi-concave in a
party's own platform---already in one dimension the payoff has a local peak on each
side of the rival, and majority premia add discontinuities in \(\nu\)---so standard
fixed-point arguments do not apply directly. The exact potential resolves both difficulties at once: its global maximizers are
pure-strategy equilibria, and its values rank the divisions that alternative directions
organize.

Let $\mathcal V$ denote the set of centrally symmetric finite voter distributions, and let
\[
D=\mathop{\times}_{k=1}^K
\left[\min_i x_i^k,\max_i x_i^k\right]
\]
be the rectangle spanned by voter bliss points. As shown in the appendix, restricting the platform choice to \(D\) is without loss for the Nash equilibrium analysis.\footnote{Under the uniform shock, the wide-support assumption of
Section~\ref{model} reads
\begin{equation}\label{wideuniform}
\phi> \sup_{a,b\in D,\ i\in I}|\Delta_i(a,b)|,
\end{equation}
for which $\phi>\operatorname{diam}(D)^2$ is sufficient. Every policy advantage arising
on \(D^2\) then lies on the affine region of the shock distribution, so
\(\pi_i = \tfrac12 + \Delta_i/(2\phi)\) throughout.}

The platform game allows for a common representation of the parties' incentives. The
direction-dependent value of separation enters both payoffs symmetrically, so it
cancels from their difference, which reduces to a contest in proximity to the
electorate's center:
\begin{equation}\label{payoffdifference}
V_A(x_A,x_B)-V_B(x_A,x_B)
=
\frac{\|x_B-\mu\|^2-\|x_A-\mu\|^2}{\phi}.
\end{equation}
Subtracting from each payoff the rival's proximity term therefore makes the two
coincide:
\begin{equation}\label{potentialdefinition}
\Phi(x_A,x_B)
:=
V_A(x_A,x_B)-\frac{\|x_B-\mu\|^2}{\phi}
=
V_B(x_A,x_B)-\frac{\|x_A-\mu\|^2}{\phi}.
\end{equation}
When \(A\) deviates, the term subtracted in the first representation stays fixed, so
\(\Phi\) changes by exactly \(A\)'s payoff change; the second representation gives the
same for \(B\). Thus \(\Phi\) is an exact potential for the platform game: one function records both parties' incentives.

\begin{proposition}[Exact potential and equilibrium]\label{potentialgame}
For every electorate in $\mathcal V$, the game on $D^2$ admits the exact potential
$\Phi$ in \eqref{potentialdefinition}: every
unilateral deviation changes the deviating party's payoff by exactly the change in
$\Phi$. Its Nash equilibrium set coincides with that of the unrestricted game on
$(\mathbb R^K)^2$. Consequently, every global maximizer of $\Phi$ on $D^2$ is an
unrestricted Nash equilibrium, and a pure-strategy Nash equilibrium exists.
\end{proposition}

Central symmetry thus places the parties' unilateral incentives on a common scale even
though political power remains constant-sum. A global maximizer of \(\Phi\) rules out
every unilateral deviation and directly compares the coalitions generated by different
platform directions, establishing pure-strategy existence.\footnote{\citet{AcharyaDuggan2026} establish pure-strategy existence in a multidimensional winner-take-all model with policy and office motives, using log-concavity and Kakutani. Here log-concavity fails, so existence is established using an exact potential; the potential also ranks endogenous cleavage directions.}

The game has multiple equilibria. I call a Nash equilibrium \textit{party-preferred} if it maximizes each party's payoff among Nash equilibria. I focus on party-preferred equilibria for the rest of the analysis.\footnote{For a finite electorate, imposing strict concavity of the
reduced-form payoff \(\nu\) turns the search for these equilibria into a finite one. For this case, Proposition~\ref{existence} in
Appendix~\ref{app:finitecharacterization} shows that the party-preferred equilibria are
exactly the global maximizers of \(\Phi\)---the equilibria with the
greatest platform distance.} The selection has a dynamic foundation: on any finite platform grid, parties revising
platforms by noisy best responses spend almost all time at the grid's potential
maximizers as mistakes vanish, and along increasingly fine grids these profiles
approach the party-preferred equilibria (Online Appendix \ref{oa:app:logitselection}).

\subsection{The endogenous cleavage}\label{conflictfunctional}
The potential makes alternative political divides comparable; the remaining question is
what makes one cleavage more valuable than another. A platform contrast over
redistribution may assemble one pair of constituencies, while a contrast over
immigration or a bundle of economic and cultural positions may assemble another.
The unidimensional analysis supplied the ingredients needed to answer this question: separation along a direction is worth more when voters lie farther apart along it, and when the groups it splits generate large payoff increments as they join each coalition.

For the characterization below, I pass to a continuous electorate: bliss points follow
an absolutely continuous distribution \(\mathcal P\) with compact support, centrally
symmetric around \(\mu\), the mean bliss point. I
retain quadratic policy loss and the uniform shock, and majority, supermajority, and
other interior threshold premia remain allowed. Beyond this, the appendix imposes only
two regularity conditions.\footnote{In particular, every one-dimensional projection of \(\mathcal P\) has a
strictly increasing cdf between the endpoints of its support, and \(\rho\) is
continuous at \(0\) and \(1\).} Let \(R\) be the radius of the smallest closed ball
around \(\mu\) containing the support, denoted \(B_R(\mu)\). The ball replaces \(D\).

Interpret \(e\) as the direction from \(B\)'s platform toward \(A\)'s. For a voter with
bliss point \(x\), the scalar \(e^\top(x-\mu)\) is the voter's signed coordinate along
this axis. Let \(Q_e(u)\) denote the \(u\)-quantile of the distribution of these
coordinates. As the popularity shock moves in \(A\)'s favor, voters enter its coalition
from highest to lowest rank, so when the rank-\(u\) voter joins, a share \(1-u\)
already supports \(A\). The measure \(M_\nu\) distributes across ranks the payoff
gains generated as \(A\)'s coalition expands.\footnote{Formally, take the
right-continuous version of \(\tilde\nu(s)=\nu(s)-\nu(0)\), let \(\Lambda_\nu\) be the
Lebesgue--Stieltjes measure satisfying
\(\Lambda_\nu((a,b])=\tilde\nu(b)-\tilde\nu(a)\) and
\(\Lambda_\nu(\{0\})=0\), and define
\(M_\nu(B)=\Lambda_\nu(\{s\in[0,1]:1-s\in B\})\). Under the normalization
\(\nu(1)-\nu(0)=1\), \(M_\nu\) is a probability measure. The appendix states the
endpoint regularity conditions.} The conflict value of direction \(e\) is
\begin{equation}\label{conflictindex}
\mathcal C(e)\equiv\mathcal C_{\nu,\mathcal P}(e)
=
\int_0^1 Q_e(u)\,dM_\nu(u).
\end{equation}
If \(\nu\) is absolutely continuous, \(dM_\nu(u)=\nu'(1-u)\,du\); a jump at vote
share \(s_0\) instead places an atom at rank \(1-s_0\). A voter at a high rank
\(u>1/2\) enters when \(A\) is weak, with support \(1-u<1/2\), whereas the voter at
the reflected rank enters when \(A\) is strong, with support \(u>1/2\). Central
symmetry gives \(Q_e(1-u)=-Q_e(u)\), while Remark~\ref{remarknu} implies that every
upper projected tail receives more \(M_\nu\)-weight than its reflected lower tail.
Integrating these tail imbalances yields \(\mathcal C(e)>0\) for every nondegenerate
direction, and its variation across directions ranks alternative divides.

This rank-weighted value reduces cleavage selection to one transparent tradeoff. For
\((x_A,x_B)\in B_R(\mu)^2\) with \(x_A\ne x_B\), let
\(m=(x_A+x_B)/2\), \(d=\|x_A-x_B\|\), and \(e=(x_A-x_B)/d\). The exact potential can
then be written as
\begin{equation}\label{potentialdecomposition}
\Phi(x_A,x_B)
=
\frac{\nu(1)+\nu(0)}{2}
-\frac{\|m-\mu\|^2}{\phi}
-\frac{d^2}{4\phi}
+\frac{d}{\phi}\mathcal C(e).
\end{equation}
The midpoint enters only as a centering cost. For fixed \(e\), the gap enters twice:
quadratically, through the electoral-positioning cost \(d^2/(4\phi)\), and linearly,
through the constituency-insurance gain \(d\,\mathcal C(e)/\phi\). Maximizing the potential therefore centers the platform pair at \(\mu\), sets the gap where the marginal insurance benefit equals the marginal positioning
cost---\(d=2\,\mathcal C(e)\)---and organizes conflict along a direction that maximizes
the conflict value.

\begin{proposition}[Endogenous political cleavage]\label{directionconflict}
Under the conditions above, define
\[
\mathcal D^\star
:=
\arg\max_{\|e\|=1}\mathcal C(e),
\qquad
\mathcal C^\star
:=
\max_{\|e\|=1}\mathcal C(e).
\]
A platform profile is a global maximizer of the potential on \(B_R(\mu)^2\) if and only
if, up to party labels, it has the form
\[
x_A^\star=\mu+\mathcal C^\star e^\star,
\qquad
x_B^\star=\mu-\mathcal C^\star e^\star
\qquad\text{for some }e^\star\in\mathcal D^\star.
\]
Every such profile is also a Nash equilibrium of the unrestricted game. More generally,
an unrestricted Nash equilibrium is party-preferred if and only if its cleavage direction
belongs to \(\mathcal D^\star\). At every party-preferred equilibrium,
\[
\|x_A-x_B\|=2\mathcal C^\star,
\qquad
\widehat V_p^\star
=
\frac{\nu(1)+\nu(0)}2
+\frac{2(\mathcal C^\star)^2}{\phi},
\qquad p=A,B.
\]
For fixed \(\nu\) and \(\phi\), both parties' maximal equilibrium payoffs therefore rise
with the maximal conflict value.
\end{proposition}

The proposition separates intensity from content: \(\mathcal D^\star\) identifies the
party-preferred cleavage directions, while \(2\mathcal C^\star\) determines platform
distance and hence the parties' common payoff. Party-preferred equilibria may also include
off-center pairs, shifted together orthogonally to the cleavage; these preserve distance
and payoffs but do not maximize the potential. Concavity of \(\nu\) rules them out, making
the party-preferred equilibria exactly the global maximizers.\footnote{The same equality holds, for every \(\rho_m\in[0,\bar\rho)\), under the proportional-with-majority-premium rule
\(\rho_{\rho_m}\) in Online Appendix \ref{oa:appendixA4}, equation
\eqref{oa:eq:rhopremfam}, even though the resulting \(\nu\) is discontinuous. For the
right-continuous version used to define \(M_\nu\), separate the payoff jump \(J_m\) at one
half from the continuous part. The latter is concave, while the jump contributes
\(J_mQ_e(1/2)=0\) to \(\mathcal C(e)\), because every centered projection of the
maintained continuous, centrally symmetric electorate has median zero. The appendix
gives the support-function argument.} Under
ellipticity, the same conclusion holds without concavity of \(\nu\), and the preferred
cleavage is a leading principal component (Theorem~\ref{pcatheorem}).

\subsection{Elliptical electorates and the dominant political cleavage}
\label{elliptical}

For an elliptical electorate, the cleavage at a party-preferred equilibrium is a
familiar empirical object: a leading principal component of voters' policy
positions---the first principal component when the largest eigenvalue is simple. Under ellipticity every centered projection has the same shape and differs only in scale, so the conflict value splits into a multiplier common to all directions and the projected standard deviation. Comparing directions then reduces to comparing projected variances, and the most valuable divide runs along the direction in which voters vary most.

Formally, suppose that $\mathcal P$ is elliptical with center $\mu$ and positive-definite
covariance matrix $\Sigma$. Every centered projection has a common standardized
distribution $H$:
\begin{equation}\label{ellipticalquantile}
Q_e(u)
=
\sqrt{e^\top\Sigma e}\,H^{-1}(u).
\end{equation}
Define
\begin{equation}\label{kappadefinition}
\kappa_{\nu,H}
=
\int_0^1H^{-1}(u)\,dM_\nu(u).
\end{equation}
Remark \ref{remarknu} implies $\kappa_{\nu,H}>0$. Substitution in
\eqref{conflictindex} yields
\[
\mathcal C_{\nu,\mathcal P}(e)
=
\kappa_{\nu,H}\sqrt{e^\top\Sigma e}.
\]

Under ellipticity, the directional characterization strengthens to a profile
characterization: the party-preferred equilibria are exactly the global potential
maximizers on \(B_R(\mu)^2\).

\begin{theorem}[Principal-component cleavage]\label{pcatheorem}
Maintain all conditions of Proposition \ref{directionconflict}, and suppose that
\(\mathcal P\) is elliptical with positive-definite covariance matrix \(\Sigma\). Let
\(\lambda_1\) be the largest eigenvalue of \(\Sigma\), and define the unit sphere in its
leading eigenspace by
\[
\mathcal L_1
=
\{e:\|e\|=1,\ \Sigma e=\lambda_1e\}.
\]
The party-preferred equilibria are exactly, up to party labels, the profiles of the form
\[
x_A^\star=\mu+\kappa_{\nu,H}\sqrt{\lambda_1}\,e_1,
\qquad
x_B^\star=\mu-\kappa_{\nu,H}\sqrt{\lambda_1}\,e_1,
\qquad e_1\in\mathcal L_1.
\]
Consequently,
\[
\|x_A^\star-x_B^\star\|
=2\kappa_{\nu,H}\sqrt{\lambda_1},
\qquad
\widehat V_p^\star
=
\frac{\nu(1)+\nu(0)}2
+\frac{2\kappa_{\nu,H}^2}{\phi}\lambda_1.
\]
If $\lambda_1$ is simple, the partisan cleavage (i.e., the axis on which the parties locate) is unique up to party labels
and equals the first principal component of voter disagreement.
\end{theorem}

The covariance matrix determines which disagreement organizes party competition and,
through \(\sqrt{\lambda_1}\), scales the distance between the platforms.
Organizational values, electoral institutions, and the standardized shape of the
electorate enter only through the common multiplier \(\kappa_{\nu,H}\). The
factorization thus separates two empirical questions: electorates that share a
covariance structure polarize along the same axis whatever their parties'
organization, while organization and institutions set the width of the divide,
amplifying the geometric scale \(\sqrt{\lambda_1}\).

\paragraph{The theorem in the data.}
Under ellipticity, Theorem~1 predicts that the parties separate along the
direction in which voter positions vary most---the electorate's first
principal component. I construct this direction in the 2024 ANES survey data as the first principal component of respondents' positions
on seven issues,\footnote{The
sample consists of respondents covered by the final combined pre-election
weight for the fresh web and in-person samples; paper-questionnaire
respondents are excluded. The seven 1--7 self-placement scales are services
and spending, defense spending, medical insurance, abortion, jobs and income,
assistance to Black Americans, and environmental regulation; services and
spending is reversed so that higher values are conservative on every scale,
and responses outside 1--7 are treated as missing.} computed from survey-weighted covariances without using party
identification or candidate placements. The component captures 58.5 percent of the
positions' total weighted variance. It is almost exactly the direction on
which voters see the candidates dividing. For each issue, each voter is asked to place Trump and Harris on the same scale; I compute the
survey-weighted mean placements of the two candidates and use that as an estimate of the corresponding party platforms.\footnote{Candidate placements
are reported by voters rather than independently coded and may partly reflect
partisan projection, so I treat the exercise as descriptive.} The direction through the platforms thus obtained forms an angle of only \(3.0^\circ\) with the voter
direction (95 percent bootstrap interval: \(2.1^\circ\)--\(4.9^\circ\)). The same direction also divides voters by party. Scoring each
respondent by her position along it, the Republican share among Democratic and
Republican identifiers, including leaners, is 10.3 percent among respondents
below the weighted mean score and 83.8 percent among those above it, and it
rises from near zero to near one across deciles of the score; a linear
regression of Republican identification on the score yields \(R^2=.555\)
(\([.511,.599]\)). A direction built from voter disagreement alone thus
recovers the two objects the theorem links: the (perceived) axis of candidate
competition and the partisan sorting of the electorate.\footnote{The PCA sample contains 1,954 respondents. Democratic
identifiers are PID categories 1--3 and Republicans categories 5--7; other
respondents remain in the PCA and cutoff calculations but are excluded from
the partisan shares, which use 1,838 respondents (splitting at the weighted
median instead of the mean gives 9.6 versus 81.6 percent). The gap vector
\(q\) collects, for each issue, the survey-weighted mean Trump-minus-Harris
placement among the 1,836 PCA-sample respondents who place both candidates on
all seven issues; the reported angle is
\(\arccos\!\left(\lvert e_1^\top q\rvert/\lVert q\rVert\right)\). Intervals
are Rao--Wu bootstrap intervals based on the ANES variance strata and PSU
variables, repeating the entire construction in each replication. Source:
ANES 2024 Time Series Study Full Release, May 19, 2026 version.}

\subsection{Correlated conflict across issues}\label{correlatedconflict}

The principal-component result distinguishes disagreement on individual issues from the
alignment of disagreement across issues. Consider two electorates with the same marginal
distribution of views on each of two issues. In one, positions on the two issues are
uncorrelated. In the other, positive correlation makes voters who stand on the same side
of one issue tend to stand together on the other. No individual issue is more polarized
in the second electorate, but a single diagonal platform contrast can separate two more
coherent blocs. The covariance matrix makes this comparison precise.
Write
\[
\Sigma
=
\begin{pmatrix}
\sigma_1^2 & \varrho\sigma_1\sigma_2\\
\varrho\sigma_1\sigma_2 & \sigma_2^2
\end{pmatrix}.
\]
Its largest eigenvalue is
\begin{equation}\label{leadingeigenvalue}
\lambda_1(\varrho)
=
\frac{\sigma_1^2+\sigma_2^2+
\sqrt{(\sigma_1^2-\sigma_2^2)^2+
4\varrho^2\sigma_1^2\sigma_2^2}}{2}.
\end{equation}

\CohesionFigure

\begin{corollary}[Correlation across issues]\label{correlationresult}
Hold fixed the center \(\mu\), the standardized projection distribution $H$, the positive
marginal variances, $\phi$, and $\nu$, with $|\varrho|<1$, and restrict attention to the
compared elliptical environments satisfying the wide-support condition
\(\phi>4R(\varrho)^2\), where \(R(\varrho)\) is the support radius. A strict
increase in $|\varrho|$ increases the distance between the party-preferred platforms and
raises both parties' equilibrium payoffs.
\end{corollary}

Absolute correlation raises the leading eigenvalue, causing the parties to move farther
apart and increasing both payoffs. Positive correlation places the two issues in the
selected direction with the same signs, whereas negative correlation places them there
with opposite signs; when the marginal variances do not coincide, correlation also changes the
precise angle of the cleavage.

Within this fixed-\(H\) elliptical family, each issue's marginal distribution is held
fixed, so the additional polarization comes entirely from correlation: the same
citizens take opposite sides issue after issue. A single platform contrast can then
protect the same constituency on several fronts at once. This is the distinction the
empirical literature draws between issue-level extremity and the alignment of opinions
into coherent bundles \citep{BaldassarriGelman2008,DellaPosta2020}; Corollary
\ref{correlationresult} supplies the strategic meaning of the second: aligned blocs
provide more valuable constituency insurance, so platform separation widens and both
parties gain.\footnote{\citet{BertrandKamenica2023} measure cultural distance between
demographic groups; \citet{Gentzkow2019} document the partisan organization of
congressional speech---the elite-side counterpart of the voter alignment studied
here.}

Outside the elliptical class, the shape of the projected distribution varies with the
direction, so different directions place different masses at the politically valuable
ranks. A four-type example in Online Appendix~\ref{oa:nonellipticexample} illustrates this point: it holds the electorate, and hence its covariance matrix, fixed while making affiliate
value more top-heavy. The first principal component does not move, yet the preferred
cleavage rotates toward a small, remote group whose loyalty is worth protecting in the
party's worst states. Internal organization can therefore determine which social
disagreement becomes political.\footnote{\citet{Mack2022} also identifies an axis of
party conflict in an electoral competition model; there, the axis is the difference
between the exogenous ideal points of policy-motivated parties.}

\subsection{Voter polarization beyond one dimension}
\label{multidimensionalspread}

The preceding results identify how the partisan cleavage emerges from a given electorate. The next comparison asks whether both parties continue to gain when citizens move farther apart
along that cleavage, even if the change ultimately makes another axis preferable. The potential allows us to establish the result without ellipticity.

\MultidimensionalSpreadFigure

\begin{definition}[\(q\)-spread] Let \(q\in\mathbb R^K\). A centrally symmetric finite electorate
\((\tilde{\mathbf x},\tilde{\mathbf s})\), with center \(\tilde\mu\), is a
\(q\)-spread of \((\mathbf x,\mathbf s)\), with center \(\mu\), if the projected
distribution
\(
\big((\tilde x_i-\tilde\mu)^\top q,\tilde s_i\big)_i
\)
is a spread, in the sense of Definition \ref{defspread}, of
\(
\big((x_i-\mu)^\top q,s_i\big)_i.
\)
\end{definition}

A \(q\)-spread sharpens disagreement along \(q\). For a finite electorate, let \(\hat V_p(\mathbf x,\mathbf s)\) denote party \(p\)'s payoff
at a party-preferred equilibrium. The finite characterization in Appendix
\ref{app:finitecharacterization} establishes that, when the reduced-form payoff \(\nu\)
is strictly concave, these equilibria are precisely the global maximizers
of \(\Phi\), and they yield the same payoff.

\begin{proposition}[Voter polarization]\label{propspreadKD}
Consider two centrally symmetric finite electorates
\((\mathbf x,\mathbf s)\) and
\((\tilde{\mathbf x},\tilde{\mathbf s})\). Hold \(\nu\) and \(\phi\) fixed, let
\(\nu\) be strictly concave, and suppose the wide-support condition
\eqref{wideuniform} holds on a common coordinate rectangle containing both centered
supports, for every voter type in either electorate. Fix a party-preferred equilibrium
\(x^\star\) of \((\mathbf x,\mathbf s)\) and let
\(q^\star=x_A^\star-x_B^\star\). If
\((\tilde{\mathbf x},\tilde{\mathbf s})\) is a \(q^\star\)-spread of
\((\mathbf x,\mathbf s)\), then
\[
\hat V_p(\tilde{\mathbf x},\tilde{\mathbf s})
>
\hat V_p(\mathbf x,\mathbf s),
\qquad p=A,B.
\]
\end{proposition}

Evaluated at the original party-preferred platforms, greater separation along the selected
projection raises the new electorate's potential by the unidimensional argument. Its own
party-preferred equilibrium attains at least that value. Both parties therefore gain when
political conflict sharpens along an equilibrium divide, even for a nonelliptical
electorate in which the preferred direction subsequently changes.

\section{Engineering Divisions, Institutions, and Welfare}\label{applications}

Competing parties mutually benefit from voter divisions. This section turns to political forces that can create or reinforce those
divisions---issue emphasis, information, and partisan identity---and to electoral
institutions that govern the return to protecting a constituency. Online Appendix
Sections \ref{oa:appendixapplications} and \ref{oa:appendixA4} provide the formal analysis.

\subsection{Engineering political divisions}

Parties help determine whether citizens understand a policy as a common-interest problem or as
a conflict between groups. When citizens share an ideal policy, both parties converge
on it; when group interests oppose one another, platforms separate, and raising the
political salience of the conflict issue raises both parties' payoffs (Online
Appendix~\ref{oa:app:issues}). An issue can therefore become politically salient without
being especially important for social welfare: its political value comes from placing citizens on opposite sides. This return differs from the priming and signaling mechanisms in models of divisive issue selection
\citep{Aragones2015,AshMorelliVanWeelden2017} and, unlike mobilization-based accounts
of strategic extremism \citep{Gleaser2005}, requires neither inherited party followings,
differential turnout, nor privately targeted messages.\footnote{Voters differ in their
concern for the common good and in their propensity toward zero-sum thinking
\citep{Kendall2022,Chinoy,ali2024}, so the distinction between common-interest and
conflict-of-interest issues is politically consequential. On cleavage formation and its
strategic use, respectively, see \citet{LipsetRokkan1967} and \citet{Riker1986}.}

The political return to public information likewise depends on the kind of
disagreement it reveals. Revealing the state of a common-interest issue moves both
parties to the same informed policy: expected voter welfare improves while party
payoffs are unchanged, because the parties continue to converge (Online
Appendix~\ref{oa:app:information}). Revealing instead which group benefits from a
policy that creates winners and losers separates the groups' perceived ideal points;
full revelation maximizes that separation, the platform distance, and both parties'
payoffs, while destabilizing policy (Section \ref{subwelfare}). The analysis identifies a political return to information
on conflicts of interest, but none to information on common-interest issues. This asymmetry parallels \citet{PeregoYuksel2022}, in which media competition
supplies relatively less common-interest information, and complements work in which
shared news can polarize beliefs \citep{Bowen2023}.

Political identity can reinforce an existing partisan divide. Following \citet{BGT21},
stronger identification shifts citizens' perceived ideal points toward the in-group
and away from the out-group. When identity is with political parties,
it constitutes a spread in the sense of Proposition~\ref{propspreadKD}: platforms
move farther apart, and both parties gain (Online Appendix~\ref{oa:app:identity}). If
either party can fund a public campaign that raises partisan salience in both camps,
an inherited platform gap magnifies the return to doing so, and sufficiently low costs
sustain mutual investment---a feedback between voter and platform polarization (Online
Appendix~\ref{oa:app:identityinvestment}). This channel complements the
preference-updating feedback in \citet{CallanderCarbajal2022} and models in which
conflict activates identity \citep{SHAYO09,BGT21,GT23}. 

Finally, experimental evidence indicates that
communication can generate correlated disagreement across issues: when a common
source bundles moral and economic messages, disagreement with its moral stance spills
over to economic opinions \citep{BonomiSpillovers}. Issue emphasis, information, and
identity can therefore produce the political conflict that profits parties.

\subsection{Institutions and the value of a political foothold}
Recall that party organization determines \(U\), how a party values the power it
obtains, while electoral institutions determine \(\rho\), how votes become power.
Together, they determine the value of a political foothold. Minority representation,
appointments, agenda access, and allied organizations preserve influence after defeat;
a majority premium instead raises the return to attracting pivotal voters.

The proportional-with-premium family of Online Appendix~\ref{oa:appendixA4} holds the
total stock of power fixed and isolates this tradeoff. Platform distance weakly
decreases with the majority premium. If one voter type is uniquely pivotal for a
majority---if cumulative voter mass crosses one half strictly at that type---the
decrease is strict, and both platforms converge to its ideal point as the premium
approaches the entire contestable power. Protecting a minority constituency then
becomes negligible relative to winning. This provides a distinct institutional channel
for the negative relationship between electoral disproportionality and platform
polarization studied by \citet{Matakos2016}.

The framework accommodates presidential systems because the strategic actor is
the broader party coalition. Executive victory creates a discontinuity through
appointments and agenda control, while the coalition's
wider electoral strength continues to shape legislative leverage, political careers,
fundraising, and allied organizations.

\subsection{Party insurance and voter welfare}\label{subwelfare}
Constituency insurance stabilizes party influence across electoral states, but it can
destabilize implemented policy. Let \(D=\bar x-\ubar x>0\) denote the equilibrium
platform distance in the unidimensional benchmark, and suppose that realized policy
averages the two platforms according to their shares of contestable power (Online
Appendix~\ref{oa:appendixA4}):
\(X^\star=\ubar x+\Lambda D\), where \(\Lambda\) is party \(A\)'s realized share.
Platforms and electoral strength then determine policy jointly
\citep{OrtunoOrtin1997,Saporiti2014}. Under this benchmark,
\[
\widehat V_p-\frac{\nu(1)+\nu(0)}{2}
=\frac{D^2}{2\phi}>0,
\qquad
\operatorname{Var}(X^\star)
=D^2\operatorname{Var}(\Lambda)>0.
\]
Corollary~\ref{oa:cor:insurancewelfare} formalizes the tension: the same differentiation
that insures the parties exposes citizens to policy volatility.

Under quadratic loss, the welfare cost separates into bias and variance. If
\(x^o=\sum_i s_i x_i\) is the utilitarian optimum, then
\[
\mathcal W^\star
=
\mathcal W(x^o)
-\bigl(\mathbb E[X^\star]-x^o\bigr)^2
-\operatorname{Var}(X^\star).
\]
For a fixed electorate, the differentiated equilibrium therefore produces a strict
welfare loss relative to certain implementation of \(x^o\). Under proportional power, a well-known result is that
power-maximizing parties offer \(x^o\); decreasing marginal
organizational value instead induces differentiation, leaving voters to bear the policy
risk it creates.

More generally, welfare depends on both policy variance and the location of mean policy
relative to \(x^o\). The majority-premium limit separates these two components. When one
voter type is uniquely pivotal, platform distance and policy variance vanish as the
premium approaches the entire stock of power, while implemented policy converges to
the median \(x_m\). Limiting welfare is
\[
\mathcal W(x^o)-(x_m-x^o)^2.
\]
The limiting institution improves welfare relative to a lower-premium rule whenever
the reduction in policy volatility outweighs any increase in squared bias; the gain is
unambiguous when the median coincides with the utilitarian optimum.\footnote{Online
Appendix~\ref{oa:appendixA4} also treats the exact-half boundary: if cumulative voter
mass reaches one half at \(x_k\) and \(x_{k+1}>x_k\), the maintained tie-sharing rule
preserves a positive platform gap and positive policy variance even in the limit.}
Welfare comparisons at intermediate premia remain ambiguous.

This comparison isolates the policy-risk channel generated by the power-weighted
implementation rule: parties value a protected constituency without internalizing the
policy risk it imposes on voters. When voter preferences are themselves uncertain,
however, differentiation can improve welfare by allowing the election to select a
policy better matched to the realized electorate
\citep{BERNHARDT_DUGGAN_SQUINTANI_2009}. Information, accountability, representation,
checks on power, and minority influence introduce further effects that a comprehensive
institutional comparison would also need to incorporate.

\section{Conclusion}\label{conclusion}

Political economy often treats social conflict as the terrain on which parties
compete. This paper shows why parties organize it. When parties attribute political power first to the highest rent-generating affiliates, returns to power are diminishing,
and the two competitors come to share an interest in building loyal
constituencies that survive adverse electoral conditions. That interest shed light on the political insurance value of voter polarization and even alienation: a party can profit when those who oppose it move farther away,
because its rival chases them.

Competition selects which of society's many disagreements becomes the partisan divide,
and what makes a divide valuable is not only how far apart citizens stand on any
single issue but whether the same citizens stand apart on all of them. When economic,
cultural, and social divisions align, one partisan cleavage creates constituencies
that resist common electoral shocks; under some conditions, such a divide runs along
a leading principal component of voter positions. Issue emphasis, information, and partisan identity can sharpen or reinforce the partisan divide through different channels.

The political return to conflict can run again voter welfare. When implemented policy
reflects the realized balance of power, platform polarization insures party influence but increases the volatility of the implemented policy---a social cost. A larger majority prize can curb both polarization
and volatility, improving welfare when convergence lands near the utilitarian optimum
and trading variance against bias otherwise. The analysis isolates a policy risk that
parties do not internalize and an electoral lever that can mitigate it.

The framework also identifies forces that can push polarization upward. More
nationalized electoral behavior makes common shocks more prominent relative to
idiosyncratic noise (Online Appendix~\ref{oa:Robustness}); greater concentration
of organizational value within parties raises the return to preserving a
foothold (Proposition~\ref{orgconcentration}); and stronger cross-issue
alignment widens the party-preferred platform gap
(Corollary~\ref{correlationresult}). Once platforms have moved apart, the
resulting payoff gain can induce both parties to invest in identity-building
communication, reinforcing the division (Online
Appendix~\ref{oa:app:identityinvestment}). A broader agenda asks who finances
such communication and how identity, party organization, and electoral
institutions shape the feedback. The claim,
however, stands: political rivals competing for power can benefit from the same intensifying conflicts.

\newpage
\vspace{1cm}
\bibliographystyle{apalike}
\begin{spacing}{1.15}
\bibliography{biblio}
\end{spacing}
\input{appendix.tex}
\fi

\end{document}

%% file: pictures.tex
\newcommand{\StaircaseFigure}{%
\begin{figure}[t]
\centering
\begin{tikzpicture}[x=.93cm,y=.88cm,>=stealth,font=\footnotesize]
  \begin{scope}
    \node[font=\small] at (3.45,4.73)
      {(a) Case \(x_A=x_B\): cliff};
    \draw[->] (.35,.65) -- (6.55,.65);
    \draw[->] (.35,.65) -- (.35,4.38);
    \draw[black!72,line width=1.35pt]
      (.52,4.00)--(3.45,4.00)--(3.45,1.05)--(6.35,1.05);
    \draw[dotted,black!42] (3.45,.65)--(3.45,4.00);
    \node[below] at (3.45,.65) {\(0\)};
    \node[left] at (.35,4.00) {\(\rho(1)\)};
    \node[left] at (.35,1.05) {\(\rho(0)\)};
  \end{scope}

  \begin{scope}[xshift=8.00cm]
    \node[font=\small] at (3.45,4.73)
      {(b) Case \(x_A>x_B\): staircase};
    \draw[->] (.35,.65) -- (6.55,.65);
    \draw[->] (.35,.65) -- (.35,4.38);

    \fill[partyrust!13] (1.25,3.25) rectangle (2.25,4.00);
    \fill[partyrust!13] (2.25,2.50) rectangle (3.45,4.00);
    \path[pattern=north east lines,pattern color=partyrust!52]
      (1.25,3.25) rectangle (2.25,4.00);
    \path[pattern=north east lines,pattern color=partyrust!52]
      (2.25,2.50) rectangle (3.45,4.00);

    \fill[partyblue!13] (3.45,1.05) rectangle (4.65,2.50);
    \fill[partyblue!13] (4.65,1.05) rectangle (5.65,1.75);
    \path[pattern=north west lines,pattern color=partyblue!52]
      (3.45,1.05) rectangle (4.65,2.50);
    \path[pattern=north west lines,pattern color=partyblue!52]
      (4.65,1.05) rectangle (5.65,1.75);

    \draw[black!58,densely dashed,line width=1.0pt]
      (.52,4.00)--(3.45,4.00)--(3.45,1.05)--(6.35,1.05);
    \draw[black!78,line width=1.45pt]
      (.52,4.00)--(1.25,4.00)
      --(1.25,3.25)--(2.25,3.25)
      --(2.25,2.50)--(4.65,2.50)
      --(4.65,1.75)--(5.65,1.75)
      --(5.65,1.05)--(6.35,1.05);

    \foreach \x/\y/\lab in
      {1.25/4.00/{\Delta_1},2.25/3.25/{\Delta_2},
       4.65/2.50/{\Delta_3},5.65/1.75/{\Delta_4}} {
        \draw[dotted,black!35] (\x,.65)--(\x,\y);
        \node[below] at (\x,.65) {\(\lab\)};
      }
    \node[below] at (3.45,.65) {\(0\)};
    \node[left] at (.35,4.00) {\(\rho(1)\)};
    \node[left] at (.35,1.05) {\(\rho(0)\)};
  \end{scope}

  \node[rotate=90] at (-.86,2.48) {\(A\)'s political power};
  \node at (7.42,-.38)
    {aggregate popularity shock \(\epsilon\) (higher values favor \(B\))};
\end{tikzpicture}
\caption{Convergence creates common exposure; differentiation creates protected
constituencies. In panel (a), identical platforms give every voter the same switching
threshold. In panel (b), voter groups switch at several thresholds \(\Delta_i\).
Relative to the dashed convergence allocation, the rust hatching marks power that party
\(A\) relinquishes in favorable states, while the blue hatching marks power it retains
in adverse states. 
}
\label{fig:staircase}
\end{figure}
}

\newcommand{\SpreadAlienationFigure}{%
\begin{figure}[H]
\centering
\begin{tikzpicture}[x=.96cm,y=1.10cm,>=stealth,font=\footnotesize]
  \begin{scope}
    \node[font=\small] at (3.65,3.18) {(a) Voter polarization};
    \draw[->] (.35,2.25) -- (7.10,2.25);
    \draw[->] (.35,.82) -- (7.10,.82);
    \node[left] at (.35,2.25) {before};
    \node[left] at (.35,.82) {after};
    \draw[dashed,black!42] (3.62,.47) -- (3.62,2.60);
    \node[above] at (3.62,2.60) {\(x_m\)};

    \foreach \x in {1.10,1.92,2.74,3.62,4.50,5.32,6.14}
      \fill[black!62] (\x,2.25) circle (2.15pt);
    \foreach \x in {.62,1.46,2.34,3.62,4.90,5.82,6.64}
      \fill[black!62] (\x,.82) circle (2.15pt);
    \foreach \a/\b in
      {1.10/.62,1.92/1.46,2.74/2.34,4.50/4.90,5.32/5.82,6.14/6.64}
      \draw[->,black!50,line width=.70pt] (\a,2.10) -- (\b,.98);

    \draw[very thick,partyblue] (2.58,2.02) -- (2.58,2.48);
    \draw[very thick,partyrust] (4.76,2.02) -- (4.76,2.48);
    \draw[very thick,partyblue] (2.06,.59) -- (2.06,1.05);
    \draw[very thick,partyrust] (5.23,.59) -- (5.23,1.05);
    \draw[->,partyblue,line width=.8pt,
      preaction={draw=white,line width=2.2pt}]
      (2.54,1.95) .. controls (2.38,1.53) .. (2.10,1.10);
    \draw[->,partyrust,line width=.8pt,
      preaction={draw=white,line width=2.2pt}]
      (4.80,1.95) .. controls (4.96,1.53) .. (5.19,1.10);
    \node[partyblue,above] at (2.58,2.48) {$x_B$};
    \node[partyrust,above] at (4.76,2.48) {$x_A$};
    \node[partyblue,below] at (2.06,.59) {$x_B'$};
    \node[partyrust,below] at (5.23,.59) {$x_A'$};
  \end{scope}

  \begin{scope}[xshift=8.45cm]
    \node[font=\small] at (3.55,3.18) {(b) Profitable alienation};
    \draw[->] (.28,2.25) -- (7.02,2.25);
    \draw[->] (.28,.82) -- (7.02,.82);
    \node[left] at (.28,2.25) {before};
    \node[left] at (.28,.82) {after};
    \foreach \x in {1.05,2.02,3.48,4.68,5.82,6.55}
      \fill[black!62] (\x,2.25) circle (2.15pt);
    \foreach \x in {.40,2.02,3.48,4.68,5.82,6.55}
      \fill[black!62] (\x,.82) circle (2.15pt);
    \draw[->,black,line width=.85pt] (1.05,2.10)
      .. controls (.78,1.57) .. (.43,.98);
    \node[above] at (1.05,2.33) {$x_1$};
    \node[below] at (.40,.69) {$x_1'$};

    \draw[very thick,partyblue] (2.38,2.02) -- (2.38,2.48);
    \draw[very thick,partyrust] (4.56,2.02) -- (4.56,2.48);
    \draw[very thick,partyblue] (1.90,.59) -- (1.90,1.05);
    \draw[very thick,partyrust] (4.42,.59) -- (4.42,1.05);
    \draw[->,partyblue,line width=.8pt,
      preaction={draw=white,line width=2.2pt}]
      (2.34,1.95) .. controls (2.17,1.50) .. (1.94,1.10);
    \draw[->,partyrust,line width=.8pt,
      preaction={draw=white,line width=2.2pt}]
      (4.52,1.95) .. controls (4.49,1.50) .. (4.44,1.10);
    \node[partyblue,above] at (2.38,2.48) {$x_B$};
    \node[partyrust,above] at (4.56,2.48) {$x_A$};
    \node[partyblue,below] at (1.90,.59) {$x_B'$};
    \node[partyrust,below] at (4.42,.59) {$x_A'$};
  \end{scope}
\end{tikzpicture}
\caption{Panel (a) spreads the two halves of the
electorate outward: the platforms separate and both parties gain. Panel (b) makes a small
leftward movement of the most left-wing group. The group supports the right-wing party
\(A\) less, but \(B\)'s platform follows it farther than \(A\)'s, protecting \(A\)'s
remaining constituency; party \(A\) gains through alienation.}
\label{fig:spreadalienation}
\end{figure}
}

\newcommand{\CohesionFigure}{%
\begin{figure}[t]
\centering
\begin{tikzpicture}[x=1cm,y=1cm,>=stealth,font=\footnotesize]
  \begin{scope}
    \node[font=\small] at (3.25,4.38)
      {(a) Uncorrelated views: \(\varrho=0\)};
    \draw[->] (.20,2.10) -- (6.30,2.10);
    \draw[->] (3.25,.28) -- (3.25,4.08);
    \begin{scope}[shift={(3.25,2.10)}]
      \clip (0,0) ellipse (2.50 and 1.60);
      \fill[partyblue!13] (-2.7,-1.8) rectangle (0,1.8);
      \fill[partyrust!13] (0,-1.8) rectangle (2.7,1.8);
      \path[pattern=north east lines,pattern color=gray!38]
        (-2.7,-1.8) rectangle (0,1.8);
      \path[pattern=north west lines,pattern color=gray!38]
        (0,-1.8) rectangle (2.7,1.8);
    \end{scope}
    \draw[gray,thick] (3.25,2.10) ellipse (2.50 and 1.60);
    \draw[gray!52] (3.25,2.10) ellipse (1.70 and 1.09);
    \draw[dashed,black!72,line width=1.0pt] (.50,2.10) -- (6.00,2.10);
    \draw[partyblue,very thick] (2.18,1.99)--(2.34,2.15)
      (2.18,2.15)--(2.34,1.99);
    \draw[partyrust,very thick] (4.16,1.99)--(4.32,2.15)
      (4.16,2.15)--(4.32,1.99);
    \node[partyblue,above left,inner sep=1pt] at (2.20,2.20)
      {\(x_B^\star\)};
    \node[partyrust,below right,inner sep=1pt] at (4.30,1.98)
      {\(x_A^\star\)};
    \node[below left,inner sep=1pt] at (6.18,2.03) {issue 1};
    \node[above right,inner sep=1pt] at (3.29,3.78) {issue 2};
    \node[below] at (3.25,-.02) {equilibrium distance \(d_0^\star\)};
  \end{scope}

  \begin{scope}[xshift=8.35cm]
    \node[font=\small] at (3.25,4.38)
      {(b) Aligned views: \(\varrho=.85\)};
    \draw[->] (.20,2.10) -- (6.30,2.10);
    \draw[->] (3.25,.28) -- (3.25,4.08);
    \begin{scope}[shift={(3.25,2.10)},rotate=30.76]
      \clip (0,0) ellipse (2.876 and .732);
      \fill[partyblue!13] (-3.1,-1.0) rectangle (0,1.0);
      \fill[partyrust!13] (0,-1.0) rectangle (3.1,1.0);
      \path[pattern=north east lines,pattern color=gray!38]
        (-3.1,-1.0) rectangle (0,1.0);
      \path[pattern=north west lines,pattern color=gray!38]
        (0,-1.0) rectangle (3.1,1.0);
    \end{scope}
    \begin{scope}[shift={(3.25,2.10)},rotate=30.76]
      \draw[gray,thick] (0,0) ellipse (2.876 and .732);
      \draw[gray!52] (0,0) ellipse (1.96 and .50);
      \draw[dashed,black!72,line width=1.0pt] (-3.05,0) -- (3.05,0);
      \draw[partyblue,very thick] (-1.23,-.10)--(-1.05,.08)
        (-1.23,.08)--(-1.05,-.10);
      \draw[partyrust,very thick] (1.05,-.10)--(1.23,.08)
        (1.05,.08)--(1.23,-.10);
    \end{scope}
    \node[partyblue,below right,inner sep=1pt] at (2.58,.88)
      {\(x_B^\star\)};
    \node[partyrust,above left,inner sep=1pt] at (3.92,3.32)
      {\(x_A^\star\)};
    \node[below left,inner sep=1pt] at (6.18,2.03) {issue 1};
    \node[above right,inner sep=1pt] at (3.29,3.78) {issue 2};
    \node[below] at (3.25,-.02)
      {greater distance: \(d_{.85}^\star>d_0^\star\)};
  \end{scope}
\end{tikzpicture}
\caption{The ellipses represent
voter-distribution contours, and the dashed line is the platform axis in the equilibrium
preferred by both parties; under ellipticity, it is a leading principal component. The
two electorates have the same marginal distribution on each issue. When views become
aligned across issues, that component separates more coherent partisan blocs; the
party-preferred platforms move farther apart and both parties gain. Negative correlation
selects the opposite-sloped diagonal.}
\label{fig:cohesion}
\end{figure}
}

\newcommand{\MultidimensionalSpreadFigure}{%
\begin{figure}[t]
\centering
\begin{tikzpicture}[x=1cm,y=1cm,>=stealth,font=\footnotesize]
  \def\spreadvoters{%
    -.812/.421/0, .710/.269/1, .564/-.653/0, -.906/-.289/1,
     .493/-.331/1, .050/-.551/0, 1.331/.791/1, -.903/.141/0,
     .190/-.379/0, -.284/-.276/1, -.555/.596/0, -.238/.940/0,
    -.153/-.265/0, -.415/.901/0, .131/.143/0, .461/.685/0,
     .812/-.421/0, -.710/-.269/1, -.564/.653/0, .906/.289/1,
    -.493/.331/1, -.050/.551/0, -1.331/-.791/1, .903/-.141/0,
    -.190/.379/0, .284/.276/1, .555/-.596/0, .238/-.940/0,
     .153/.265/0, .415/-.901/0, -.131/-.143/0, -.461/-.685/0}

  \def\OldQx{.8747}
  \def\OldQy{.4846}
  \def\OldNx{-.4846}
  \def\OldNy{.8747}

  \def\CoeffScale{1.35}
  \def\CoeffShift{.50}
  \def\Shear{.35}
  \def\NewQx{.6428}
  \def\NewQy{.7660}

  \def\panelaxes{%
    \draw[black!22,line width=.48pt] (-2.78,0)--(2.78,0);
    \draw[black!22,line width=.48pt] (0,-2.18)--(0,2.18);
    \node[font=\scriptsize,black!55,anchor=west] at (2.22,-1.96)
      {issue 1};
    \node[font=\scriptsize,black!55,rotate=90,anchor=south]
      at (-2.57,1.48) {issue 2};}
  \def\oldcleavage##1{%
    \draw[##1]
      ({-2.78*\OldQx},{-2.78*\OldQy})--
      ({ 2.78*\OldQx},{ 2.78*\OldQy});}
  \def\newcleavage##1{%
    \draw[##1]
      ({-2.64*\NewQx},{-2.64*\NewQy})--
      ({ 2.64*\NewQx},{ 2.64*\NewQy});}
  \def\platformcross##1##2##3##4{%
    \begin{scope}[shift={(##1,##2)}]
      \draw[##3,line width=##4,line cap=round]
        (-.060,-.060)--(.060,.060)
        (-.060,.060)--(.060,-.060);
    \end{scope}}

  \begin{scope}
    \node[font=\small] at (3.25,4.58)
      {(a) Initial equilibrium};
    \begin{scope}[shift={(3.25,2.18)}]
      \panelaxes
      \foreach \u/\v/\mark in \spreadvoters
        \fill[partyblue!34] (\u,\v) circle (1.85pt);
      \platformcross{-.296}{-.164}{partyblue}{.85pt}
      \platformcross{ .296}{ .164}{partyrust}{.85pt}
    \end{scope}
  \end{scope}

  \begin{scope}[xshift=7.90cm]
    \node[font=\small] at (3.25,4.58)
      {(b) Projection on \(q^\star\)};
    \begin{scope}[shift={(3.25,2.18)}]
      \panelaxes
      \oldcleavage{densely dashed,black!58,line width=.82pt}
      \foreach \u/\v/\mark in \spreadvoters {
        \pgfmathsetmacro{\alpha}{\u*\OldQx+\v*\OldQy}
        \pgfmathsetmacro{\xp}{\alpha*\OldQx}
        \pgfmathsetmacro{\yp}{\alpha*\OldQy}
        \draw[densely dashed,black!25,line width=.42pt]
          (\u,\v)--(\xp,\yp);
        \fill[partyblue!34] (\u,\v) circle (1.75pt);
        \draw[black!68,line width=.72pt]
          ({\xp-.045*\OldNx},{\yp-.045*\OldNy})--
          ({\xp+.045*\OldNx},{\yp+.045*\OldNy});
      }
      \platformcross{-.296}{-.164}{partyblue}{.75pt}
      \platformcross{ .296}{ .164}{partyrust}{.75pt}
      \node[black!62,fill=white,inner sep=1pt]
        at ({2.20*\OldQx},{2.20*\OldQy+.12}) {\(q^\star\)};
    \end{scope}
  \end{scope}

  \begin{scope}[yshift=-5.35cm]
    \node[font=\small] at (3.25,4.58)
      {(c) After a \(q^\star\)-spread};
    \begin{scope}[shift={(3.25,2.18)}]
      \panelaxes
      \oldcleavage{densely dashed,black!47,line width=.78pt}

      \foreach \u/\v/\mark in \spreadvoters {
        \pgfmathsetmacro{\alpha}{\u*\OldQx+\v*\OldQy}
        \pgfmathsetmacro{\beta}{\u*\OldNx+\v*\OldNy}
        \pgfmathsetmacro{\side}{ifthenelse(\alpha>=0,1,-1)}
        \pgfmathsetmacro{\alphat}
          {\CoeffScale*\alpha+\CoeffShift*\side}
        \pgfmathsetmacro{\betat}{\beta+\Shear*\alphat}
        \pgfmathsetmacro{\ut}{\alphat*\OldQx+\betat*\OldNx}
        \pgfmathsetmacro{\vt}{\alphat*\OldQy+\betat*\OldNy}
        \pgfmathsetmacro{\xpt}{\alphat*\OldQx}
        \pgfmathsetmacro{\ypt}{\alphat*\OldQy}
        \draw[densely dashed,black!20,line width=.36pt]
          (\ut,\vt)--(\xpt,\ypt);
        \fill[black!42] (\xpt,\ypt) circle (.68pt);
        \fill[partyblue!27] (\ut,\vt) circle (1.85pt);
      }
      \platformcross{-.296}{-.164}{partyblue}{.75pt}
      \platformcross{ .296}{ .164}{partyrust}{.75pt}
      \node[font=\scriptsize,black!50,anchor=south east,inner sep=0pt]
        at ({2.14*\OldQx},{2.14*\OldQy+.10}) {\(q^\star\)};
    \end{scope}
  \end{scope}

  \begin{scope}[xshift=7.90cm,yshift=-5.35cm]
    \node[font=\small] at (3.25,4.58)
      {(d) Updated equilibrium};
    \begin{scope}[shift={(3.25,2.18)}]
      \panelaxes
      \oldcleavage{densely dashed,black!30,line width=.68pt}
      \newcleavage{dash pattern=on 5pt off 2.4pt,black!72,line width=.92pt}

      \foreach \u/\v/\mark in \spreadvoters {
        \pgfmathsetmacro{\alpha}{\u*\OldQx+\v*\OldQy}
        \pgfmathsetmacro{\beta}{\u*\OldNx+\v*\OldNy}
        \pgfmathsetmacro{\side}{ifthenelse(\alpha>=0,1,-1)}
        \pgfmathsetmacro{\alphat}
          {\CoeffScale*\alpha+\CoeffShift*\side}
        \pgfmathsetmacro{\betat}{\beta+\Shear*\alphat}
        \pgfmathsetmacro{\ut}{\alphat*\OldQx+\betat*\OldNx}
        \pgfmathsetmacro{\vt}{\alphat*\OldQy+\betat*\OldNy}
        \fill[partyblue!38] (\ut,\vt) circle (1.85pt);
      }

      \platformcross{-.296}{-.164}{black!38}{.58pt}
      \platformcross{ .296}{ .164}{black!38}{.58pt}
      \draw[->,partyblue,line width=.72pt]
        (-.31,-.15) to[out=-145,in=25] (-.86,-1.02);
      \draw[->,partyrust,line width=.72pt]
        (.31,.15) to[out=35,in=205] (.86,1.02);
      \platformcross{-.868}{-1.034}{partyblue}{.85pt}
      \platformcross{ .868}{ 1.034}{partyrust}{.85pt}
      \node[font=\scriptsize,black!42,anchor=north west,inner sep=0pt]
        at ({2.10*\OldQx+.04},{2.10*\OldQy-.10}) {\(q^\star\)};
      \node[font=\scriptsize,black!72,anchor=south east,inner sep=0pt]
        at ({2.22*\NewQx-.02},{2.22*\NewQy+.10})
        {\(\widetilde q^\star\)};
    \end{scope}
  \end{scope}
\end{tikzpicture}
\caption{Let \(\widehat q^\star=q^\star/\lVert q^\star\rVert\) and
\(\alpha_i=(x_i-\mu)^\top\widehat q^\star\). Colored crosses mark the
equilibrium platforms (blue for \(B\), rust for \(A\)); in panel (d), gray
crosses mark the initial platforms and arrows trace their adjustment. Panel (b) projects the original
electorate onto this cleavage. Panel (c) replaces \(\{\alpha_i\}\) by a strict
spread \(\{\widetilde\alpha_i\}\) and shows one admissible centrally symmetric
lift; orthogonal coordinates are unrestricted. Re-optimization may therefore
select a different cleavage, as in panel (d), while increasing platform
distance.}
\label{fig:multidimensionalspread}
\end{figure}
}

%% file: online_appendix_content.tex
\newpage
\appendix
\renewcommand{\thesection}{OA.\arabic{section}}
\renewcommand{\thesubsection}{\thesection.\arabic{subsection}}
\renewcommand{\thesubsubsection}{\thesubsection.\arabic{subsubsection}}
\section{Additional Equilibrium Results}\label{appendixextensions}

\subsection{Convex transformations and convergence}\label{app:convexconverse}
Suppose that a
convergent profile is an equilibrium when parties maximize expected political power
directly. Let $r_0=\rho(0)$ and $r_1=\rho(1)$. At convergence, each party receives the
endpoint lottery that assigns probability one half to $r_0$ and probability one half to
$r_1$. Any deviation generates a random power allocation $R\in[r_0,r_1]$ satisfying
\[
\mathbb E[R]\le\frac{r_0+r_1}{2},
\]
because convergence is optimal under expected power. For any increasing convex
transformation $U$, the chord bound gives
\[
U(R)
\le
\frac{r_1-R}{r_1-r_0}U(r_0)
+
\frac{R-r_0}{r_1-r_0}U(r_1).
\]
Taking expectations and using the bound on $\mathbb E[R]$ yields
\[
\mathbb E[U(R)]
\le
\frac{U(r_0)+U(r_1)}2,
\]
which is the transformed payoff at convergence. Thus an increasing convex transformation
preserves any convergent equilibrium of the risk-neutral game.

\subsection{Logit revision and potential selection}\label{app:logitselection}

The potential game structure allows for the following dynamic selection result.  Let
\(\mathcal G\subset D\) be a finite platform grid and restrict the
platform game to \(\mathcal G^2\).  Time is discrete.  At each date, one
party \(p\in\{A,B\}\) is selected to revise with fixed probability
\(\chi_p>0\), where \(\chi_A+\chi_B=1\).  Conditional on the opponent's
current platform \(x_{-p}\), the revising party chooses
\(y\in\mathcal G\) according to
\begin{equation}\label{logitchoice}
L_{p,\tau}(y\mid x_{-p})
=
\frac{\exp\{V_p(y,x_{-p})/\tau\}}
{\displaystyle\sum_{z\in\mathcal G}
\exp\{V_p(z,x_{-p})/\tau\}},
\qquad \tau>0.
\end{equation}
The current platform is among the available choices, and \(\tau\)
measures noise in platform revision.

\begin{proposition}[Logit selection]
\label{logitselection}
Maintain the assumptions of Proposition \ref{main:potentialgame}.  On every
finite grid \(\mathcal G\subset D\), the asynchronous logit process in
\eqref{logitchoice} has the unique stationary distribution
\begin{equation}\label{gibbsdistribution}
\Pi_{\tau}^{\mathcal G}(x_A,x_B)
=
\frac{\exp\{\Phi(x_A,x_B)/\tau\}}
{\displaystyle\sum_{(y_A,y_B)\in\mathcal G^2}
\exp\{\Phi(y_A,y_B)/\tau\}}.
\end{equation}
Let
\[
\mathcal M_{\mathcal G}
:=
\arg\max_{(x_A,x_B)\in\mathcal G^2}\Phi(x_A,x_B).
\]
As \(\tau\downarrow0\),
\[
\Pi_{\tau}^{\mathcal G}(x)
\longrightarrow
\begin{cases}
|\mathcal M_{\mathcal G}|^{-1},&x\in\mathcal M_{\mathcal G},\\
0,&x\notin\mathcal M_{\mathcal G}.
\end{cases}
\]
Thus the stochastically stable profiles are exactly the global
potential maximizers on the grid.

Moreover, let \((\mathcal G_n)_{n\ge1}\) satisfy
\[
\sup_{x\in D}\min_{y\in\mathcal G_n}\|x-y\|\longrightarrow0.
\]
If \(x^n\in\mathcal M_{\mathcal G_n}\), every accumulation point of
\((x^n)\) belongs to
\(\arg\max_{x\in D^2}\Phi(x)\).  Under the strict-concavity condition
of Proposition \ref{main:existence}, every such limiting profile is a
party-preferred Nash equilibrium.
\end{proposition}

The result selects globally rather than through a neighborhood-based
refinement. Rare platform mistakes concentrate play on the profiles
that maximize the exact potential that orders every unilateral deviation.
A fixed grid selects its own potential maximizers. As the policy grid
becomes fine, every accumulation point is a global potential maximizer
of the continuous-action finite-type game. Under the additional
strict-concavity condition of Proposition \ref{main:existence}, these limiting
profiles are precisely the equilibria that both parties prefer.
\ 
\paragraph{Proof of Proposition \ref{logitselection}}
Because the logit rule assigns strictly positive probability to every
grid action, every profile can be reached through a finite sequence of
unilateral revisions.  The chain is therefore irreducible.  It is
aperiodic because a revising party chooses its current platform with
strictly positive probability.

Consider two profiles \(x,x'\in\mathcal G^2\) that differ only in party
\(p\)'s platform.  The opponent's platform, and hence the denominator
in \eqref{logitchoice}, is the same in the two directions.  Exact
potentiality gives
\[
\frac{\Pr_\tau(x\to x')}{\Pr_\tau(x'\to x)}
=
\exp\left\{
\frac{V_p(x'_p,x_{-p})-V_p(x_p,x_{-p})}{\tau}
\right\}
=
\exp\left\{
\frac{\Phi(x')-\Phi(x)}{\tau}
\right\}.
\]
It follows that \eqref{gibbsdistribution} satisfies detailed balance:
\[
\Pi_\tau^{\mathcal G}(x)\Pr_\tau(x\to x')
=
\Pi_\tau^{\mathcal G}(x')\Pr_\tau(x'\to x).
\]
Transitions between profiles differing in both coordinates have zero
probability in both directions.  Thus
\(\Pi_\tau^{\mathcal G}\) is stationary; irreducibility makes it unique.

For a fixed grid, let
\[
\Phi_{\mathcal G}^\star:=\max_{x\in\mathcal G^2}\Phi(x).
\]
Divide the numerator and denominator in \eqref{gibbsdistribution} by
\(\exp\{\Phi_{\mathcal G}^\star/\tau\}\).  Every nonmaximizing term
converges to zero, while every maximizing term equals one.  This proves
the zero-noise result.

Finally, continuity of \(\Phi\) on the compact set \(D^2\), together
with
\[
\sup_{x\in D}\min_{y\in\mathcal G_n}\|x-y\|\longrightarrow0,
\]
implies
\[
\max_{\mathcal G_n^2}\Phi\longrightarrow\max_{D^2}\Phi.
\]
If a subsequence of grid maximizers converges to \(x^\infty\),
continuity gives
\(\Phi(x^\infty)=\max_{D^2}\Phi\).  Proposition
\ref{main:potentialgame} makes every such profile a Nash equilibrium, and
Proposition \ref{main:existence} identifies the global potential maximizers
with \(\mathcal E^\star\) under strict concavity of \(\nu\).
\(\blacksquare\)

\begin{remark}[Joint limit]\label{jointgridnoise}
The last conclusion first lets noise vanish on a fixed grid and then
refines the grid.  A joint limit obtains if
\(\tau_n\downarrow0\), the grids become dense, and
\[
\tau_n\log|\mathcal G_n|\longrightarrow0.
\]
For every neighborhood \(\mathcal O\) of the continuous global
maximizers, compactness gives a positive potential gap on
\(D^2\setminus\mathcal O\).  The number of grid profiles is at most
\(|\mathcal G_n|^2\), while each profile outside \(\mathcal O\)
receives the corresponding exponential penalty.  The displayed
condition therefore implies
\(\Pi_{\tau_n}^{\mathcal G_n}(\mathcal O)\to1\).
\end{remark}

\subsection{A parametric affiliate-concentration example}

A convenient parametric illustration is
\[
T_\theta(t)=\frac{t}{\theta+(1-\theta)t},
\qquad
q_\theta(k)=T_\theta'(k)=\frac{\theta}{[\theta+(1-\theta)k]^2},
\qquad \theta\in(0,1).
\]
Total affiliate value is one for every \(\theta\), while a lower \(\theta\) makes value
more top-heavy and therefore weakly increases equilibrium differentiation.

\subsection{Beyond ellipticity: organizational value selects the divide}
\label{nonellipticexample}

The principal-component result separates the geometry of voter disagreement from the
organization's valuation of political power: within an elliptical family, geometry
selects the direction of differentiation and organizational value determines its
magnitude. Outside that family, the separation need not hold. The following example keeps
voter views fixed and changes only the distribution of organizational value.

There are four voter types,
\[
X_-=(-2,0),\qquad X_+=(2,0),\qquad
Y_-=(0,-1),\qquad Y_+=(0,1),
\]
with respective masses \(0.05,0.05,0.45,0.45\). The electorate is centered at the
origin and has covariance matrix
\[
\Sigma=
\begin{pmatrix}
0.4&0\\
0&0.9
\end{pmatrix},
\]
so its first principal component is vertical. Maintain the wide-support condition; for
this support, \(\phi>\operatorname{diam}(D)^2=20\) is sufficient. Political power is proportional to vote
share, and the top-share curve for organizational value is
\[
T_\theta(s)=\frac{s}{\theta+(1-\theta)s},
\qquad \theta\in(0,1).
\]
Thus \(\nu_\theta=T_\theta\). Lower values of \(\theta\) leave total organizational value
unchanged but place a larger share of it among the highest-value affiliates.

Up to coordinate reflections and party labels, only two strict projection orderings are
possible. In the horizontal-outer ordering,
\(R_H=(X_-,Y_-,Y_+,X_+)\), the two low-mass horizontal types occupy the outer ranks. In
the vertical-outer ordering, \(R_V=(Y_-,X_-,X_+,Y_+)\), the high-mass vertical types do.
The associated party-\(A\) candidates are
\begin{align}
z_H(\theta)
&=\left(
2[T_\theta(.05)+T_\theta(.95)-1],
\,2T_\theta(.50)-T_\theta(.05)-T_\theta(.95)
\right),\label{eq:zH}\\
z_V(\theta)
&=\left(
2[2T_\theta(.50)-T_\theta(.45)-T_\theta(.55)],
\,T_\theta(.45)+T_\theta(.55)-1
\right).\label{eq:zV}
\end{align}
The opposing platform is the negative of the relevant candidate. Among the
self-consistent candidate classes, both parties prefer the class with the
larger squared norm.

At \(\theta=0.5\), both orderings are self-consistent, but the vertical-outer candidate
dominates. Up to reflections and party labels, the platforms are
\(\pm(0.0059,0.3304)\), at distance \(0.6608\): the political divide is almost the first
principal component. At \(\theta=0.03\), both orderings remain self-consistent, but the
horizontal-outer candidate dominates. The platforms become
\(\pm(1.2707,0.3064)\), at distance \(2.6143\). The top five percent of affiliates now
account for \(63.7\) percent of organizational value rather than \(9.5\) percent. A party
therefore places exceptional value on retaining the remote, low-mass type that remains
with it in the most adverse electoral states, and the preferred divide rotates toward the
horizontal axis.

The electorate and its covariance matrix have not changed. What changes is which
projected tail is most valuable to protect. Ellipticity makes every projection a common
standardized distribution scaled by its standard deviation; in this example, different
directions instead place different masses and locations at the politically valuable
ranks. The direction of conflict is then chosen jointly by voter geometry and
organizational value.

\section{Formal Political Applications}\label{appendixapplications}

The main analysis characterizes changes in voter conflict that raise both parties'
equilibrium payoffs. I now show how three familiar political channels---issue emphasis,
information provision, and party identity---can generate such changes. The first two applications
identify the parties' gross returns to changes in political conflict; the identity application also
places a costly communication choice before electoral competition and shows when an
inherited division becomes self-reinforcing.

\subsection{Common-Interest versus Conflict-of-Interest Issues}\label{app:issues}

A recent stream of literature stresses that politicians often focus on divisive, zero-sum
issues over common-interest policies, even when the latter could benefit a broad share of
voters \citep[e.g.,][]{fiorina2011,Kendall2022,Dzuida23}. This tendency is potentially
amplified by voters' propensity toward zero-sum thinking \citep[e.g.,][]{Carvalho2023,Chinoy,ali2024}.
The model identifies a common party-payoff incentive that can contribute to these
patterns.

Consider a stylized version of the baseline model with two equally sized voter types, one
religious ($R$) and one secular ($S$), and one policy issue $x$ (e.g., government spending
aimed at tackling climate change). In the common-interest scenario, preferences are aligned, so that
$x_R=x_S$. In the conflict-of-interest scenario, preferences differ, so that $x_R<x_S$. One
interpretation is that the religious voter views such spending as conflicting with divine
control over nature \citep{Pew2022}.

The baseline logic applies immediately. If $x_R=x_S$, there is no heterogeneity in bliss
points, so parties converge on the common ideal policy and obtain expected payoff
$\frac{\nu(1)+\nu(0)}{2}$. If instead $x_R<x_S$, the electorate is diverse and the equilibrium
platforms satisfy $\ubar{x}<\bar{x}$, generating expected payoff
\[
\frac{\nu(1)+\nu(0)}{2}+\frac{(\bar{x}-\ubar{x})^2}{2\phi}
>\frac{\nu(1)+\nu(0)}{2}.
\]
Thus both parties obtain higher equilibrium payoffs when voters perceive the policy as
defining a conflict of interest.

\begin{remark}
Holding electoral institutions fixed, both parties obtain a higher equilibrium payoff
when voters perceive an issue as separating their ideal policies than when they agree on
the ideal policy.
\end{remark}

The same reasoning implies that parties benefit when conflict-of-interest issues are more
salient than common-interest issues. To illustrate, consider two policy instruments, $(x,q)$,
and voter utility
\begin{equation}\label{application}
v(x,q,x_i,q_i)=-\alpha(x_i-x)^2-(1-\alpha)(q_i-q)^2,
\end{equation}
where $\alpha\in(0,1)$ captures the relative payoff importance of issue $x$. Assume $x_R\neq
x_S$ and $q_R=q_S=q^\star$. In equilibrium, parties converge on the common-interest policy,
$q_A=q_B=q^\star$, but diverge on $x$. Equilibrium payoffs equal
\[
\frac{\nu(1)+\nu(0)}{2}+\alpha\frac{(\bar{x}-\ubar{x})^2}{2\phi},
\]
which is increasing in $\alpha$ over comparisons that maintain the wide-support
condition.

\begin{remark}
Both parties' equilibrium payoffs rise with the relative electoral weight of the
conflict-of-interest issue, provided the wide-support condition continues to hold.
\end{remark}

\subsection{Information and Perceived Conflict}\label{app:information}

The model also distinguishes information that resolves a common-interest uncertainty from
information that sharpens perceived conflict. The following calculation isolates parties'
private returns to public information. It also provides a welfare comparison for
common-interest information; without an implementation rule, it does not rank the welfare
effects of information about conflicting interests.

Maintain utility \eqref{application}, and assume that the wide-support condition holds in
every revealed and unrevealed environment compared below. Let the common-interest state be
\(q^\star\in\{0,1\}\), with
\(\Pr(q^\star=1)=\pi_q\). If it is revealed, both parties converge to \(q^\star\); if it is
not, both converge to \(\pi_q\). Party payoffs are identical in the two cases because the
parties do not differentiate on \(q\). Under quadratic loss, however, revelation improves
expected voter welfare on that issue by
\((1-\alpha)\pi_q(1-\pi_q)\) per voter. Thus the electoral model supplies no private return
that would finance a positive disclosure cost for purely common-interest information.

Now let \(\omega_x\in\{0,1\}\) determine which group benefits from the conflict issue:
\(\omega_x=1\) gives \((x_R,x_S)=(1,0)\), and \(\omega_x=0\) reverses the pair. If
\(\pi_x=\Pr(\omega_x=1)\) is the prior and \(\hat\pi_x\) is the posterior probability of
\(\omega_x=1\), the groups' perceived bliss points are \(\hat\pi_x\) and
\(1-\hat\pi_x\). Write
\[
a=\nu(\tfrac12)-\nu(0),\qquad
b=\nu(1)-\nu(\tfrac12),\qquad
\gamma=a-b>0.
\]
The two equilibrium platforms, indexed by the group they target, are
\[
x^R=a\hat\pi_x+b(1-\hat\pi_x),
\qquad
x^S=a(1-\hat\pi_x)+b\hat\pi_x,
\]
and their distance is
\[
|x^R-x^S|=\gamma|2\hat\pi_x-1|.
\]
Full revelation sets \(|2\hat\pi_x-1|=1\), maximizing platform distance and therefore
both parties' equilibrium payoffs. Relative to no revelation, the gross payoff gain for
each party is
\[
\frac{\alpha\gamma^2}{2\phi}
\left[1-(2\pi_x-1)^2\right],
\]
which is strictly positive unless the prior is degenerate. Whether voters share that
preference depends on how electoral power translates platforms into implemented policy. Common-interest information creates
no differentiation return, whereas information that clarifies opposing group interests
can create one.

\subsection{Party Identity}\label{app:identity}

My last application concerns party identity. By associating policy views with group
membership, identity shapes voters' perceptions and behavior \citep{Greene99}. Formal
models show that identification can shift beliefs and preferences toward positions that
distinguish the in-group from the outgroup \citep{SHAYO09,BGT21,GT23}. In this
environment, stronger party identity amplifies disagreement along the existing party
divide and increases equilibrium differentiation.

Following \citet{BGT21}, suppose that when voter $i$ identifies with group $G\in\{A,B\}$, her
identity-adjusted bliss point takes the reduced-form
\begin{equation}\label{identity}
\tilde{x}_i=x_i+\theta(x_G-x_{\bar{G}}),
\end{equation}
where $x_G$ is the policy typical of group $G$, $x_{\bar{G}}$ is the policy typical of the
outgroup, and $\theta>0$ captures the strength of identity effects.

To connect \eqref{identity} to the baseline model, start from a status quo voter
distribution \((\mathbf{x},\mathbf{s})\in\mathcal V\) and an associated party-preferred
equilibrium \(\hat x\in\mathcal E^\star(\mathbf{x},\mathbf{s})\). For every voter strictly
closer to one status quo platform, let \(x_G\) be that platform and \(x_{\bar G}\) the
other; leave voters equidistant from the two platforms unchanged. Conjugate voter types
then move by equal amounts in opposite directions along
\(d=\hat x_A-\hat x_B\). The transformed electorate remains centrally symmetric and is a
strict \(d\)-spread of the status quo electorate. Provided \(\nu\) is strictly concave
and the common wide-support condition holds for the union of the original and
identity-transformed supports, Proposition \ref{main:propspreadKD} therefore implies
\[
\hat V_p(\tilde{\mathbf x},\mathbf s)>
\hat V_p(\mathbf x,\mathbf s),
\qquad p=A,B.
\]
This comparison identifies the gross return to stronger identity alignment.

\begin{remark}
Stronger identity raises both parties' equilibrium payoffs when it aligns disagreements
along the existing party divide.
\end{remark}

The key effect is not dispersion on each issue separately. Party identity shift spreads
voter positions along the existing party divide, increases equilibrium platform
distance, and raises both parties' payoffs.

\subsubsection{Identity investment and polarization feedbacks}
\label{app:identityinvestment}

The following exercise adds a communication cost and endogenizes parties' investment in
identity salience, generating a feedback loop between voter polarization and party polarization similar to the one of \citet{CallanderCarbajal2022}. Consider two voter groups of equal mass with baseline ideals
\(x_L<x_R\), and let \(d=x_R-x_L\). Define
\[
\gamma=2\nu(\tfrac12)-\nu(1)-\nu(0)>0,
\]
the excess payoff gain from securing the first half of the electorate rather than the
second. With no identity effect, the two-group equilibrium has platform distance
\(D_0=\gamma d\).

Suppose elections recur. At the start of election \(t\geq1\), the distance
\(D_{t-1}\) between the preceding platforms defines partisan group contrast inherited by the
two groups. Each party simultaneously makes an identity campaign investment decision \(a_p^t\in\{0,1\}\), paying \(c>0\) if it
invests in communication that makes party identity more salient (\(a_p^t\)). The campaign is public:
by sharpening the partisan contrast, either party can strengthen identification among
voters in both camps, including the rival camp. Let
\[
\theta_t=\eta(a_A^t+a_B^t),\qquad \eta>0.
\]
The identity-adjusted ideals are
\begin{equation}\label{eq:identityinvestment}
\tilde x_L^t=x_L-\theta_tD_{t-1},
\qquad
\tilde x_R^t=x_R+\theta_tD_{t-1}.
\end{equation}
Thus one campaign raises common identity salience, and a second raises it by the same
amount.\footnote{If a campaign instead moves only the investing party's affiliates away
from the opposing camp by \(\eta_oD\), one investment increases voter distance by
\(\eta_oD\). A unilateral campaign also moves the electorate's midpoint, but translation
invariance makes the investment game payoff-equivalent to the common-salience model with
\(\eta=\eta_o/2\), provided policy-space boundaries do not bind.} After investment, parties play
the current electoral game. They maximize its payoff net of the current campaign cost,
which isolates the contemporaneous incentive to invest without adding forward-looking
motives.

For an inherited distance \(D>0\), let \(n\in\{0,1,2\}\) be the total number of
campaigns. Identity then increases the voter gap to \(d+2n\eta D\). Take
\(\mathcal X_D=[x_L-2\eta D,x_R+2\eta D]\) as a common feasible platform
interval across the three investment realizations; choices outside the current voter hull
are dominated. Proposition \ref{main:NE1D} gives current platform distance and each party's
gross equilibrium payoff as
\begin{equation}\label{eq:identitycontinuation}
D_n'(D)=\gamma(d+2n\eta D),
\qquad
\Pi_n(D)
=\frac{\nu(1)+\nu(0)}2
+\frac{\gamma^2(d+2n\eta D)^2}{2\phi}.
\end{equation}
Assume the wide-support condition holds for each of these three continuation
electorates. The private gross return to investing when the other party chooses
\(m\in\{0,1\}\) is
\begin{equation}\label{eq:Bmidentity}
B_m(D)
:=\Pi_{m+1}(D)-\Pi_m(D)
=\frac{2\gamma^2\eta D}{\phi}
  \bigl[d+(2m+1)\eta D\bigr].
\end{equation}
Both returns rise with the inherited platform gap, and
\[
B_1(D)-B_0(D)=\frac{4\gamma^2\eta^2D^2}{\phi}>0.
\]
The second campaign therefore has a larger private return than the first.

\begin{proposition}[Identity investment and feedback]\label{prop:identityinvestment}
Let \(D>0\) be the inherited platform distance, away from the boundary costs
\(B_0(D)\) and \(B_1(D)\). If \(c<B_0(D)\), investment is strictly dominant and
both parties invest in the unique investment equilibrium. If
\(B_0(D)<c<B_1(D)\), mutual investment and mutual noninvestment are both
pure-strategy investment equilibria. If \(c>B_1(D)\), noninvestment is strictly
dominant and neither party invests in the unique investment equilibrium.

Moreover, suppose
\[
c<B_0(D_0),\qquad
4\gamma\eta<1,\qquad
\phi>\left(\frac{d}{1-4\gamma\eta}\right)^2.
\]
Take as the common feasible policy interval
\[
\mathcal X
=\left[\mu-\frac{\bar d}{2},\mu+\frac{\bar d}{2}\right],
\qquad
\mu=\frac{x_L+x_R}{2},
\qquad
\bar d=\frac{d}{1-4\gamma\eta}.
\]
At every state generated by investment choices followed by equilibrium platform play,
investment is strictly dominant. Along the unique investment path, both parties invest
in every election and the equilibrium platform distance satisfies
\[
D_t
=D_0\sum_{s=0}^{t}(4\gamma\eta)^s
\longrightarrow
\frac{\gamma d}{1-4\gamma\eta}.
\]
The identity-adjusted distance between the two voter groups increases from \(d\) toward
\(\bar d\).
\end{proposition}

\begin{proof}
Equation \eqref{eq:Bmidentity} is the payoff gain from investing when the other party
chooses action \(m\). Since \(B_1(D)>B_0(D)\), comparison with \(c\) gives the three
pure-strategy regions.\footnote{At an intermediate cost, the action game also has the
standard symmetric mixed equilibrium in which each party invests with probability
\((c-B_0(D))/(B_1(D)-B_0(D))\). At the boundary costs, the corresponding best response
is weak.} Both returns are strictly increasing in \(D\), and every state generated by
investment choices followed by equilibrium platform play has platform distance at least
\(\gamma d=D_0\). Hence \(c<B_0(D_0)\) makes investment strictly dominant at every such
state. Mutual investment implies
\[
D_t=\gamma d+4\gamma\eta D_{t-1}.
\]
Solving this recurrence gives the stated geometric sum. The contraction condition keeps
every identity-adjusted electorate inside \(\mathcal X\), and
\(\phi>\bar d^2\) is a sufficient wide-support condition throughout.
\end{proof}

A public campaign benefits its sponsor even though the separation it creates also
protects the rival's constituency. Because an inherited platform gap raises both
\(B_0(D)\) and \(B_1(D)\), temporary polarization can move politics into the
mutual-investment region and thereby reinforce voter and platform polarization. In the
two-group proportional benchmark, concentrating organizational value among the
highest-return affiliates weakly raises \(\gamma\), strengthening the incentive to
invest and the feedback multiplier. With linear organizational value and proportional
power, \(\gamma=0\), so the electoral return to identity investment disappears.

\section{Additional Foundations, Welfare, and Robustness}

\subsection{Alternative Organizational Foundations}\label{AppConc}

Section \ref{main:organization} develops the paper's leading organizational foundation:
political power is allocated first to the highest-value affiliates or uses. This appendix
records two complementary mechanisms. Both are symmetric across parties and are compatible
with a constant-sum allocation of aggregate political power.

\paragraph{Rent Sharing and Diminishing Marginal Utility of Insiders}
A second, independent source of concavity arises because political rents are ultimately
consumed by individuals associated with the party
\citep{Strom1990BehavioralTheory,MullerStrom1999PolicyOfficeVotes,KatzMair1995CartelParty}.
Suppose that total political rents \(P\ge0\) are shared among \(n\ge1\) party insiders
to maximize
\[
U(P) \;=\; \max_{x_1,\ldots,x_n\ge0} \sum_{i=1}^n u(x_i)
\quad \text{s.t.} \quad \sum_{i=1}^n x_i = P,
\]
where \(u(\cdot)\) is strictly increasing and strictly concave. By symmetry, the optimal
allocation satisfies \(x_i=P/n\) for all \(i\), so that
\[
U(P) = n\,u\left(\frac{P}{n}\right).
\]
Because \(u\) is strictly concave, \(U(P)\) is strictly concave in \(P\). This
channel reflects diminishing marginal utility of monetary and quasi-monetary benefits
accruing to party members---including salaries, discretionary budgets, fundraising
capacity, and post-office opportunities. Concavity here comes from insiders' diminishing
marginal utility, without adding uncertainty or a separate party-level taste for
insurance.

\paragraph{Convex Organizational and Governance Costs}
Concavity may also arise from organizational and governance frictions internal to
parties. As political power expands, parties must coordinate a broader agenda, manage
larger coalitions, monitor more officeholders, and enforce discipline among more agents
\citep{McCubbinsNollWeingast1987AdministrativeProcedures,HuberShipan2002DeliberateDiscretion,MartinVanberg2011ParliamentsCoalitions}.
These activities entail real costs that may increase more than proportionally with the
scale of power.

To capture this channel, suppose that gross political rents equal \(P\), but
maintaining control entails a cost \(c(P)\), with \(0<c'(P)<1\) and
\(c''(P)>0\) on the relevant interval.
Net party payoffs are then
\[
U(P) = P - c(P),
\]
which is strictly concave in \(P\). This formulation captures the increasing
complexity of governing larger organizations and coalitions without relying on
uncertainty or heterogeneity in preferences.

These mechanisms capture distinct but complementary aspects of political organizations.
Ranked appointments and affiliated placements emphasize heterogeneity among insiders and
selection by value; rent sharing reflects diminishing marginal utility at the individual
level; and convex governance costs capture increasing organizational complexity. Each
mechanism is sufficient to generate strict concavity. Together, they show why Assumption
\ref{main:concavity} can represent how political power translates into effective benefits
within party organizations, rather than an ad hoc taste for insurance.

\subsection{Welfare and Majority Premia}\label{appendixA4}\label{welfare}

This appendix evaluates voter welfare under a
\emph{contestable-power-weighted implementation rule}.
Throughout, maintain the unidimensional assumptions of
section \ref{main:Conflict1D} (uniform shock $\epsilon\sim U(-\phi,\phi)$ and quadratic loss).

\paragraph{Implemented Policy}
Fix any platform profile $(x_A,x_B)$ and any shock realization $\epsilon$. Let
\[
s_A(x_A,x_B,\epsilon)\;:=\;\sum_{i=1}^N s_i\,\mathbf{1}\{\Delta_i(x_A,x_B)\ge \epsilon\}
\]
denote party \(A\)'s realized vote share. Let \(a=\rho(0)\), \(b=\rho(1)\), and
recall the contestable-power share
\[
r(s):=\frac{\rho(s)-a}{b-a}.
\]
Assumption \ref{main:officerents} implies \(r(s)\in[0,1]\) and
\(r(s)+r(1-s)=1\). The political process implements the
contestable-power-weighted compromise
\begin{equation*}
x^{\star}(x_A,x_B,\epsilon)
:=r\!\left(s_A(x_A,x_B,\epsilon)\right)x_A
+r\!\left(1-s_A(x_A,x_B,\epsilon)\right)x_B.
\end{equation*}
The weights are nonnegative and sum to one, so
\(x^{\star}(x_A,x_B,\epsilon)\in[\min\{x_A,x_B\},\max\{x_A,x_B\}]\).
Following the paper's convention, write the equilibrium platforms as
\(x_A=\bar{x}\) and \(x_B=\ubar{x}\). The induced implemented-policy random variable is
\[
X^{\star}(\epsilon):=x^{\star}(\bar{x},\ubar{x},\epsilon).
\]
For convenience, also define party \(A\)'s realized
\emph{share of contestable power}
\[
\Lambda(\epsilon)\;:=\;r\!\left(s_A(\bar{x},\ubar{x},\epsilon)\right),
\qquad
D\;:=\;\bar{x}-\ubar{x}>0,
\]
so that
\begin{equation}\label{eq:Xlambda}
X^{\star}(\epsilon)=\ubar{x}+\Lambda(\epsilon)\,D.
\end{equation}

\paragraph{Utilitarian Benchmark}
Let $x^o\in\arg\max_x\sum_i s_i v(x,x_i)$ be the utilitarian optimum.
Under quadratic loss, $x^o=\sum_i s_i x_i$.

\begin{remark}[Welfare benchmark]\label{rem1D}
If the voter distribution is symmetric or political power is proportional to vote shares,
then $\ubar{x}<x^o<\bar{x}$.
\end{remark}

\begin{proof}
If political power is proportional to vote share, the risk-neutral weights in the proof
of Proposition \ref{main:NE1D} satisfy
\[
r_i
=
\frac{\rho(\bar s_i)-\rho(\bar s_{i+1})}{\rho(1)-\rho(0)}
=s_i,
\]
and hence \(x^{RN}=x^o\). If instead the voter distribution is symmetric around \(c\),
every type is paired with its reflection around \(c\), with the same mass. Constant-sum
power gives paired types the same risk-neutral weight, so \(x^{RN}=c=x^o\). In either
case, Proposition \ref{main:NE1D}(i) yields \(\ubar x<x^o<\bar x\).
\end{proof}

\begin{lemma}[Decomposition]\label{lem:welfdecompB}
Let $\mathcal{W}^\star:=\mathbb{E}_\epsilon\big[\sum_i s_i v(X^{\star}(\epsilon),x_i)\big]$ be ex-ante utilitarian welfare. Under quadratic loss,
\begin{equation}\label{eq:welfdecompB}
\mathcal{W}^\star
=
\mathcal{W}(x^o)
-\Big(\mathbb{E}[X^{\star}]-x^o\Big)^2
-\mathrm{Var}(X^{\star}),
\end{equation}
where $\mathcal{W}(x^o):=\sum_i s_i v(x^o,x_i)$ is the first-best welfare level. Moreover, using \ref{eq:Xlambda},
\begin{equation}\label{eq:welfdecompB2}
\mathbb{E}[X^{\star}] = \ubar{x}+D\,\mathbb{E}[\Lambda],
\qquad
\mathrm{Var}(X^{\star}) = D^2\,\mathrm{Var}(\Lambda).
\end{equation}
\end{lemma}

\begin{proof}
Write $\sum_i s_i v(x,x_i)=C-\sum_i s_i(x-x_i)^2$ for a constant $C\in\mathbb{R}$. Under quadratic loss,
$\sum_i s_i(x-x_i)^2=(x-x^o)^2+\sum_i s_i(x_i-x^o)^2$, hence
$\sum_i s_i v(x,x_i)=\mathcal{W}(x^o)-(x-x^o)^2$. Taking expectations at $x=X^{\star}$ yields
$\mathcal{W}^\star=\mathcal{W}(x^o)-\mathbb{E}\big[(X^{\star}-x^o)^2\big]$, and the standard identity
$\mathbb{E}[(X-x^o)^2]=(\mathbb{E}X-x^o)^2+\mathrm{Var}(X)$ gives \ref{eq:welfdecompB}.
Finally, \ref{eq:welfdecompB2} follows from $X^{\star}=\ubar{x}+\Lambda D$.
\end{proof}

\begin{corollary}[Party insurance and voter welfare]\label{cor:insurancewelfare}
At the differentiated equilibrium, under the contestable-power-weighted implementation rule,
\[
\widehat V_p-\frac{\nu(1)+\nu(0)}{2}
=\frac{D^2}{2\phi}>0,
\qquad
\mathrm{Var}(X^\star)=D^2\mathrm{Var}(\Lambda)>0,
\qquad
\mathcal W^\star<\mathcal W(x^o).
\]
\end{corollary}

\begin{proof}
Proposition \ref{main:NE1D} and its equilibrium-payoff identity give the first statement.
At a differentiated equilibrium, voter policy advantages vary across voter types.
The wide-support condition places their switching thresholds in the interior of the
shock support, so realized vote share is nonconstant. Strict monotonicity of \(r\)
then makes \(\Lambda\) nondegenerate. Hence
\(\mathrm{Var}(X^\star)=D^2\mathrm{Var}(\Lambda)>0\). The strict welfare inequality
follows from Lemma \ref{lem:welfdecompB}.
\end{proof}

\paragraph{Varying the Majority Premium in a Proportional System}
To interpret comparative statics in $\rho_m$, it is useful to consider a one-parameter family of
\emph{proportional-with-premium} power mappings (holding $\bar\rho$ fixed):
\begin{equation}\label{eq:rhopremfam}
\rho_{\rho_m}(s) :=
\begin{cases}
(\bar\rho - \rho_m)\, s, & s < \tfrac12, \\[6pt]
\tfrac{\bar\rho}{2}, & s = \tfrac12, \\[6pt]
(\bar\rho - \rho_m)\, s + \rho_m, & s > \tfrac12,
\end{cases}
\end{equation}
for \(\rho_m \in [0,\bar\rho)\). This family satisfies Assumption
\ref{main:officerents} and has jump \(\rho_m\) at \(1/2\).
Write \(\ubar{x}(\rho_m)\) and \(\bar{x}(\rho_m)\) for the equilibrium platforms
under \(\rho_{\rho_m}\) (see Proposition \ref{main:NE1D}). Let \(X^\star_{\rho_m}\)
be the resulting implemented-policy random variable.

\Needspace{12\baselineskip}
\begin{lemma}[Majority-premium comparative statics and maximal-premium limit]
\label{lem:limitplatforms}
Under \eqref{eq:rhopremfam},
\[
D(\rho_m):=\bar{x}(\rho_m)-\ubar{x}(\rho_m)
\]
is weakly decreasing in \(\rho_m\). If
\[
\sum_{i<m}s_i<\frac12<\sum_{i\le m}s_i,
\]
then \(D(\rho_m)\) is strictly decreasing and
\[
\ubar{x}(\rho_m)\to x_m,
\qquad
\bar{x}(\rho_m)\to x_m,
\qquad
D(\rho_m)\to0
\quad\text{as }\rho_m\to\bar\rho^-.
\]
\end{lemma}

\begin{proof}
Write
\[
\ubar s_i=\sum_{j\le i}s_j,
\qquad
\bar s_i=\sum_{j\ge i}s_j,
\qquad
\Delta U=U(\bar\rho)-U(0)>0,
\]
and let \(\ubar w_i(\rho_m)\) and \(\bar w_i(\rho_m)\) denote the
equilibrium weights on the left- and right-positioned platforms. Define
\[
H(z):=
U(z)+U(\bar\rho-z)-U(0)-U(\bar\rho),
\qquad z\in[0,\bar\rho/2].
\]
For each \(i=2,\ldots,N\), Proposition~\ref{main:NE1D}(ii) and
telescoping give
\begin{align*}
\sum_{j=i}^N\bar w_j(\rho_m)
&=
\frac{U(\rho_{\rho_m}(\bar s_i))-U(0)}{\Delta U},\\
\sum_{j=i}^N\ubar w_j(\rho_m)
&=
\frac{U(\bar\rho)-U(\rho_{\rho_m}(\ubar s_{i-1}))}{\Delta U}.
\end{align*}
Because
\[
\bar s_i=1-\ubar s_{i-1}
\quad\text{and}\quad
\rho_{\rho_m}(s)+\rho_{\rho_m}(1-s)=\bar\rho,
\]
define
\[
z_i(\rho_m):=
\min\left\{
\rho_{\rho_m}(\ubar s_{i-1}),
\bar\rho-\rho_{\rho_m}(\ubar s_{i-1})
\right\}.
\]
The difference between the two preceding tail sums is therefore
\(H(z_i(\rho_m))/\Delta U\). Since both weight vectors sum to one,
summation by parts yields
\begin{align*}
D(\rho_m)
&=
\sum_{i=1}^N
\bigl(\bar w_i(\rho_m)-\ubar w_i(\rho_m)\bigr)x_i\\
&=
\sum_{i=2}^N
(x_i-x_{i-1})
\sum_{j=i}^N
\bigl(\bar w_j(\rho_m)-\ubar w_j(\rho_m)\bigr)\\
&=
\frac{1}{\Delta U}
\sum_{i=2}^N
(x_i-x_{i-1})H(z_i(\rho_m)).
\end{align*}

Strict concavity of \(U\) implies that \(H\) is strictly increasing on
\([0,\bar\rho/2]\). Indeed, for
\(0\le a<b\le\bar\rho/2\), strict concavity gives
\[
U(b)-U(a)
>
U(\bar\rho-a)-U(\bar\rho-b),
\]
and hence \(H(b)>H(a)\). Equation \eqref{eq:rhopremfam} implies
\[
z_i(\rho_m)=
\begin{cases}
(\bar\rho-\rho_m)
\min\{\ubar s_{i-1},1-\ubar s_{i-1}\},
& \ubar s_{i-1}\ne\tfrac12,\\[4pt]
\bar\rho/2,
& \ubar s_{i-1}=\tfrac12.
\end{cases}
\]
Thus each summand in the representation of \(D(\rho_m)\) is weakly
decreasing in \(\rho_m\), and it is strictly decreasing whenever
\(\ubar s_{i-1}\ne\tfrac12\). This proves weak monotonicity. Under
strict median crossing, no cumulative share equals one half. Since
equal bliss points have been consolidated, \(x_i-x_{i-1}>0\), and the
decrease is therefore strict.

It remains to establish the limits under strict median crossing. The
condition can be written as
\[
\ubar s_{m-1}<\frac12<\ubar s_m,
\]
and it also implies
\[
\bar s_{m+1}=1-\ubar s_m<\frac12
<1-\ubar s_{m-1}=\bar s_m.
\]
As \(\rho_m\to\bar\rho^-\), equation \eqref{eq:rhopremfam} gives
\[
\rho_{\rho_m}(s)\longrightarrow
\begin{cases}
0,&s<\tfrac12,\\
\bar\rho/2,&s=\tfrac12,\\
\bar\rho,&s>\tfrac12.
\end{cases}
\]

Consider first the weights on the left-positioned platform:
\[
\ubar w_i(\rho_m)
=
\frac{
U(\rho_{\rho_m}(\ubar s_i))
-
U(\rho_{\rho_m}(\ubar s_{i-1}))
}{\Delta U}.
\]
For \(i<m\), both cumulative shares are below one half, so
\(\ubar w_i(\rho_m)\to0\). For \(i>m\), both are above one half, so again
\(\ubar w_i(\rho_m)\to0\). At \(i=m\), the interval crosses one half
strictly, and hence \(\ubar w_m(\rho_m)\to1\).

The weights on the right-positioned platform satisfy
\[
\bar w_i(\rho_m)
=
\frac{
U(\rho_{\rho_m}(\bar s_i))
-
U(\rho_{\rho_m}(\bar s_{i+1}))
}{\Delta U}.
\]
The inequalities for \(\bar s_m\) and \(\bar s_{m+1}\) give
\(\bar w_m(\rho_m)\to1\), while
\(\bar w_i(\rho_m)\to0\) for every \(i\ne m\). Proposition
\ref{main:NE1D}(ii) therefore yields
\[
\ubar x(\rho_m)
=
\sum_i\ubar w_i(\rho_m)x_i
\longrightarrow x_m,
\qquad
\bar x(\rho_m)
=
\sum_i\bar w_i(\rho_m)x_i
\longrightarrow x_m.
\]
Their difference converges to zero.
\end{proof}

\begin{corollary}[Full Majority Limit]\label{cor:limitwelfareB}
Under the strict median-crossing condition of Lemma~\ref{lem:limitplatforms},
\[
X^\star_{\rho_m}\xrightarrow[\rho_m\to\bar\rho^-]{}\;x_m\quad\text{a.s.},\qquad
\mathrm{Var}(X^\star_{\rho_m})\to 0,\qquad
\mathcal{W}^\star(\rho_m)\to \mathcal{W}(x^o)-(x_m-x^o)^2.
\]
Hence, if $x_m=x^o$ (e.g.\ under a symmetric voter distribution),
then $\lim_{\rho_m\to\bar\rho^-}\mathcal{W}^\star(\rho_m)=\mathcal{W}(x^o)$ (first-best welfare).
\end{corollary}

\begin{proof}
$X^\star_{\rho_m}(\epsilon)\in[\ubar{x}(\rho_m),\bar{x}(\rho_m)]$ for all $\epsilon$,
so $|X^\star_{\rho_m}(\epsilon)-\ubar{x}(\rho_m)|\le D(\rho_m)$. Lemma \ref{lem:limitplatforms} implies
$\ubar{x}(\rho_m)\to x_m$ and $D(\rho_m)\to 0$, hence $X^\star_{\rho_m}\to x_m$ a.s.\ and $\mathrm{Var}(X^\star_{\rho_m})\to 0$.
The welfare limit follows from lemma \ref{lem:welfdecompB} and $\mathbb{E}[X^\star_{\rho_m}]\to x_m$.
\end{proof}

\begin{remark}[Exact-half case]\label{rem:exacthalflimit}
Strict median crossing is essential.
Suppose instead that, for some \(k<N\),
\[
\ubar s_k:=\sum_{i\le k}s_i=\frac12,
\qquad x_k<x_{k+1},
\]
and define
\[
\theta
:=
\frac{U(\bar\rho/2)-U(0)}
{U(\bar\rho)-U(0)}.
\]
Strict concavity of \(U\) implies \(\theta>\tfrac12\). Under
\eqref{eq:rhopremfam},
\[
\ubar x(\rho_m)\longrightarrow
\theta x_k+(1-\theta)x_{k+1},
\qquad
\bar x(\rho_m)\longrightarrow
(1-\theta)x_k+\theta x_{k+1}.
\]
Thus
\[
D(\rho_m)\longrightarrow
(2\theta-1)(x_{k+1}-x_k)>0.
\]
Under the maintained contestable-power-weighted implementation rule and wide-support
condition, implemented-policy variance also remains bounded away from zero as
\(\rho_m\to\bar\rho^-\).
\end{remark}

\begin{proof}
In the maximal-premium limit, all weights on the left-positioned platform vanish except
those on \(x_k\) and \(x_{k+1}\). Their respective limits are
\[
\frac{U(\bar\rho/2)-U(0)}{U(\bar\rho)-U(0)}
=\theta
\quad\text{and}\quad
\frac{U(\bar\rho)-U(\bar\rho/2)}{U(\bar\rho)-U(0)}
=1-\theta.
\]
The weights on the right-positioned platform reverse these two values. Proposition
\ref{main:NE1D}(ii) gives the displayed platform limits. Strict concavity places
\(U(\bar\rho/2)\) strictly above
the chord joining \(U(0)\) and \(U(\bar\rho)\), so \(\theta>\tfrac12\).

Finally, let
\[
M_\Delta:=
\sup_{a,b\in[x_1,x_N],\,i}|\Delta_i(a,b)|<\phi,
\qquad
c:=\frac{\phi-M_\Delta}{2\phi}>0.
\]
For every \(\rho_m\), the probability that \(A\) receives all votes and implemented
policy equals \(\bar x(\rho_m)\), and the probability that \(A\) receives no votes and
policy equals \(\ubar x(\rho_m)\), are each at least \(c\). If \(X'\) is an independent
copy of \(X^\star_{\rho_m}\), then
\[
\operatorname{Var}(X^\star_{\rho_m})
=\frac12\mathbb E[(X^\star_{\rho_m}-X')^2]
\ge c^2D(\rho_m)^2.
\]
The positive limit of \(D(\rho_m)\) therefore implies that implemented-policy variance
is bounded away from zero.
\end{proof}

\paragraph{Discussion}
Lemma \ref{lem:welfdecompB} shows that voter welfare depends on
(i) the \emph{bias} of the mean implemented policy relative to $x^o$ and (ii) the
\emph{variance} of the implemented policy. Increasing \(\rho_m\) affects both terms
through the endogenous platform responses
\((\ubar{x}(\rho_m),\bar{x}(\rho_m))\) and the induced distribution of the power share
\(\Lambda(\epsilon)\) under \(\rho_{\rho_m}\). If \(x_m\ne x^o\), designing the majority
premium to maximize welfare can therefore entail a trade-off between reducing variance
and reducing bias.

\subsection{Algorithmic Equilibrium Search}\label{appendixA2}
The finite candidate set $\mathcal C$ can be constructed directly from voter rankings.
Let $\mathcal C_0=\emptyset$, construct $\mathcal R(I)$ as the set of permutations of
$I$, and denote its elements by $R^1,\ldots,R^{|\mathcal R(I)|}$. For each $r$, (i)
compute $\bar x(R^r)$ and $\ubar x(R^r)$; (ii) compute
$\mathcal r(\bar x(R^r),\ubar x(R^r))$; and (iii), if
$R^r\in\mathcal r(\bar x(R^r),\ubar x(R^r))$, set
\[
\mathcal C_r=\mathcal C_{r-1}\cup\{(\bar x(R^r),\ubar x(R^r))\},
\]
and otherwise set $\mathcal C_r=\mathcal C_{r-1}$. Then
$\mathcal C_{|\mathcal R(I)|}=\mathcal C$. Under the hypotheses of Proposition
\ref{main:existence}---in particular, for a centrally symmetric finite electorate and a
strictly concave reduced-form payoff \(\nu\)---the party-preferred equilibria are the
elements of $\mathcal C$ that maximize platform distance.

\subsection{Robustness to Voter-Level Idiosyncratic Shocks}\label{Robustness}

In the environment of section \ref{main:Conflict1D}, treat each voter type as an atomless
continuum. In addition to the aggregate shock
\(\epsilon\sim U(-\phi,\phi)\), voters have idiosyncratic relative-valence shocks
\(\eta\) whose exact cross-sectional distribution within every type is
\(U[-\kappa,\kappa]\), independently of \(\epsilon\). A type-\(i\) voter supports
\(A\) if and only if
\[
\Delta_i(x_A,x_B)+\eta\ge \epsilon.
\]
For \(\kappa>0\), let \(G_\kappa\) denote the cdf of this uniform distribution:
\[
G_\kappa(t)=
\begin{cases}
0 & \text{if } t\le -\kappa,\\
\frac{t+\kappa}{2\kappa} & \text{if } -\kappa<t<\kappa,\\
1 & \text{if } t\ge \kappa.
\end{cases}
\]
and define \(G_0(t)=\mathbf 1\{t\ge0\}\). By symmetry of the uniform distribution,
the exact fraction of type \(i\) supporting \(A\), conditional on \(\epsilon\), is
\(G_\kappa(\Delta_i-\epsilon)\). Party \(A\)'s vote share is therefore
\[
S_A^\kappa(\epsilon;x_A,x_B)=\sum_{i=1}^N s_i G_\kappa(\Delta_i(x_A,x_B)-\epsilon),
\]
so that
\[
V_A^\kappa(x_A,x_B)=\frac{1}{2\phi}\int_{-\phi}^{\phi}\nu(S_A^\kappa(\epsilon;x_A,x_B))\,d\epsilon.
\]
Define $V_B^\kappa$ analogously.

\begin{proposition}[Small voter-level noise]\label{prop:idio}
Fix a voter distribution $(\mathbf{x},\mathbf{s})$ and let $(\ubar{x},\bar{x})$ be the equilibrium characterized in proposition \ref{main:NE1D}, with $\ubar{x}<\bar{x}$. There exist $\bar{\kappa}>0$, a neighborhood $\mathcal{N}^+$ of $(\bar{x},\ubar{x})$, and a neighborhood \(\mathcal N^-\) of \((\ubar x,\bar x)\) such that, for every $\kappa\in[0,\bar{\kappa})$, the game with idiosyncratic shocks admits $(\bar{x},\ubar{x})$ and $(\ubar{x},\bar{x})$ as strict pure-strategy equilibria.

Moreover, for every \(\kappa\in[0,\bar\kappa)\) and every
\((x_A,x_B)\in\mathcal{N}^+\) with $x_A>x_B$,
\[
V_A^\kappa(x_A,x_B)
=
C_\kappa(\mathbf{s})
+\nu(0)
+\sum_{i=1}^N F(\Delta_i(x_A,x_B))\big(\nu(\bar{s}_i)-\nu(\bar{s}_{i+1})\big),
\]
for a constant $C_\kappa(\mathbf{s})$ independent of $(x_A,x_B)$. Hence
\[
\frac{\partial V_A^\kappa(x_A,x_B)}{\partial x_A}
=
\frac{1}{\phi}\sum_{i=1}^N (x_i-x_A)\big(\nu(\bar{s}_i)-\nu(\bar{s}_{i+1})\big),
\]
and symmetrically, for every \(\kappa\in[0,\bar\kappa)\) and every
\((x_A,x_B)\in\mathcal N^-\) with $x_A<x_B$,
\[
\frac{\partial V_A^\kappa(x_A,x_B)}{\partial x_A}
=
\frac{1}{\phi}\sum_{i=1}^N (x_i-x_A)\big(\nu(\ubar{s}_i)-\nu(\ubar{s}_{i-1})\big).
\]
Consequently, the local first-order conditions coincide with those in proposition \ref{main:NE1D}.
\end{proposition}

\begin{proof}
Consolidate voter types with identical bliss points, so that
\(x_1<\cdots<x_N\) without changing the game.
Let $x^\star=(\bar{x},\ubar{x})$, and note that
\[
\Delta_{i+1}(x^\star)-\Delta_i(x^\star)=2(\bar{x}-\ubar{x})(x_{i+1}-x_i)>0
\]
for each $i=1,\ldots,N-1$, while $-\phi<\Delta_1(x^\star)<\Delta_N(x^\star)<\phi$ by the support assumption used in section \ref{main:Conflict1D}. Hence,
\[
g:=\min\left\{\phi-\Delta_N(x^\star),\ \Delta_1(x^\star)+\phi,\ \min_{i<N}\big(\Delta_{i+1}(x^\star)-\Delta_i(x^\star)\big)\right\}>0.
\]
Fix \(\bar{\kappa}_1<g/8\). By continuity of
\((\Delta_i)_{i\in I}\), choose a single neighborhood \(\mathcal N^+\) of
\(x^\star\), independent of \(\kappa\), on which
\[
\phi-\Delta_N>\frac g2,\qquad
\Delta_1+\phi>\frac g2,\qquad
\Delta_{i+1}-\Delta_i>\frac g2\quad(i<N),
\]
and shrink it so that \(x_A>x_B\) throughout. Then, for every
\(0<\kappa<\bar\kappa_1\) and every
\((x_A,x_B)\in\mathcal N^+\),
\begin{equation}\label{eq:nonoverlap}
-\phi<\Delta_1-\kappa<\Delta_1+\kappa<\cdots<\Delta_N-\kappa<\Delta_N+\kappa<\phi.
\end{equation}
For \(\kappa=0\), the baseline rank-weight identity applies directly. For
\(0<\kappa<\bar\kappa_1\), as $\epsilon$ increases, voter groups peel off one at a time. Therefore,
\[
S_A^\kappa(\epsilon;x_A,x_B)=
\begin{cases}
1, & \epsilon\in[-\phi,\Delta_1-\kappa],\\
\bar{s}_{i+1}+s_i\frac{\Delta_i+\kappa-\epsilon}{2\kappa}, & \epsilon\in[\Delta_i-\kappa,\Delta_i+\kappa],\ i=1,\ldots,N,\\
\bar{s}_{i+1}, & \epsilon\in[\Delta_i+\kappa,\Delta_{i+1}-\kappa],\ i=1,\ldots,N-1,\\
0, & \epsilon\in[\Delta_N+\kappa,\phi].
\end{cases}
\]
Integrating and performing the change of variables $u=(\Delta_i+\kappa-\epsilon)/(2\kappa)$ on each ramp yields
\[
V_A^\kappa(x_A,x_B)
=
C_\kappa(\mathbf{s})
+\nu(0)
+\sum_{i=1}^N F(\Delta_i(x_A,x_B))\big(\nu(\bar{s}_i)-\nu(\bar{s}_{i+1})\big),
\]
where $C_\kappa(\mathbf{s})$ depends only on $(\mathbf{s},\kappa,\nu,\phi)$.
Set \(C_0(\mathbf s)=0\); equation \eqref{main:eq:rankidentity} then gives the same
representation at \(\kappa=0\). Differentiating and using
\[
\frac{\partial \Delta_i(x_A,x_B)}{\partial x_A}=2(x_i-x_A)
\]
gives the displayed derivative formula. Since these weights sum to $\nu(1)-\nu(0)=1$, the corresponding second derivative is $-1/\phi<0$, so the local best response is unique. Applying the same uniform construction at the swapped profile
\((\ubar x,\bar x)\) gives a neighborhood \(\mathcal N^-\), independent of
\(\kappa\), and yields the formula with $(\ubar{s}_i)_{i\le N}$ whenever
\(x_A<x_B\). Party-label symmetry permits the same bound
\(\bar\kappa_1\) for both neighborhoods.

Since proposition \ref{main:NE1D} implies that $x^\star=(\bar{x},\ubar{x})$ is a strict equilibrium of the baseline game on the compact action set $[x_1,x_N]^2$, choose neighborhoods $U_A\ni\bar{x}$ and $U_B\ni\ubar{x}$ small enough that
\[
U_A\times U_B\subseteq\mathcal N^+,
\qquad
U_B\times U_A\subseteq\mathcal N^-.
\]
Compactness and strictness then give \(\delta>0\) such that
\[
V_A^0(\bar{x},\ubar{x})\ge V_A^0(x_A,\ubar{x})+3\delta \quad\text{for all }x_A\notin U_A,
\]
\[
V_B^0(\bar{x},\ubar{x})\ge V_B^0(\bar{x},x_B)+3\delta \quad\text{for all }x_B\notin U_B.
\]
By party-label symmetry, the corresponding two inequalities hold at
\((\ubar x,\bar x)\), with \(U_B\) and \(U_A\) as the respective own-action
neighborhoods and with the same \(\delta\).
Finally, for every $(x_A,x_B)\in[x_1,x_N]^2$,
\[
|V_p^\kappa(x_A,x_B)-V_p^0(x_A,x_B)|\le \frac{N\kappa}{\phi},
\]
because, for each type $i$, $G_\kappa(\Delta_i-\epsilon)$ differs from $\mathbf{1}\{\Delta_i\ge \epsilon\}$ only when $|\Delta_i-\epsilon|\le \kappa$, a set of $\epsilon$-measure at most $2\kappa$, and $\nu(1)-\nu(0)=1$. Choosing $\bar{\kappa}_2=\delta\phi/N$, the above strict inequalities continue to hold under $V_p^\kappa$ for every $\kappa<\bar{\kappa}_2$. Inside $U_A\times U_B$, the first-order conditions are unchanged, so $\bar{x}$ and $\ubar{x}$ remain the unique local best responses to one another. Hence $(\bar{x},\ubar{x})$ is a strict pure-strategy equilibrium of the perturbed game for every $\kappa<\bar{\kappa}:=\min\{\bar{\kappa}_1,\bar{\kappa}_2\}$. The same argument on
\(U_B\times U_A\subseteq\mathcal N^-\) establishes the symmetric profile
\((\ubar{x},\bar{x})\) under the same bound \(\bar\kappa\).
\end{proof}

\begin{remark}
For general voter distributions, the non-overlap condition used in the proof of
Proposition \ref{prop:idio} is a sufficient device ensuring that the perturbed
first-order conditions coincide exactly with those of Proposition \ref{main:NE1D}. Conditional
on non-overlap in the relevant neighborhoods, the baseline alienation result and other
bliss-point comparative statics that hold voter shares and their ideological ordering
fixed remain unchanged along these persisting equilibrium branches:
\(C_\kappa(\mathbf s)\) is independent of platforms and bliss points. Non-overlap is not
generally necessary, but once the ramps overlap, the first-order conditions acquire
additional terms and require a separate analysis.
\end{remark}

\subsubsection{When idiosyncratic noise restores convergence}
\label{app:largeridio}

The preceding result follows the differentiated equilibrium from zero voter-level noise.
At the other extreme, sufficiently dispersed idiosyncratic shocks eliminate the
electoral value of sorting voters into durable blocs. The next result does not require a
symmetric electorate or concavity of the payoff from votes.

Let
\[
\mu_x=\sum_{i=1}^N s_i x_i,
\qquad
M_\Delta
=
\max_{\substack{x_A,x_B\in[x_1,x_N]\\i\in I}}
|\Delta_i(x_A,x_B)|.
\]

\begin{proposition}[Large idiosyncratic noise]
\label{prop:largeidio}
If \(\kappa>\phi+M_\Delta\), each party has the unique strictly dominant platform
\(\mu_x\). Hence the unique equilibrium is \((\mu_x,\mu_x)\).
\end{proposition}

\begin{proof}
The condition implies
\(-\kappa<\Delta_i(x_A,x_B)-\epsilon<\kappa\) for every voter type, platform
profile, and aggregate-shock realization. Every type is therefore on the affine portion
of \(G_\kappa\), and
\[
S_A^\kappa(\epsilon;x_A,x_B)
=
\frac12+
\frac{\sum_i s_i\Delta_i(x_A,x_B)-\epsilon}{2\kappa}.
\]
Under quadratic policy loss,
\[
\sum_i s_i\Delta_i(x_A,x_B)
=
(x_B-\mu_x)^2-(x_A-\mu_x)^2.
\]
For every \(x_B\) and every \(\epsilon\), party \(A\)'s vote share is thus uniquely
maximized by \(x_A=\mu_x\). Because \(\nu\) is strictly increasing, the same platform
uniquely maximizes its expected payoff. The symmetric argument applies to party \(B\).
\end{proof}

The bound in Proposition \ref{prop:largeidio} is sufficient rather than necessary: it
requires voter-level noise to be large enough that all voters remain on the linear part
of their response for every relevant platform and aggregate shock. It nevertheless
establishes a global conclusion. Once voter-specific considerations dominate both common
electoral news and policy utility differences, ideological sorting no longer insures
parties against aggregate swings, and convergence to the electorate's mean bliss point
is the unique outcome.

\subsubsection{The transition: an exact two-bloc phase diagram}
\label{app:idiophasediagram}

A two-group example characterizes what happens between the small- and large-noise
limits. Consider two equally sized voter groups with bliss points \(-a\) and \(a\),
where \(a>0\), and let both parties choose platforms in \([-a,a]\). Normalize
\(\nu(1)-\nu(0)=1\), and suppose that \(\nu\) is twice continuously differentiable,
strictly increasing, and satisfies \(\nu''<0\). Define
\[
\gamma
=
2\nu\left(\frac12\right)-\nu(1)-\nu(0)>0,
\qquad
\chi
=
\nu'(0)-\nu'(1)>0.
\]
This smooth, concave specialization obtains, for example, under proportional power and a
twice continuously differentiable, strictly concave organizational value function.
Here \(\gamma\) is the insurance value of splitting the electorate into two
non-overlapping blocs, while \(\chi\) measures how sharply the marginal value of
electoral support declines across vote shares and governs the local return to separating
coincident switching ramps. For \(s\in[0,1/2]\), set
\[
\Gamma(s)
=
\nu(s)+\nu(1-s)-\nu(0)-\nu(1).
\]
Strict concavity implies that \(\Gamma\) is strictly increasing and strictly concave,
\(\Gamma(0)=0\), \(\Gamma(1/2)=\gamma\), and
\(\Gamma'(0)=\chi\). Strict monotonicity of \(\nu\) also gives \(\gamma<1\), and
concavity gives \(\chi\geq2\gamma\).

For \(\kappa>0\), define
\[
J_\kappa(h)
=
\int_0^h
\Gamma\left(\min\left\{\frac{t}{2\kappa},\frac12\right\}\right)dt,
\qquad h\geq0,
\]
and let \(J_0(h)=\gamma h\). Fix
\(\bar\kappa>2a^2\chi\) and assume
\(\phi>\bar\kappa+4a^2\), so the aggregate-shock support is wide enough for every
\(\kappa\in[0,\bar\kappa]\) considered below.

\begin{proposition}[Noise and equilibrium polarization]
\label{prop:idiophase}
Define
\[
q=x_B^2-x_A^2,
\qquad
h=2a|x_A-x_B|
\]
for any platforms \(x_A,x_B\in[-a,a]\).
For a constant \(C_\kappa\) independent of the platforms, expected payoffs satisfy
\begin{align}
V_A^\kappa(x_A,x_B)
&=
C_\kappa+\frac{q+J_\kappa(h)}{2\phi},
\label{eq:idiopayoffA}\\
V_B^\kappa(x_A,x_B)
&=
C_\kappa+\frac{-q+J_\kappa(h)}{2\phi}.
\label{eq:idiopayoffB}
\end{align}
The game is an exact potential game with potential
\begin{equation}
\Phi_\kappa(x_A,x_B)
=
C_\kappa+
\frac{-x_A^2-x_B^2+J_\kappa(2a|x_A-x_B|)}{2\phi}.
\label{eq:idiopotential}
\end{equation}

The pure-strategy equilibrium set has three regions:
\begin{enumerate}
\item If \(0\leq\kappa<a^2\chi\), the only equilibria, up to party labels, are
      \((z_\kappa,-z_\kappa)\).
\item If \(a^2\chi\leq\kappa<2a^2\chi\), both
      \((z_\kappa,-z_\kappa)\), up to party labels, and convergence at \((0,0)\)
      are equilibria. Both parties strictly prefer the differentiated equilibria.
\item If \(2a^2\chi\leq\kappa\leq\bar\kappa\), convergence at \((0,0)\) is the
      unique equilibrium.
\end{enumerate}
Here \(z_0=a\gamma\); for \(0<\kappa<2a^2\chi\), \(z_\kappa\) is the unique positive
solution to
\begin{equation}
z_\kappa
=
a\Gamma\left(
\min\left\{\frac{2az_\kappa}{\kappa},\frac12\right\}
\right).
\label{eq:idiofixedpoint}
\end{equation}
Whenever the differentiated equilibrium exists, it is party-preferred and its platform
distance is \(2z_\kappa\). That distance remains equal to its
no-idiosyncratic-noise value \(2a\gamma\) for
\(0\leq\kappa\leq4a^2\gamma\), then decreases strictly and continuously to zero as
\(\kappa\) approaches \(2a^2\chi\); for larger \(\kappa\), the party-preferred distance
is zero.
\end{proposition}

\begin{proof}
The two groups' switching thresholds are \(q-h\) and \(q+h\). The support condition
places both transition intervals strictly inside \([-\phi,\phi]\). Translating both
thresholds by \(q\) lengthens the region in which \(A\) receives all votes by \(q\) and
shortens the region in which it receives none by the same amount. Since
\(\nu(1)-\nu(0)=1\), this adds \(q\) to the integral defining \(2\phi V_A^\kappa\).

It remains to measure the effect of separating the two thresholds around zero. Suppose
first that \(0\leq h\leq\kappa\), and write \(r=h/(2\kappa)\). Across the five
successive regions of the aggregate shock, party \(A\)'s support equals one, falls from
one to \(1-r\), falls from \(1-r\) to \(r\), falls from \(r\) to zero, and equals zero.
A change of variables from the aggregate shock to vote share gives the full
aggregate-shock integral at \(q=0\):
\[
\begin{split}
I_\kappa(h)
={}&
(\phi-h-\kappa)\{\nu(1)+\nu(0)\}\\
&+4\kappa\left\{
\int_{1-r}^1\nu(s)\,ds+\int_0^r\nu(s)\,ds
\right\}
+2\kappa\int_r^{1-r}\nu(s)\,ds .
\end{split}
\]
Differentiation yields
\[
I_\kappa'(h)
=
\nu(r)+\nu(1-r)-\nu(0)-\nu(1)
=
\Gamma(r).
\]
If \(h\geq\kappa\), the two transition ramps no longer overlap. Increasing \(h\)
expands the interval on which each party receives half the votes and contracts the two
unanimous-vote tails, so
\[
I_\kappa'(h)
=
2\nu\left(\frac12\right)-\nu(1)-\nu(0)
=
\gamma.
\]
Consequently,
\(I_\kappa(h)-I_\kappa(0)=J_\kappa(h)\). The same calculation at
\(\kappa=0\) gives \(J_0(h)=\gamma h\). This proves
\eqref{eq:idiopayoffA}; party symmetry gives \eqref{eq:idiopayoffB}. Each unilateral
change in \eqref{eq:idiopotential} therefore equals the corresponding change in that
party's payoff, proving the potential claim.

We next characterize convergence. A common platform other than zero cannot be an
equilibrium: a party can move marginally toward zero, while the separation term has
zero derivative at convergence when \(\kappa>0\); at \(\kappa=0\), an arbitrarily small
separation is strictly profitable. Against a rival at zero, a deviation to \(x\) changes
party \(A\)'s payoff by
\[
\frac{-x^2+J_\kappa(2a|x|)}{2\phi}.
\]
For \(\kappa>0\), concavity gives \(\Gamma(s)\leq\chi s\). Moreover,
\(\gamma\leq\chi/2\), so for every \(h\geq0\),
\[
J_\kappa'(h)
\leq
\frac{\chi h}{2\kappa},
\qquad
J_\kappa(h)
\leq
\frac{\chi h^2}{4\kappa}.
\]
It follows that \((0,0)\) is an equilibrium if \(\kappa\geq a^2\chi\).
If \(0<\kappa<a^2\chi\), the expansion
\[
J_\kappa(2a|x|)
=
\frac{\chi a^2}{\kappa}x^2+o(x^2)
\]
makes a sufficiently small deviation profitable. Thus convergence is an equilibrium
if and only if \(\kappa\geq a^2\chi\); at \(\kappa=0\), the linear gain
\(J_0(2a|x|)=2a\gamma|x|\) makes convergence unprofitable.

Consider next a differentiated equilibrium with \(x_A>x_B\). Since
\(J_\kappa'(h)\leq\gamma<1\), neither party can be best responding at the boundary of
\([-a,a]\). The interior first-order conditions from
\eqref{eq:idiopayoffA}--\eqref{eq:idiopayoffB} are
\[
x_A=aJ_\kappa'(h),
\qquad
x_B=-aJ_\kappa'(h).
\]
Hence every differentiated equilibrium is centered: write
\((x_A,x_B)=(z,-z)\). Since \(h=4az\), the first-order condition is precisely
\eqref{eq:idiofixedpoint}. For \(\kappa>0\), the right side of that equation minus
\(z\) is strictly concave before it reaches its plateau, is zero at \(z=0\), and has
right derivative
\[
\frac{2a^2\chi}{\kappa}-1
\]
there. It has a unique positive zero if \(\kappa<2a^2\chi\), and none if
\(\kappa\geq2a^2\chi\). The same conclusion at \(\kappa=0\) follows from
\(z=a\gamma\).

To verify sufficiency and exclude further equilibria, parameterize platform profiles by
their midpoint \(m=(x_A+x_B)/2\) and half-distance
\(z=|x_A-x_B|/2\). Apart from the positive factor \(1/(2\phi)\) and an additive
constant, the potential is
\[
-2m^2-2z^2+J_\kappa(4az).
\]
It is uniquely maximized at \(m=0\). When the positive solution \(z_\kappa\) exists,
the derivative in \(z\) is positive before \(z_\kappa\) and negative after it, so the
two label-swapped profiles are global potential maximizers and hence equilibria. The
first-order argument above exhausts all possible differentiated equilibria. Combining
this fact with the convergence condition gives the three stated regions. At a
differentiated profile, each party gains
\(J_\kappa(4az_\kappa)/(2\phi)>0\) relative to convergence, proving the
party-preference claim.

Finally, \(z=a\gamma\) solves \eqref{eq:idiofixedpoint} exactly when
\(\kappa\leq4a^2\gamma\). This cutoff is no larger than \(2a^2\chi\).
Above it, the solution is interior to the overlapping-ramp region and satisfies
\[
z_\kappa=a\Gamma\left(\frac{2az_\kappa}{\kappa}\right).
\]
Strict concavity implies
\(\Gamma(r)>r\Gamma'(r)\) for \(r>0\). Implicit differentiation therefore gives
\(dz_\kappa/d\kappa<0\). Continuity and the preceding existence cutoff imply
\(z_\kappa\to0\) as \(\kappa\uparrow2a^2\chi\).
\end{proof}

The phase diagram separates two effects that a small-noise robustness statement cannot.
As long as the two groups' switching ramps do not overlap at the party-preferred
platforms, voter-level noise changes neither the platforms nor their marginal incentives.
Once the ramps overlap, voter-specific considerations erode the value of maintaining
distinct partisan blocs, and the preferred platforms move together. Convergence becomes
sustainable before differentiation disappears, creating an intermediate region in which
history or adjustment dynamics can matter. Eventually, only convergence remains. The
two-bloc benchmark thus predicts stronger platform polarization when electoral movement
is organized by common swings rather than predominantly idiosyncratic voting.

\paragraph{The most attractive direction away from convergence.}
The same calculation links voter-level noise to cleavage selection in several
dimensions. Let a finite or compactly supported electorate have mean \(\mu\) and
positive-definite covariance matrix \(\Sigma\), retain quadratic Euclidean loss, and
suppose the feasible platform domain contains a neighborhood of \(\mu\). For
\(\kappa>0\), assume
\(\phi>\kappa+M_\Delta\), where \(M_\Delta\) is the largest absolute policy advantage
on that domain, so the aggregate-shock support covers every idiosyncratic transition
interval. At
\((x_A,x_B)=(\mu,\mu)\), the second derivative of party \(A\)'s payoff along a
unilateral deviation \(tv\) is
\[
\left.
\frac{d^2}{dt^2}
V_A^\kappa(\mu+tv,\mu)
\right|_{t=0}
=
\frac1\phi
\left\{
\frac{\chi}{\kappa}v^\top\Sigma v-\|v\|^2
\right\}.
\]
To obtain the expression, write a voter's policy advantage after the deviation as
\[
\Delta(t)
=
2t\,v^\top(X-\mu)-t^2\|v\|^2.
\]
At \(t=0\), the first derivative of aggregate support vanishes because
\(\mathbb E[X-\mu]=0\). Applying the Leibniz rule to the two endpoints of the uniform
idiosyncratic-shock density in the second derivative contributes
\(\chi v^\top\Sigma v/(\kappa\phi)\), while the direct loss in mean policy appeal
contributes \(-\|v\|^2/\phi\).
Thus convergence is locally strict when
\(\kappa>\chi\lambda_1(\Sigma)\), and it is locally unstable when
\(\kappa<\chi\lambda_1(\Sigma)\). In the latter case, the most profitable incipient
deviation follows a leading principal component. This second-order diagnostic does not
replace the global equilibrium characterization in Proposition \ref{prop:idiophase};
it identifies the leading principal component as the locally most attractive direction
of departure from convergence. For elliptical electorates, that direction coincides
with the globally selected cleavage characterized in Theorem \ref{main:pcatheorem}.

%% file: appendix.tex
\newpage
\appendix
\vspace{1cm}
\section{Proofs and Core Auxiliary Results}\label{appendixproofs}
\vspace{1cm}

\subsection{Preliminaries and divergence}
\paragraph{Rank-weight identity}
The following identity will be used repeatedly. At any platform profile, let \(R\) be a ranking of voter
types such that
\(\Delta_{R_1}\le\cdots\le\Delta_{R_N}\), and write
\[
\bar s_{R_i}:=\sum_{j=i}^N s_{R_j},
\qquad
\bar s_{R_{N+1}}:=0.
\]
Then
\begin{equation}\label{eq:rankidentity}
V_A(x_A,x_B)
=
\nu(0)+
\sum_{i=1}^N
F(\Delta_{R_i}(x_A,x_B))
\bigl[\nu(\bar s_{R_i})-\nu(\bar s_{R_{i+1}})\bigr].
\end{equation}
Indeed, as the common shock crosses \(\Delta_{R_i}\), type \(R_i\) leaves party \(A\)
and its vote share falls from \(\bar s_{R_i}\) to \(\bar s_{R_{i+1}}\). Integrating
this ``staircase'' gives \eqref{eq:rankidentity}. If several thresholds coincide,
the corresponding terms telescope, so the identity is independent of how tied types
are ordered. It therefore remains valid when \(\rho\), and hence \(\nu\), has jumps.
Because \(F\) and the thresholds are continuous and adjacent formulas agree at every
ranking tie, expected payoffs are continuous in platforms. Throughout the proofs,
\(\nu(1)-\nu(0)=1\) by normalization.

\paragraph{Proof of Remark \ref{remarknu}}
Let \(a=\rho(s)\), \(b=\rho(s')\), and \(d=b-a>0\). Assumption
\ref{officerents} implies \(\rho(\tfrac12)=\bar\rho/2\), so
\[
a<b\le \frac{\bar\rho}{2}\le \bar\rho-b.
\]
It also implies
\[
\rho(1-s')=\bar\rho-b,
\qquad
\rho(1-s)=\bar\rho-a.
\]
The two payoff increments in Remark \ref{remarknu} are therefore increments of \(U\)
over the equal-length intervals \([a,b]\) and
\([\bar\rho-b,\bar\rho-a]\). The first interval lies strictly to the left of the
second because \(a<\bar\rho-b\). Under strict concavity, equal-length secant increments
strictly decrease as their interval moves to the right. Hence
\[
U(b)-U(a)>U(\bar\rho-a)-U(\bar\rho-b),
\]
which is the claimed inequality.
\(\blacksquare\)

\paragraph{Proof of Proposition \ref{divergence}}
Suppose, toward a contradiction, that a pure-strategy equilibrium
\(x^\star\in\mathbb{R}^{2K}\) is convergent: \(x_A^\star=x_B^\star\). Because the electorate is not perfectly homogeneous, there is an issue \(k\) on which at least two voter types have
different bliss points. Relabel the types so that
\(x_1^k\le x_2^k\le\cdots\le x_N^k\), with at least one strict inequality.

The function \(h^k(|a-z|)\) has strictly increasing differences in \((a,z)\): its
derivative with respect to \(a\) is strictly increasing in \(z\), by strict concavity of
\(h^k\) and \(h^{k\prime}(0)=0\). Let \(x_A(t)\) coincide with \(x_A^\star\)
except in coordinate \(k\), where \(x_A^k(t)=x_A^{k\star}+t\), and define
\[
g_i
:=
\left.
\frac{\partial h^k(|a-x_i^k|)}{\partial a}
\right|_{a=x_A^{k\star}}.
\]
Then \(g_1\le\cdots\le g_N\), with at least one strict inequality, and
differentiability gives
\[
\Delta_i(x_A(t),x_B^\star)=g_i t+o(t).
\]
Because \(F\) has density \(f\) continuous at zero,
\[
F(z)=F(0)+f(0)z+o(z)\qquad\text{as }z\to0.
\]

For \(t>0\), increasing differences order the thresholds from type \(1\) to type
\(N\). Equation \eqref{eq:rankidentity} therefore gives
\[
V_A(x_A(t),x_B^\star)
=
\nu(0)+
\sum_{i=1}^N
F\!\left(\Delta_i(x_A(t),x_B^\star)\right)\bar w_i,
\]
where
\[
\bar w_i=\nu(\bar s_i)-\nu(\bar s_{i+1}),
\qquad
\bar s_i=\sum_{j\ge i}s_j,
\qquad
\bar s_{N+1}=0.
\]
For \(t<0\), the ordering reverses, and the same identity assigns type \(i\) the
weight
\[
\ubar w_i=\nu(1-\bar s_{i+1})-\nu(1-\bar s_i).
\]
It follows directly from the preceding first-order expansions that the right and left
derivatives of \(V_A(\,\cdot\,,x_B^\star)\) at convergence are, respectively,
\[
D_+V_A(x^\star)=\sum_{i=1}^N d_i\bar w_i,
\qquad
D_-V_A(x^\star)=\sum_{i=1}^N d_i\ubar w_i,
\qquad
d_i:=f(0)g_i.
\]
The sequence \(d_1,\ldots,d_N\) is weakly increasing, with at least one strict
increase.

We now show that \(D_-V_A(x^\star)-D_+V_A(x^\star)<0\), which proves that
\(x_A^\star\) is not a best response to \(x_B^\star\). The difference is
\begin{gather*}
\sum_{i=1}^N d_i(\ubar{w}_i - \bar{w}_i)
=
\sum_{i=1}^N d_i\left[(\nu(1-\bar{s}_{i+1}) + \nu(\bar{s}_{i+1})) - (\nu(1-\bar{s}_{i}) + \nu(\bar{s}_{i}))\right] \\
=
(d_N-d_1)(\nu(1) + \nu(0))
+
\sum_{i=2}^N(d_{i-1}-d_i)\left(\nu(1-\bar{s}_i) + \nu(\bar{s}_i)\right).
\end{gather*}
Because \(\bar s_i\in(0,1)\) for \(i\ge2\), applying Remark
\ref{remarknu} to \(\min\{\bar s_i,1-\bar s_i\}\) gives
\(\nu(1-\bar s_i)+\nu(\bar s_i)>\nu(1)+\nu(0)\).
Since \(d_{i-1}-d_i\le0\), with a strict inequality for at least one \(i\), it
follows that
\begin{equation*}
\sum_{i=1}^N d_i(\ubar{w}_i - \bar{w}_i)
<
\underbrace{\left[(d_N-d_1)+\sum_{i=2}^N(d_{i-1}-d_i)\right]}_{=0}
\left(\nu(1) + \nu(0)\right).
\end{equation*}
Thus the left derivative is strictly smaller than the right derivative. At a local
maximum, the left derivative must be weakly nonnegative and the right derivative
weakly nonpositive, so the former must be weakly larger than the latter. This
contradiction proves that \(x_A^\star\) is not a best response to \(x_B^\star\). Hence
\(x^\star\) is not a Nash equilibrium. \(\blacksquare\)

\begin{remark}\label{remark_best}
Because the argument above uses only the equality \(x_A^\star=x_B^\star\), it establishes
a stronger statement. For any \(x\in\mathbb{R}^K\), if one party offers \(x\), then \(x\)
is not a best response for the opponent.
\end{remark}

\subsection{Unidimensional equilibrium and comparative statics}

\paragraph{Proof of Proposition \ref{NE1D}}

We start by proving equilibrium existence and part (i). Fix a distribution
\((\mathbf{x},\mathbf{s})\) and order types so that
\(x_1\le x_2\le\cdots\le x_N\). Fix \(p\in\{A,B\}\) and
\(x_{-p}\in[x_1,x_N]\). As a first step we show that, if
\(x_p\in[x_1,x_N]\) is a best response to \(x_{-p}\), then either
\(x_p>x_{-p}\) and
\(x_p=\sum_{i=1}^{N}\bar{w}_i x_i\), or \(x_p<x_{-p}\) and
\(x_p=\sum_{i=1}^{N}\ubar{w}_i x_i\), for the weights in the proposition.

Assume that \(x_p\) is a best response to \(x_{-p}\). By Remark
\ref{remark_best}, \(x_p\neq x_{-p}\). The best response is interior. Indeed, if
\(x_p=x_1\), the rival must lie strictly above it, and the feasible right derivative is
\(\phi^{-1}\sum_i\ubar w_i(x_i-x_1)>0\). If \(x_p=x_N\), the rival must lie strictly
below it, and the feasible left derivative is
\(\phi^{-1}\sum_i\bar w_i(x_i-x_N)<0\).
There are therefore two interior cases. The first is \(x_p>x_{-p}\). Subject to this
inequality, \(V_p\) is strictly concave in \(x_p\), so the first-order condition holds.
When \(p=A\), it is
\begin{equation}\label{FOC1D}
\frac{\partial V_A(x_A,x_B)}{\partial x_A} = \sum_{i=1}^N \frac{\partial\pi_i(x_A, x_B)}{\partial x_A}\bar{w}_i = 0.
\end{equation}
Uniformity of the aggregate shock gives
\(\pi_i(x_A, x_B) = \frac{1}{2} + \frac{\Delta_i}{2\phi}\), and hence
\begin{equation*}
\frac{\partial\pi_i(x_A, x_B)}{\partial x_A} = \frac{x_i - x_A}{\phi} \text{ for } i = 1,..., N.
\end{equation*}
Substituting this derivative into \eqref{FOC1D} yields
\begin{align*}
&\frac{\partial V_A(x_A,x_B)}{\partial x_A} = 0
\iff \sum_{i=1}^N\frac{x_i - x_A}{\phi}\bar{w}_i = 0
\iff x_A = \sum_{i=1}^N\bar{w}_ix_i
\end{align*}
Next, consider the second case, when the best response to \(x_{-p}\) is strictly less than
\(x_{-p}\). Conditional on \(x_p<x_{-p}\), \(V_p\) is strictly concave in party \(p\)'s
platform. Thus \(x_p\) must satisfy the first-order condition
\(\frac{\partial V_p}{\partial x_p}=0\). When \(p=A\), this requires
\begin{equation}\label{FOC1D2}
\frac{\partial V_A(x_A,x_B)}{\partial x_A} = \sum_{i=1}^N \frac{\partial\pi_i(x_A, x_B)}{\partial x_A}\ubar{w}_i = 0.
\end{equation}
Substituting for \(\frac{\partial\pi_i(x_A, x_B)}{\partial x_A} = \frac{x_i - x_A}{\phi}\) in \ref{FOC1D2}, we obtain,
\begin{align*}
&\frac{\partial V_A(x_A,x_B)}{\partial x_A} = 0 \iff \sum_{i=1}^N\frac{x_i - x_A}{\phi}\ubar{w}_i = 0 \iff x_A = \sum_{i=1}^N\ubar{w}_ix_i.
\end{align*}
The same calculation applies to party \(B\), because the parties have identical
primitives. Thus, for either party,
\(x_p = \sum_{i=1}^N\bar{w}_ix_i\) when \(x_p>x_{-p}\), and
\(x_p = \sum_{i=1}^N\ubar{w}_ix_i\) when \(x_p<x_{-p}\).
The only possible best-response locations are therefore
\(\ubar{x} = \sum_{i=1}^N\ubar{w}_ix_i\) and
\(\bar{x} = \sum_{i=1}^N\bar{w}_ix_i\).

For each \(p\in\{A,B\}\) and \(x_{-p}\in[x_1,x_N]\), a best response exists:
payoffs are continuous on the compact feasible interval, so the extreme value theorem
applies.

Because the best-response correspondence is nonempty, it remains to show that
\(\ubar{x}<\bar{x}\). Define the risk-neutral weights
\[
r_i:=\frac{\rho(\bar{s}_i)-\rho(\bar{s}_{i+1})}{\rho(1)-\rho(0)}
=
\frac{\rho(\ubar{s}_i)-\rho(\ubar{s}_{i-1})}{\rho(1)-\rho(0)},
\qquad i=1,\ldots,N,
\]
where the equality follows from \(1-\bar{s}_i=\ubar{s}_{i-1}\),
\(1-\bar{s}_{i+1}=\ubar{s}_i\), and assumption \ref{officerents} (ii). Then
\[
x^{RN}=\sum_{i=1}^N r_i x_i.
\]

For each \(i=1,\ldots,N\), let
\(
\lambda_i:=\frac{\rho(\bar{s}_i)-\rho(0)}{\rho(1)-\rho(0)}\in[0,1].
\)
Since
\(
\rho(\bar{s}_i)=(1-\lambda_i)\rho(0)+\lambda_i\rho(1),
\)
strict concavity of \(U\) implies
\(
\nu(\bar{s}_i)=U(\rho(\bar{s}_i))
\ge (1-\lambda_i)\nu(0)+\lambda_i\nu(1),
\)
with strict inequality whenever \(0<\lambda_i<1\), i.e. whenever \(i>1\). Using
telescoping,
\[
\sum_{j\ge i}^N \bar{w}_j
=
\frac{\nu(\bar{s}_i)-\nu(0)}{\nu(1)-\nu(0)}
\ge
\lambda_i
=
\sum_{j\ge i}^N r_j,
\]
strictly for every \(i>1\).

Next, assumption \ref{officerents} (ii) gives
\(
\rho(1-\bar{s}_i)=\rho(1)+\rho(0)-\rho(\bar{s}_i),
\)
so that
\(
1-\lambda_i=\frac{\rho(1-\bar{s}_i)-\rho(0)}{\rho(1)-\rho(0)}.
\)
Applying the same concavity argument to \(U(\rho(1-\bar{s}_i))\) yields
\(
\nu(1-\bar{s}_i)\ge \lambda_i\nu(0)+(1-\lambda_i)\nu(1),
\)
hence
\[
\sum_{j\ge i}^N r_j
=
\lambda_i
\ge
\frac{\nu(1)-\nu(1-\bar{s}_i)}{\nu(1)-\nu(0)}
=
\sum_{j\ge i}^N \ubar{w}_j,
\]
again strictly for every \(i>1\).

Hence, \((\bar{w}_i,x_i)_{i\in I}\) strictly first-order stochastically dominates
\((r_i,x_i)_{i\in I}\), which strictly first-order stochastically dominates
\((\ubar{w}_i,x_i)_{i\in I}\). Since all three distributions have common support
\(\{x_1,\ldots,x_N\}\), it follows that
\[
\ubar{x}<x^{RN}<\bar{x}.
\]
Since the only best-response candidates are \(\ubar{x}\) and \(\bar{x}\), and
\(\ubar{x}<\bar{x}\), a party facing \(\ubar{x}\) cannot choose the equal platform and
therefore chooses \(\bar{x}\); symmetrically, a party facing \(\bar{x}\) chooses
\(\ubar{x}\). The pure-strategy equilibrium profiles are exactly
\((\bar{x},\ubar{x})\) and \((\ubar{x},\bar{x})\). Moreover,
\(\ubar{w}_i,\bar{w}_i>0\) for all \(i\) because \(\nu\) is strictly increasing and
\(s_i>0\), so both \(\ubar{x}\) and \(\bar{x}\) are strict interior convex combinations
of \(x_1,\ldots,x_N\). Hence
\[
x_1<\ubar{x}<x^{RN}<\bar{x}<x_N.
\]
This concludes the proof of equilibrium existence and of part (i).

For part (ii), fix a voter type \(i\). When the party at \(\bar{x}\) gains this
type, it already receives the votes of all types \(j>i\), whose total mass is
\(\bar{s}_{i+1}\). Its vote share therefore rises from
\(\bar{s}_{i+1}\) to \(\bar{s}_i=\bar{s}_{i+1}+s_i\), producing the
payoff increment
\[
\bar{w}_i=\nu(\bar{s}_i)-\nu(\bar{s}_{i+1}).
\]
For the party at \(\ubar{x}\), gaining type \(i\) raises its vote share from
\(1-\bar{s}_i\) to \(1-\bar{s}_{i+1}\), so the corresponding increment is
\[
\ubar{w}_i=\nu(1-\bar{s}_{i+1})-\nu(1-\bar{s}_i).
\]
The equilibrium platforms are therefore weighted averages of voter ideal points,
where each weight is the party's payoff gain from adding that voter type to its
existing coalition.

Part (iii) follows from the location of the interval $[\bar{s}_{i+1},\bar{s}_i]$
relative to $\frac12$. Recall
\[
\bar{w}_i=\nu(\bar{s}_i)-\nu(\bar{s}_{i+1}),\qquad
\ubar{w}_i=\nu(1-\bar{s}_{i+1})-\nu(1-\bar{s}_i).
\]
By Remark \ref{remarknu}, for any $a>b$ with $a\le \frac12$,
\[
\nu(a)-\nu(b)>\nu(1-b)-\nu(1-a),
\]
and for any $a>b$ with $b\ge \frac12$,
\[
\nu(a)-\nu(b)<\nu(1-b)-\nu(1-a).
\]
Applying this with $a=\bar{s}_i$ and $b=\bar{s}_{i+1}$ yields
\[
\bar{s}_i\le \frac12 \implies \bar{w}_i>\ubar{w}_i,\qquad
\bar{s}_{i+1}\ge \frac12 \implies \bar{w}_i<\ubar{w}_i.
\]
If $x_i>x_m$, then at most half of voters lie weakly to the right of $x_i$, so
$\bar{s}_i\le\frac12$. If $x_i<x_m$, then at least half of voters lie strictly to the right
of $x_i$, so $\bar{s}_{i+1}\ge \frac12$. Therefore \(\bar{w}_i>\ubar{w}_i\) for
\(x_i>x_m\) and \(\bar{w}_i<\ubar{w}_i\) for \(x_i<x_m\).

Finally, assume without loss of generality that \(x_A=\bar{x}\) in equilibrium. Then
\begin{align*}
V_A(\bar{x},\ubar{x}) = \nu(0) + \sum_{i=1}^N \pi_i(\bar{x},\ubar{x})\bar{w}_i &= \frac{\nu(1)+\nu(0)}{2}+\sum_{i=1}^N\frac{\ubar{x}^2-\bar{x}^2 - 2x_i(\ubar{x}-\bar{x})}{2\phi}\bar{w}_i \\
&= \frac{\nu(1)+\nu(0)}{2}+\frac{(\bar{x}-\ubar{x})^2}{2\phi}
\end{align*}
and 
\begin{align*}
V_B(\bar{x},\ubar{x}) = \nu(1) - \sum_{i=1}^N \pi_i(\bar{x},\ubar{x})\ubar{w}_i &= \frac{\nu(1)+\nu(0)}{2}+\sum_{i=1}^N\frac{\bar{x}^2-\ubar{x}^2 - 2x_i(\bar{x}-\ubar{x})}{2\phi}\ubar{w}_i \\
&= \frac{\nu(1)+\nu(0)}{2} + \frac{(\bar{x}-\ubar{x})^2}{2\phi},
\end{align*}
which concludes the proof of the proposition. \(\blacksquare\)

\paragraph{Proof of Proposition \ref{polarization}} 
Fix two voter distributions $(\textbf{x},\textbf{s})$ and $(\textbf{x}',\textbf{s}')$ such that $(\textbf{x}'-\textbf{x}'_m,\textbf{s}')$ is a spread of
$(\textbf{x}-\textbf{x}_m,\textbf{s})$ in the sense of definition \ref{defspread}. Translating all bliss points by a constant
translates both equilibrium platforms by the same constant, so equilibrium platform
distance and equilibrium payoffs are translation invariant. Hence, without loss of
generality, recenter and assume $x_m=x'_m=0$.

If either distribution assigns positive mass to the median bliss point $0$, split that mass
into two virtual types $x_{m^-}$ and $x_{m^+}$, with $x_{m^-}=x_{m^+}=0$, and choose their
masses so that exactly half of the population lies weakly to each side of the split.
Concretely, for $(\textbf{x},\textbf{s})$ set
\[
s_{m^-}=\tfrac12-\sum_{x_i<0}s_i,
\qquad
s_{m^+}=\tfrac12-\sum_{x_i>0}s_i,
\]
and keep all other masses unchanged; define \(s'_{m^+}\) and \(s'_{m^-}\) using
$(\textbf{x}',\textbf{s}')$ analogously. This operation leaves equilibrium platforms
unchanged: every tail mass away from the split is unchanged, while the two new terms in
each rank-weighted average are multiplied by the common bliss point \(0\).

Let $\bar x>\underline x$ denote the equilibrium platforms under $(\textbf{x},\textbf{s})$, and
$\bar x'>\underline x'$ those under $(\textbf{x}',\textbf{s}')$. We compare $\bar x-\underline x$ to
$\bar x'-\underline x'$. 

Define the common grid \(\mathcal{X}\) as follows. Let \(
z_1 \le \cdots \le z_M
\)
be the weakly ordered list of all distinct bliss point values appearing in either
distribution, except that if the median is split we include the value $0$ twice, as
$z_{m^-}=z_{m^+}=0$, where $z_{m^-}$ immediately precedes $z_{m^+}=0$ in the ordering. All other bliss point values appear only once in the grid. Thus, the grid is weakly
ordered, with possible equality only at value \(0\):
\[
z_1 \le z_2 \le \cdots \le z_M,
\]
and $z_k=z_{k+1}$ can occur only when $(z_k,z_{k+1}) = (z_{m^-},z_{m^+}) = (0,0)$.
For each grid point $z_k\neq 0$, define $s(z_k)=\sum_{i: x_i=z_k}s_i$ and
$s'(z_k)=\sum_{i: x'_i=z_k}s'_i$. If \(0\) belongs to the grid, then the median was split, so it appears twice. In this case, set $s(z_{m^-})=s_{m^-}$, $s(z_{m^+})=s_{m^+}$, $s'(z_{m^-})=s'_{m^-}$ and $s'(z_{m^+})=s'_{m^+}$.

For each $k=1,..., M$, define tail masses
\[
\bar s(z_k):=\sum_{\ell\ge k}s(z_\ell),
\qquad
\bar s'(z_k):=\sum_{\ell\ge k}s'(z_\ell).
\]

By proposition \ref{NE1D} (ii), equilibrium platforms can be written as weighted averages of bliss
points. On the grid $\{z_k\}$, define the equilibrium weights for $(\textbf{x},\textbf{s})$ by
\(\bar w(z_k)=\nu(\bar s(z_k))-\nu(\bar s(z_{k+1}))\) and
\(\underline w(z_k)=\nu(1-\bar s(z_{k+1}))-\nu(1-\bar s(z_k)),
\)
with $\bar s(z_{M+1})=0$. Then
\[
\bar x=\sum_{k=1}^M \bar w(z_k)\,z_k,
\qquad
\underline x=\sum_{k=1}^M \underline w(z_k)\,z_k,
\qquad
\bar x-\underline x=\sum_{k=1}^M\big(\bar w(z_k)-\underline w(z_k)\big)z_k.
\]
Let \(A_k:=\sum_{\ell\ge k}\big(\bar w(z_\ell)-\underline w(z_\ell)\big)\) and \(
\Gamma(s):=\nu(s)+\nu(1-s)-\nu(1)-\nu(0)\). Note that
\(A_k =\Gamma(\bar s(z_k))\). Hence,
\[
\bar x-\underline x=\sum_{k=1}^M \Gamma(\bar s(z_k))\,(z_k-z_{k-1}),
\qquad z_0:=z_1.
\]
The same construction for $(\textbf{x}',\textbf{s}')$ gives
\(
\bar x'-\underline x'=\sum_{k=1}^M \Gamma(\bar s'(z_k))\,(z_k-z_{k-1})
\).

We now use remark \ref{remarknu}. For any $0\le s<s'\le \tfrac12$, it implies \(
\nu(s')-\nu(s)>\nu(1-s)-\nu(1-s')
\). Equivalently,
\begin{gather*}
\Gamma(s')>\Gamma(s)\quad\text{for all }0\le s<s'\le\tfrac12, \\
\Gamma(s')>\Gamma(s)\quad\text{for all }\tfrac12\le s'<s\le 1.
\end{gather*}

Because $(\textbf{x}',\textbf{s}')$ is a spread of $(\textbf{x},\textbf{s})$ around the common median $0$, the right conditional
distribution shifts to the right (first-order stochastic dominance) and the left conditional
distribution shifts to the left. Evaluated on the common grid, this implies that for all $k$,
\[
z_k>0 \implies \bar s'(z_k)\ge \bar s(z_k),
\qquad
z_k<0 \implies \bar s'(z_k)\le \bar s(z_k),
\]
Moreover, whenever the common grid contains the two virtual median points, their
construction gives
\[
\bar s(z_{m^+})=\bar s'(z_{m^+})=\tfrac12,
\qquad
\bar s(z_{m^-})\ge\tfrac12, \qquad\bar s'(z_{m^-})\ge\tfrac12.
\]

Fix $k$. If $z_k>0$, then by construction $\bar s(z_k)\le\tfrac12$ and
$\bar s'(z_k)\le\tfrac12$. Since $(\mathbf{x}',\mathbf{s}')$ is a spread of
$(\mathbf{x},\mathbf{s})$, we have $\bar s'(z_k)\ge \bar s(z_k)$. This implies \(
\Gamma(\bar s'(z_k))\ge \Gamma(\bar s(z_k)),
\) with strict inequality whenever $\bar s'(z_k)>\bar s(z_k)$. If $z_k<0$, then $\bar s(z_k)\ge\tfrac12$ and $\bar s'(z_k)\ge\tfrac12$, and the spread
property implies $\bar s'(z_k)\le \bar s(z_k)$. This again
implies \(
\Gamma(\bar s'(z_k))\ge \Gamma(\bar s(z_k)),
\) with strict inequality whenever $\bar s'(z_k)<\bar s(z_k)$. If the virtual median
points are present and $z_k=z_{m^-}$, the spread condition implies \(s_{m^-}\ge s'_{m^-}\), because of the first order dominance relation between left conditional distributions.\footnote{FOSD of the left conditional distributions implies that their cumulative distribution
functions satisfy \(\lim_{t\to 0^-}F_L(t)\le \lim_{t\to 0^-}F'_L(t)\). Since \(\lim_{t\to 0^-}F_L(t) = 1- 2s_{m^-}\) and \(\lim_{t\to 0^-}F'_L(t) = 1- 2s'_{m^-}\), this implies \(s_{m^-}\ge s'_{m^-}\).} Hence $\bar s(z_k)\ge \bar s'(z_k)\ge\tfrac12$, which implies \(
\Gamma(\bar s'(z_k))\ge \Gamma(\bar s(z_k))
\). If the virtual median points are present and $z_k=z_{m^+}$, then $\bar s(z_k)=\bar s'(z_k)=\tfrac12$ by construction, so the
coefficients coincide. 

Because at least one conditional dominance relation is strict, consider the right and
left halves separately. If the right-half comparison is strict, finiteness and equality
of total right-half mass imply that
\(\bar s'(z_k)>\bar s(z_k)\) at some positive grid point \(z_k>0\).
Such a point has \(k\ge2\) and \(z_k-z_{k-1}>0\). If instead the left-half comparison
is strict, either
\(\bar s'(z_k)<\bar s(z_k)\) at some negative grid point, in which case \(k>1\)
(tail mass equals one at the lowest grid point) and \(z_k-z_{k-1}>0\), or all
negative-grid tail masses coincide. Unless the virtual median points are present, the
two finite left-half distributions would then coincide, contradicting strictness. Thus
the virtual points are present, and strictness can only come from a different mass at
the virtual median point:
\(s_{m^-}>s'_{m^-}\). At least one distribution then has negative support, so
\(m^- > 1\) and \(z_{m^-}-z_{m^--1}>0\). Moreover,
\(\bar s(z_{m^-})>\bar s'(z_{m^-})\ge\tfrac12\).
Thus, in every case, there is \(k^\star\ge2\) with a positive grid increment and a
strict tail comparison in the spread direction. By the preceding strict monotonicity of
\(\Gamma\),
\[
\Gamma(\bar s'(z_{k^\star}))>\Gamma(\bar s(z_{k^\star})).
\]
Since for every \(k\) we already established
\(\Gamma(\bar s'(z_k))\ge \Gamma(\bar s(z_k))\) and every increment
\(z_k-z_{k-1}\) is nonnegative, it follows that
\[
\bar x'-\underline x'
=\sum_{k=1}^M \Gamma(\bar s'(z_k))\,(z_k-z_{k-1})
>
\sum_{k=1}^M \Gamma(\bar s(z_k))\,(z_k-z_{k-1})
=\bar x-\underline x.
\]

Finally, proposition \ref{NE1D} (iv) implies that equilibrium expected payoffs satisfy
\[
\hat{V}_p(\textbf{x},\textbf{s})=\frac{\nu(1)+\nu(0)}{2}+\frac{(\bar x-\underline x)^2}{2\phi},
\qquad p\in\{A,B\},
\]
which is strictly increasing in $\bar x-\underline x$. Therefore
$\hat V_p(\textbf{x}',\textbf{s}')>\hat V_p(\textbf{x},\textbf{s})$ for $p=A,B$. \(\blacksquare\)

\paragraph{Proof of Proposition \ref{alienation}}
Order types so that \(x_1\le\cdots\le x_N\), and let party \(A\) offer \(\bar x\).
Differentiating the equilibrium-payoff identity in Proposition \ref{NE1D} gives,
for either party \(p\),
\[
\frac{d\hat V_p}{dx_i}
=
\frac{\bar x-\ubar x}{\phi}\bigl(\bar w_i-\ubar w_i\bigr).
\]
By Proposition \ref{NE1D}(iii), this derivative is positive when \(x_i>x_m\) and
negative when \(x_i<x_m\).

Let \(\pi_i^A\) denote type \(i\)'s equilibrium probability of supporting \(A\).
Using \(d\bar x/dx_i=\bar w_i\), \(d\ubar x/dx_i=\ubar w_i\), and the quadratic-uniform
specification,
\[
\frac{d\pi_i^A}{dx_i}
=
\frac{
\ubar x\,\ubar w_i-\bar x\,\bar w_i
+(\bar x-\ubar x)+x_i(\bar w_i-\ubar w_i)
}{\phi}.
\]
If \(x_i>\max\{\ubar x,x_m\}\), then \(\bar w_i>\ubar w_i\), and the numerator is
strictly larger than its value at \(x_i=\ubar x\), namely
\((1-\bar w_i)(\bar x-\ubar x)>0\). Thus both \(A\)'s payoff and the group's support
for \(A\) rise with \(x_i\): \(A\) benefits from attraction. If instead
\(x_i<\min\{\ubar x,x_m\}\), then \(\bar w_i<\ubar w_i\). The same numerator is again
strictly positive, now because replacing \(x_i\) by \(\ubar x\) lowers it, while
\(d\hat V_A/dx_i<0\). Hence support and payoff move in opposite directions, so \(A\)
benefits from alienation. Applying the same argument to
\(\pi_i^B=1-\pi_i^A\) and interchanging \(\bar x\) and \(\ubar x\) gives the stated
conditions for party \(B\).

Finally, all equilibrium weights lie in \((0,1)\), so
\(x_1<\ubar x<\bar x<x_N\). If an extreme group is nonmedian, the stated conditions
therefore apply directly. If instead \(x_N=x_m\), then \(s_N>\tfrac12\) and
\[
\bar w_N-\ubar w_N
=\nu(s_N)+\nu(1-s_N)-\nu(1)-\nu(0)>0
\]
by Remark \ref{remarknu}. Since \(x_N>\bar x>\ubar x\), the preceding support-derivative
calculation implies that raising \(x_N\) attracts group \(N\) to \(A\), alienates it from
\(B\), and raises both parties' payoffs. The case \(x_1=x_m\) is symmetric:
\(s_1>\tfrac12\) and
\[
\bar w_1-\ubar w_1
=\nu(1)+\nu(0)-\nu(s_1)-\nu(1-s_1)<0,
\]
so lowering \(x_1\) alienates the group from \(A\), attracts it to \(B\), and raises both
payoffs. Thus the claims for the most extreme groups also hold when an extreme carries a
majority atom.
\(\blacksquare\)

\subsection{Affiliate-value concentration}\label{orgcomparative}

The ranked-assignment foundation permits a comparative static on party organization
itself. Consider two affiliate-value distributions, \(G_L\) and \(G_H\), supported on
strictly positive values and with the same positive finite mean. For any such
distribution \(G\), let \(Q_G\) denote its ascending quantile function. If the party can
empower a fraction \(t\) of its affiliates, optimal ranked assignment gives
\[
W_G(t)=\int_{1-t}^{1}Q_G(u)\,du,
\qquad
T_G(t)=\frac{W_G(t)}{W_G(1)}.
\]
Thus \(T_G(t)\) is the share of total organizational value generated by the
highest-value fraction \(t\) of affiliates. When \(G\) has a positive continuous density
\(g\) on an interval, for \(t\in(0,1)\),
\[
W_G'(t)=Q_G(1-t)>0,
\qquad
W_G''(t)=-\frac{1}{g(Q_G(1-t))}<0,
\]
which supplies the strict concavity used in the main analysis.

\begin{definition}[Affiliate-value concentration]\label{deforgconcentration}
Affiliate value is more concentrated under \(H\) than under \(L\) if
\[
T_{G_H}(t)\ge T_{G_L}(t)
\qquad\text{for every }t\in[0,1].
\]
\end{definition}

This is the upper-Lorenz, or top-share, order. With equal means, it is equivalent to
convex-order dominance; a comparison of variances alone is insufficient. For
\(\ell\in\{L,H\}\), let
\[
\nu_\ell(s)=T_{G_\ell}(r(s))
\]
and denote the corresponding equilibrium platforms by
\((\ubar{x}^{\ell},\bar{x}^{\ell})\).

\begin{proposition}[Affiliate-value concentration]\label{orgconcentration}
Fix the electorate and the power mapping \(\rho\). Suppose both distributions have
positive continuous densities on compact intervals contained in \((0,\infty)\). If affiliate value is more concentrated
under \(H\), then
\[
\ubar{x}^{H}\le \ubar{x}^{L}<\bar{x}^{L}\le \bar{x}^{H}.
\]
Consequently, equilibrium platform distance and both parties' normalized equilibrium
payoffs are weakly larger under \(H\). Platform distance increases strictly if there
exists \(i\in\{2,\ldots,N\}\) with \(x_i>x_{i-1}\) such that, for
\(r_i=r(\bar{s}_i)\),
\[
\big[T_{G_H}(r_i)-T_{G_L}(r_i)\big]
+
\big[T_{G_H}(1-r_i)-T_{G_L}(1-r_i)\big]>0.
\]
\end{proposition}

\begin{proof}
For \(\ell\in\{L,H\}\), normalization of total affiliate value gives
\[
\nu_\ell(s)=T_{G_\ell}(r(s)),\qquad
\nu_\ell(0)=0,\qquad
\nu_\ell(1)=1,
\]
and constant-sum power implies \(r(1-s)=1-r(s)\). Using the equilibrium weights in
Proposition \ref{NE1D} and summation by parts,
\begin{align}
\bar{x}^{\ell}
&=
x_1+\sum_{i=2}^{N}(x_i-x_{i-1})
T_{G_\ell}(r(\bar s_i)),\label{upperplatformtopshares}\\
\ubar{x}^{\ell}
&=
x_1+\sum_{i=2}^{N}(x_i-x_{i-1})
\left[1-T_{G_\ell}(1-r(\bar s_i))\right].
\label{lowerplatformtopshares}
\end{align}
Definition \ref{deforgconcentration} and
\eqref{upperplatformtopshares} imply
\(\bar{x}^{H}\ge\bar{x}^{L}\); the same definition and
\eqref{lowerplatformtopshares} imply
\(\ubar{x}^{H}\le\ubar{x}^{L}\). Subtracting the two expressions gives
\[
\bar{x}^{\ell}-\ubar{x}^{\ell}
=
\sum_{i=2}^{N}(x_i-x_{i-1})
\left[
T_{G_\ell}(r_i)+T_{G_\ell}(1-r_i)-1
\right].
\]
Every coefficient \(x_i-x_{i-1}\) is nonnegative. Top-share dominance therefore implies
the weak comparison in platform distance, with strictness under the condition in the
proposition. Finally, Proposition \ref{NE1D} gives
\[
\hat V_p^\ell
=
\frac12+
\frac{(\bar{x}^{\ell}-\ubar{x}^{\ell})^2}{2\phi},
\]
so the payoff comparison follows.
\end{proof}

\subsection{Multidimensional equilibrium and cleavage selection}

\paragraph{Proof of Proposition \ref{potentialgame}}
Translate coordinates so that $\mu=0$. Fix a platform profile that induces a strict
ranking $R$ of voter types, ordered from the type least likely to support $A$ to the type
most likely to support $A$. Let \(\bar x(R)\) and \(\ubar x(R)\) be the rank-weighted
averages obtained by applying Proposition \ref{NE1D}'s weights to the order \(R\).
Direct integration over the uniform shock gives
\begin{align}
V_A(x_A,x_B)
&=
c_0+\frac{1}{2\phi}
\left[
\|x_B\|^2-\|x_A\|^2
+2\bar x(R)^\top(x_A-x_B)
\right],\label{finitepayoffA}\\
V_B(x_A,x_B)
&=
c_0-\frac{1}{2\phi}
\left[
\|x_B\|^2-\|x_A\|^2
+2\ubar x(R)^\top(x_A-x_B)
\right],\label{finitepayoffB}
\end{align}
where $c_0=(\nu(1)+\nu(0))/2$. Central symmetry reverses the ranking and pairs every
type with its reflection, which implies
\begin{equation}\label{rankreflection}
\ubar x(R)=-\bar x(R).
\end{equation}
Subtracting \eqref{finitepayoffB} from \eqref{finitepayoffA} therefore yields
\[
V_A(x_A,x_B)-V_B(x_A,x_B)
=
\frac{\|x_B\|^2-\|x_A\|^2}{\phi}.
\]
Translating coordinates back yields the payoff-difference identity stated in
Section \ref{NEKD}.

Strict-ranking profiles are dense in $D^2$. At a tie, changing the ordering affects
payoffs only on shock intervals whose length converges to zero as the profile approaches
the tie. Expected payoffs are therefore continuous, even when $\nu$ inherits a jump from
$\rho$. Taking limits extends that payoff-difference identity to all of $D^2$.
Substituting the identity into the definition of \(\Phi\) in Section \ref{NEKD} then
implies, for every unilateral deviation by $A$,
\[
V_A(x_A',x_B)-V_A(x_A,x_B)
=
\Phi(x_A',x_B)-\Phi(x_A,x_B),
\]
and analogously for $B$. Thus $\Phi$ is an exact potential on $D^2$.

The potential is continuous on compact $D^2$, so it has a global maximizer. Every
equilibrium on $D^2$ is unrestricted: outside deviations are weakly dominated by their
projections. Conversely, take an unrestricted equilibrium $(x_A,x_B)$ with $x_A\notin D$
and its projection $(a,b)$ on \(D^2\). \mbox{Projection gives}
$\Delta_i(a,x_B)>\Delta_i(x_A,x_B)$ and
$\Delta_i(a,x_B)\geq\Delta_i(a,b)>-\phi$ for every $i$. If some
$\Delta_i(x_A,x_B)<\phi$, moving to $a$ raises $A$'s support on a positive-measure set
of shocks and never lowers it, strictly raising its payoff. Otherwise $A$ wins every vote
almost surely, so $B$ profitably imitates $x_A$ and obtains $c_0>\nu(0)$. This is a
contradiction; symmetrically, $x_B\in D$. Hence, the game on \(D^2\) has the same Nash
equilibrium set as the unrestricted game. \(\blacksquare\)

\paragraph{Rank weights and proof of Proposition \ref{directionconflict}}
The precise conditions for the continuous characterization are as follows. Voter bliss
points have a compact, absolutely continuous distribution \(\mathcal P\), centrally
symmetric around \(\mu\), and every projected cdf is strictly increasing between the
extrema of its support. Let
\[
R=\sup_{x\in\operatorname{supp}\mathcal P}\|x-\mu\|.
\]
Platforms may be chosen from \(\mathbb R^K\), and the uniform shock satisfies
\(\phi>4R^2\). The power rule \(\rho\) is continuous at \(0\) and \(1\), although
interior threshold jumps remain allowed. The proof first works on \(B_R(\mu)\) and then
shows that no deviation outside that ball is profitable.

Because the power rule can jump, first define the payoff-increment measure precisely.
At every interior discontinuity replace \(\nu\) by its right-continuous version; this
changes expected payoffs only on a measure-zero set of popularity shocks. Endpoint
continuity leaves \(\nu(0)\) and \(\nu(1)\) unchanged. Let \(\Lambda_\nu\) be the
Lebesgue--Stieltjes measure generated by \(\nu-\nu(0)\):
\[
\Lambda_\nu((a,b])=\nu(b)-\nu(a)
\quad (0\leq a<b\leq1),
\qquad
\Lambda_\nu(\{0\})=0.
\]
Since \(\nu(1)-\nu(0)=1\), \(\Lambda_\nu\) is a probability measure. Define its
reflection across one half by
\[
M_\nu(B)
=
\Lambda_\nu\bigl(\{s\in[0,1]:1-s\in B\}\bigr)
\]
for every Borel set \(B\subseteq[0,1]\). This is the rank-weight measure used in
\eqref{conflictindex}. If \(\nu\) is absolutely continuous,
\(dM_\nu(u)=\nu'(1-u)\,du\). If the one-sided limits of \(\nu\) differ at an interior
threshold \(s_0\), then
\[
M_\nu(\{1-s_0\})
=
\nu(s_0+)-\nu(s_0-).
\]
Thus the construction includes the full payoff increment at a majority,
supermajority, or other interior threshold, regardless of the value assigned to
\(\rho(s_0)\) at the threshold itself.

We first establish that the conflict functional is strictly positive. Let \(F_e\) be the
cdf of the centered projection, and use the quantile convention
\(Q_e(u)=\inf\{z:F_e(z)\geq u\}\), with the support endpoints assigned at \(u=0,1\).
Fix \(z>0\) in the interior of the positive support. Set
\(t=F_e(z)\in(1/2,1)\) and \(a=1-t=F_e(-z)\), where the second equality follows from
central symmetry. Then
\(\{u:Q_e(u)>z\}=(t,1]\) and
\(\{u:Q_e(u)<-z\}=[0,a)\), so exactly
\begin{align*}
&M_\nu\{u:Q_e(u)>z\}
-M_\nu\{u:Q_e(u)<-z\}\\
&\hspace{2cm}
=
\Lambda_\nu([0,a))-\Lambda_\nu((t,1])
=
\nu(a-)+\nu(t)-\nu(0)-\nu(1).
\end{align*}
Here \(\nu(t)\) denotes the right-continuous version and hence the right-hand limit of
the original payoff. Assumption \ref{officerents} implies
\[
\rho(a-)+\rho(t+)=\bar\rho.
\]
Both terms lie strictly between \(\rho(0)\) and \(\rho(1)\). Strict concavity of \(U\)
therefore gives
\[
U(\rho(a-))+U(\rho(t+))
>
U(\rho(0))+U(\rho(1)).
\]
The preceding tail difference is consequently strictly positive. The layer-cake
identity
\[
\mathcal C_{\nu,\mathcal P}(e)
=
\int_0^\infty
\left[
M_\nu\{u:Q_e(u)>z\}
-M_\nu\{u:Q_e(u)<-z\}
\right]dz
\]
then implies \(\mathcal C_{\nu,\mathcal P}(e)>0\). Notice that the argument uses
concavity of \(U\) over political power, not concavity of \(\nu=U\circ\rho\) in vote
share; \(\rho\) may be nonlinear and may have interior jumps.

We next derive the potential. For a unit direction \(e\), write
\(Z_e=e^\top(X-\mu)\), and let \(Q_e\) be its quantile function. If
\(x_A-x_B=de\) and \(m=(x_A+x_B)/2\), the policy advantage of party \(A\) for a voter
with projection \(z\) is
\[
\Delta(z)=2d\left[z-e^\top(m-\mu)\right].
\]
The wide-support condition and the uniform popularity shock imply A-vote probability
\[
\pi(z)
=
\frac12+\frac d\phi
\left[z-e^\top(m-\mu)\right].
\]
As the common shock varies, voters enter \(A\)'s coalition in projected-rank order.
Tonelli's theorem applied to the positive payoff-increment measure therefore gives
\[
V_A
=
\nu(0)+\int_0^1\pi(Q_e(u))\,dM_\nu(u).
\]
Using the definition of the conflict functional and central symmetry yields
\begin{align}
V_A(x_A,x_B)
&=
\frac{\nu(1)+\nu(0)}2
+\frac d\phi
\left[
\mathcal C_{\nu,\mathcal P}(e)-e^\top(m-\mu)
\right],\label{continuouspayoffA}\\
V_B(x_A,x_B)
&=
\frac{\nu(1)+\nu(0)}2
+\frac d\phi
\left[
\mathcal C_{\nu,\mathcal P}(e)+e^\top(m-\mu)
\right].\label{continuouspayoffB}
\end{align}
These expressions extend continuously to \(d=0\). They imply
\[
V_A-\frac{\|x_B-\mu\|^2}{\phi}
=
V_B-\frac{\|x_A-\mu\|^2}{\phi}
=
\Phi,
\]
so \(\Phi\) is an exact potential, and the identity
\[
\frac{\|x_A-\mu\|^2+\|x_B-\mu\|^2}{2}
=
\|m-\mu\|^2+\frac{d^2}{4}
\]
gives \eqref{potentialdecomposition}.

For completeness, \(\mathcal C_{\nu,\mathcal P}\) is continuous on the unit sphere.
Indeed, if \(e_n\to e\), the compactness of the voter support and the assumed continuity
and strict increase of projected cdfs imply pointwise convergence
\(Q_{e_n}(u)\to Q_e(u)\) for \(u\in(0,1)\). Endpoint continuity of \(\nu\) gives
\(M_\nu(\{0,1\})=0\), and \(|Q_{e_n}(u)|\leq R\), so dominated convergence applies.
Hence the set
\[
\mathcal D^\star
=
\arg\max_{\|e\|=1}\mathcal C_{\nu,\mathcal P}(e)
\]
is nonempty. Write
\(\mathcal C^\star=\max_{\|e\|=1}\mathcal C_{\nu,\mathcal P}(e)\).

Let \(c_0=[\nu(1)+\nu(0)]/2\). For a fixed direction \(e\), complete the square in
\eqref{potentialdecomposition}:
\[
\Phi
=
c_0-\frac{\|m-\mu\|^2}{\phi}
+\frac{\mathcal C_{\nu,\mathcal P}(e)^2}{\phi}
-\frac{\bigl(d-2\mathcal C_{\nu,\mathcal P}(e)\bigr)^2}{4\phi}.
\]
Thus \(m=\mu\) and
\(d=2\mathcal C_{\nu,\mathcal P}(e)\) maximize the potential conditional on \(e\).
Because \(|Q_e(u)|\leq R\) and \(M_\nu\) is a probability measure,
\(\mathcal C_{\nu,\mathcal P}(e)\leq R\); the resulting platforms lie in
\(B_R(\mu)\). It follows that the full set of global maximizers on
\(B_R(\mu)^2\) is obtained by taking \(m=\mu\), \(e\in\mathcal D^\star\), and
\(d=2\mathcal C^\star\). Equivalently, up to party labels, every such maximizer and
only such a maximizer has the form
\[
x_A^\star=\mu+\mathcal C^\star e^\star,
\qquad
x_B^\star=\mu-\mathcal C^\star e^\star
\qquad\text{for some }e^\star\in\mathcal D^\star.
\]
Every global maximizer of an exact potential is a Nash equilibrium on \(B_R(\mu)^2\).
We next compare these profiles with arbitrary unrestricted equilibria. First, every
unrestricted equilibrium lies in \(B_R(\mu)^2\). Let \(P_Ry\) be the metric projection
of \(y\notin B_R(\mu)\) onto \(B_R(\mu)\). For every
\(X\in\operatorname{supp}\mathcal P\), the projection inequality gives
\[
\|y-X\|^2
\ge
\|P_Ry-X\|^2+\|y-P_Ry\|^2.
\]
Replacing \(y\) by \(P_Ry\) therefore strictly improves that party's policy comparison
for every voter. The policy advantage is affine in a one-dimensional projection of voter
bliss points. Because every projected cdf is strictly increasing between its support
extrema, the advantage distribution has interval support and positive mass in every
interior subinterval. Under the uniform shock, the strict improvement therefore raises
expected payoff unless the party wins surely or loses surely throughout the shock
support. Sure defeat cannot be a best response: imitating the opponent yields
\(c_0=[\nu(1)+\nu(0)]/2>\nu(0)\). A profile at which one party wins surely leaves the
opponent surely defeated and therefore cannot be an equilibrium. Hence no unrestricted
equilibrium contains an outside platform. The same projection argument shows that no
deviation outside the ball can profit from a global maximizer on the ball, so every
maximizer characterized above is an unrestricted Nash equilibrium.

Now fix any unrestricted equilibrium \(x\). It cannot be convergent. Indeed, if
\(x_A=x_B=x_0\), choose the sign of a unit vector \(e\) so that
\(e^\top(x_0-\mu)\leq0\). For sufficiently small \(\delta>0\), a deviation by \(A\) to
\(x_0+\delta e\) changes its payoff by
\[
\frac{\delta}{\phi}
\left[
\mathcal C_{\nu,\mathcal P}(e)-e^\top(x_0-\mu)-\frac{\delta}{2}
\right]
>0,
\]
where strict positivity follows from
\(\mathcal C_{\nu,\mathcal P}(e)>0\). Write therefore
\(x_A-x_B=de\), with \(d>0\). Since \(x\in B_R(\mu)^2\) and
\(\phi>4R^2\) strictly, \eqref{continuouspayoffA}--\eqref{continuouspayoffB}
remain valid in a neighborhood of \(x\), even if a platform lies on the boundary of the
ball. Holding \(x_B\) fixed and varying \(x_A=x_B+te\) gives
\[
V_A
=
c_0+\frac1\phi
\left\{
t\left[\mathcal C_{\nu,\mathcal P}(e)-e^\top(x_B-\mu)\right]
-\frac{t^2}{2}
\right\}.
\]
The first-order condition at \(t=d\) implies
\[
e^\top(x_A-\mu)=\mathcal C_{\nu,\mathcal P}(e).
\]
Similarly, holding \(x_A\) fixed and varying \(x_B=x_A-te\) implies
\[
e^\top(x_A-\mu)=d-\mathcal C_{\nu,\mathcal P}(e).
\]
Consequently, every unrestricted equilibrium satisfies
\[
d=2\mathcal C_{\nu,\mathcal P}(e),
\qquad
e^\top(m-\mu)=0,
\]
and \eqref{continuouspayoffA}--\eqref{continuouspayoffB} yield
\[
V_A(x_A,x_B)=V_B(x_A,x_B)
=
c_0+\frac{2\mathcal C_{\nu,\mathcal P}(e)^2}{\phi}
\leq
c_0+\frac{2(\mathcal C^\star)^2}{\phi}.
\]
The global potential maximizers characterized above attain the final bound. Thus an
unrestricted equilibrium is party-preferred if and only if
\(e\in\mathcal D^\star\), and every global potential maximizer is party-preferred. In
particular, at every party-preferred equilibrium,
\[
\widehat V_p^\star
=
c_0+\frac{2(\mathcal C^\star)^2}{\phi},
\qquad p=A,B.
\]
\(\blacksquare\)

\paragraph{When global maximization and party preference coincide}
Suppose in addition that the reduced-form payoff \(\nu\) is concave, and define
\[
h(q)
=
\begin{cases}
\|q\|\mathcal C_{\nu,\mathcal P}(q/\|q\|),&q\neq0,\\
0,&q=0.
\end{cases}
\]
The reflected marginal weights generated by a concave \(\nu\) are nondecreasing in voter
rank. The quantile representation and the rearrangement inequality therefore make \(h\)
an even, convex, positively homogeneous function. Hence \(h\) is the support function
\(\sigma_K\) of a compact, centrally symmetric convex set \(K\). Since
\(h(e)=\mathcal C_{\nu,\mathcal P}(e)\leq R\) for every unit vector \(e\), this set
satisfies \(K\subseteq B_R(0)\).

Write \(a=x_A-\mu\) and \(c=\mu-x_B\), so \(x_A-x_B=a+c\). Up to constants and the
positive factor \(1/\phi\), party \(A\) maximizes
\(h(a+c)-\|a\|^2/2\) with respect to \(a\), and party \(B\) maximizes the same expression
with respect to \(c\). Since \(h=\sigma_K\),
\[
\max_{a\in B_R(0)}
\left\{h(a+c)-\frac{\|a\|^2}{2}\right\}
=
\max_{z\in K}
\left\{z^\top c+\frac{\|z\|^2}{2}\right\},
\]
and every best response satisfies \(a=z\) for a maximizer on the right-hand side. The
analogous statement holds for \(c\). At an equilibrium, therefore, \(a,c\in K\), and
optimality relative to the feasible choices \(z=c\) and \(z=a\), respectively, gives
\[
a^\top c+\frac{\|a\|^2}{2}\geq\frac{3\|c\|^2}{2},
\qquad
a^\top c+\frac{\|c\|^2}{2}\geq\frac{3\|a\|^2}{2}.
\]
Adding these inequalities implies \(\|a-c\|^2\leq0\), so \(a=c\) and every equilibrium
is centered at \(\mu\). The party-preferred characterization above then implies that
every party-preferred equilibrium has
\(a=c=\mathcal C^\star e^\star\) for some \(e^\star\in\mathcal D^\star\). Thus the
party-preferred equilibria and the global potential maximizers coincide.

The same conclusion applies to the proportional-with-majority-premium family in Online
Appendix \ref{oa:appendixA4}, equation \eqref{oa:eq:rhopremfam}. For the right-continuous
version used in the definition of \(M_\nu\), write
\[
\nu(s)=g(s)+J_m\mathbf 1\{s\geq1/2\},
\qquad
J_m=\nu(1/2+)-\nu(1/2-).
\]
Assumption \ref{concavity} and the two affine branches of
\(\rho_{\rho_m}\) imply that \(g\) is increasing and concave. The payoff jump places an
atom of mass \(J_m\) at rank \(1/2\), but its contribution to the conflict functional is
\(J_mQ_e(1/2)=0\), because every centered projection of the maintained continuous,
centrally symmetric electorate has median zero. The resulting function \(h\) is therefore
the same sublinear support function generated by \(g\), and the preceding argument
continues to apply for every \(\rho_m\in[0,\bar\rho)\).

\paragraph{Proof of Theorem \ref{pcatheorem}}
The defining property of an elliptical distribution implies
\[
Q_e(u)=\sqrt{e^\top\Sigma e}\,H^{-1}(u).
\]
Therefore
\[
\mathcal C_{\nu,\mathcal P}(e)
=
\kappa_{\nu,H}\sqrt{e^\top\Sigma e}.
\]
By the Rayleigh-quotient inequality,
\[
e^\top\Sigma e\le\lambda_1
\]
for every unit $e$, with equality exactly when \(e\in\mathcal L_1\). Thus
\(\mathcal D^\star=\mathcal L_1\), and Proposition \ref{directionconflict} gives the
full set of global-maximizing profiles together with the distance and payoff formulas.

It remains to verify the party-preferred qualification. Define, for
$q=x_A-x_B$,
\[
\mathcal H(q)=\kappa_{\nu,H}\sqrt{q^\top\Sigma q}.
\]
Up to terms independent of its own platform, party $A$ maximizes
$\mathcal H(x_A-x_B)-\|x_A-\mu\|^2/2$, while party $B$ maximizes
$\mathcal H(x_A-x_B)-\|x_B-\mu\|^2/2$.

By Proposition \ref{directionconflict} and its proof, every unrestricted equilibrium
lies in \(B_R(\mu)^2\) and is differentiated. Since \(\phi>4R^2\), the preceding payoff
representation holds on an open neighborhood of every such equilibrium. The function
\(\mathcal H\) is differentiable at every \(q\ne0\), so the unrestricted first-order
conditions apply:
\[
x_A-\mu=\nabla\mathcal H(q),
\qquad
x_B-\mu=-\nabla\mathcal H(q).
\]
Thus the midpoint is $\mu$, and
\[
q
=
\frac{2\kappa_{\nu,H}\Sigma q}
{\sqrt{q^\top\Sigma q}}.
\]
Every equilibrium direction is therefore an eigenvector of $\Sigma$. If its eigenvalue is
$\lambda_j$, platform distance is $2\kappa_{\nu,H}\sqrt{\lambda_j}$ and each party
obtains
\[
c_0+\frac{2\kappa_{\nu,H}^2\lambda_j}{\phi}.
\]
Every profile generated by \(e_1\in\mathcal L_1\) is a global potential maximizer and
hence a Nash equilibrium, and all such profiles give both parties the same payoff. They
strictly Pareto dominate every equilibrium associated with a lower eigenvalue. Conversely,
the first-order characterization exhausts all unrestricted equilibria, so no other
profile is party-preferred. Simplicity of $\lambda_1$ gives uniqueness up to sign and
party labels. \(\blacksquare\)

\paragraph{Proof of Corollary \ref{correlationresult}}
Equation \eqref{leadingeigenvalue} depends on $\varrho$ only through $\varrho^2$ and is
strictly increasing in $|\varrho|$ for positive marginal variances. The distance formula
in Theorem \ref{pcatheorem} is increasing in $\sqrt{\lambda_1}$, while the payoff formula
is increasing in $\lambda_1$. Both conclusions follow. \(\blacksquare\)

\subsection{Finite-type characterization}\label{app:finitecharacterization}

Throughout this subsection, voter types with identical bliss points are consolidated by
adding their population shares. This leaves the game unchanged and makes the remaining
bliss points distinct.

For each strict ranking \(R\) of voter types, let \(\bar x(R)\) and
\(\ubar x(R)\) be the two rank-weighted averages obtained by applying the weights in
Proposition \ref{NE1D} to \(R\). Define the set of strict rankings generated by a
platform profile as
\[
R\in\mathcal r(x_A,x_B)
\quad\Longleftrightarrow\quad
\Delta_{R_1}(x_A,x_B)<\cdots<\Delta_{R_N}(x_A,x_B),
\]
and define
\[
\mathcal C
=
\left\{
(\bar x(R),\ubar x(R)):
R\in\mathcal r(\bar x(R),\ubar x(R))
\right\}.
\]
A ranking generates an element of \(\mathcal C\) only if the platforms produced by its
organizational weights reproduce that same ranking.

\begin{proposition}[Finite-type characterization]\label{existence}
Suppose \(\nu\) is strictly concave. For every centrally symmetric finite electorate,
define each party's preferred equilibrium set by
\[
\mathcal E_p^\star
:=
\arg\max_{x\in\mathcal E}V_p(x),
\qquad p=A,B,
\]
where \(\mathcal E\) is the equilibrium set of the unrestricted game. Then both sets are
nonempty, coincide, and satisfy
\[
\mathcal E_A^\star=\mathcal E_B^\star
=
\arg\max_{(x_A,x_B)\in D^2}\Phi(x_A,x_B)
=
\arg\max_{(x_A,x_B)\in\mathcal C}
\|x_A-x_B\|^2
\neq\varnothing.
\]
Denote this common set by \(\mathcal E^\star\).
\end{proposition}

Strict concavity rules out voter-rank ties at a best response. Each party's first-order
condition then yields its rank-weighted platform, reducing the search to the finite set
\(\mathcal C\). Among those self-consistent political divisions, the parties agree on the
ones with greatest platform separation. Online Appendix \ref{oa:appendixA2} records the resulting
finite algorithm. The additional strict-concavity condition is needed for this finite
self-consistency characterization, not for Proposition \ref{potentialgame} or for the
continuous results, which continue to allow threshold premia.
\subsubsection*{Direct finite-type characterization}

Maintain the assumptions and definitions of Proposition
\ref{existence}: quadratic Euclidean policy loss, the wide uniform
shock, central symmetry around \(\mu\), the normalization
\(\nu(1)-\nu(0)=1\), strict concavity of \(\nu\), and the consolidation
of voter types with identical bliss points.

\begin{lemma}[Interior best responses]\label{lemma:interiorbr}
Every best response in \(D\) lies in the relative interior of \(D\). Coordinates on which
all voter bliss points coincide can be suppressed.
\end{lemma}

\paragraph{Proof.}
Fix a nondegenerate coordinate \(k\), with
\(\ell_k=\min_i x_i^k<u_k=\max_i x_i^k\). If party \(A\) chose
\(x_A^k=\ell_k\), the feasible right derivative, evaluated under the strict refinement
of any ties induced by a small inward movement, would be
\[
\frac{1}{\phi}\sum_i w_i^+(x_i^k-\ell_k)>0.
\]
All rank-increment weights \(w_i^+\) are positive and at least one voter has
\(x_i^k>\ell_k\). Thus the lower boundary cannot be a best response. At
\(x_A^k=u_k\), the analogous feasible left derivative is
\[
\frac{1}{\phi}\sum_i w_i^-(x_i^k-u_k)<0,
\]
so a small negative movement is profitable. The argument for party \(B\) is identical.
\(\blacksquare\)

\begin{lemma}[No ties at a best response]
\label{lemma:notiebr}
If \(x_A\) is a best response to \(x_B\) in \(D\), then
\[
\Delta_i(x_A,x_B)\ne\Delta_j(x_A,x_B)
\qquad\text{for every }i\ne j.
\]
The analogous statement holds for a best response by party \(B\).
\end{lemma}

\paragraph{Proof.}
Lemma \ref{lemma:interiorbr} makes both sufficiently small coordinate perturbations
feasible. Suppose that a best response \(x_A\) nevertheless induces a tie. Partition
the tied types into maximal blocks according to their common value of
\(\Delta_i(x_A,x_B)\).  Since bliss points are distinct, there is a tie
block \(G\) and a coordinate \(k\) on which its types do not all have
the same bliss point.  Order that block so that
\[
x_{g_1}^k\le\cdots\le x_{g_m}^k,
\]
with at least one strict inequality.  Let \(t\) be the mass of types
strictly more likely to support \(A\) than the types in \(G\), and let
\(S_G=\sum_{j=1}^ms_{g_j}\).

A small positive change in \(x_A^k\) ranks the types in \(G\) from low
to high \(x_i^k\), while a small negative change reverses that ranking.
The payoff-increment weights assigned to \(g_j\) on the two sides of
the tie are
\[
w_{g_j}^+
=
\nu\left(t+\sum_{\ell=j}^m s_{g_\ell}\right)
-\nu\left(t+\sum_{\ell=j+1}^m s_{g_\ell}\right)
\]
and
\[
w_{g_j}^-
=
\nu\left(t+\sum_{\ell=1}^j s_{g_\ell}\right)
-\nu\left(t+\sum_{\ell=1}^{j-1}s_{g_\ell}\right).
\]
For every \(r\in\{1,\ldots,m\}\), setting
\(S_r=\sum_{j=r}^m s_{g_j}\) gives
\[
\sum_{j=r}^m w_{g_j}^+
=\nu(t+S_r)-\nu(t)
\]
and
\[
\sum_{j=r}^m w_{g_j}^-
=\nu(t+S_G)-\nu(t+S_G-S_r).
\]
These are increments of \(\nu\) over intervals of equal length.  The
first interval lies weakly to the left of the second, so strict
concavity implies
\[
\sum_{j=r}^m w_{g_j}^+
\ge
\sum_{j=r}^m w_{g_j}^-,
\]
with strict inequality for a cutoff separating two distinct values of
\(x_{g_j}^k\).  Hence
\begin{equation}\label{tiederivativejump}
\sum_{j=1}^m(w_{g_j}^+-w_{g_j}^-)x_{g_j}^k>0.
\end{equation}

Under quadratic loss and the uniform shock,
\[
\frac{\partial F(\Delta_i(x_A,x_B))}{\partial x_A^k}
=\frac{x_i^k-x_A^k}{\phi}.
\]
The weights within a tie block have the same total on the two sides,
so the term involving \(x_A^k\) cancels.  Equation
\eqref{tiederivativejump} is therefore the block's contribution,
multiplied by \(1/\phi\), to
\[
D_+V_A(x_A,x_B)-D_-V_A(x_A,x_B).
\]
Every other tie block contributes weakly positively by the same
argument, and the selected block contributes strictly positively.
Consequently,
\[
D_+V_A(x_A,x_B)>D_-V_A(x_A,x_B).
\]
At a local maximum, however, the left derivative must be weakly
nonnegative and the right derivative weakly nonpositive, requiring
\(D_+V_A\le D_-V_A\).  This contradiction proves the claim for
party \(A\).  The proof for party \(B\) is symmetric.
\(\blacksquare\)

\begin{lemma}[Candidates]\label{lemma:necandidates}
Every pure-strategy Nash equilibrium belongs to \(\mathcal C\).
\end{lemma}

\paragraph{Proof.}
Let \(x=(x_A,x_B)\) be a Nash equilibrium. Lemma \ref{lemma:interiorbr} makes the
ordinary first-order conditions applicable, and Lemma
\ref{lemma:notiebr} implies that \(x\) generates a unique strict
ranking \(R\).  In a neighborhood preserving \(R\), direct integration
over the uniform shock gives
\begin{align}
V_A(x_A,x_B)
&=
c_0+\frac{1}{2\phi}
\left[
\|x_B\|^2-\|x_A\|^2
+2\bar x(R)^\top(x_A-x_B)
\right],
\label{directfinitepayoffA}\\
V_B(x_A,x_B)
&=
c_0-\frac{1}{2\phi}
\left[
\|x_B\|^2-\|x_A\|^2
+2\ubar x(R)^\top(x_A-x_B)
\right],
\label{directfinitepayoffB}
\end{align}
where \(c_0=(\nu(1)+\nu(0))/2\).  Conditional on \(R\), the two
own-platform gradients are
\[
\nabla_{x_A}V_A(x_A,x_B)
=\frac{\bar x(R)-x_A}{\phi},
\qquad
\nabla_{x_B}V_B(x_A,x_B)
=\frac{\ubar x(R)-x_B}{\phi}.
\]
Each own payoff is a strictly concave quadratic within the ranking
cell.  Since a best response is also a local best response, the
first-order conditions require
\[
x_A=\bar x(R),
\qquad
x_B=\ubar x(R).
\]
The ranking \(R\) is generated by this pair, so
\[
x=(\bar x(R),\ubar x(R))\in\mathcal C.
\]
\(\blacksquare\)

\paragraph{Direct proof of Proposition \ref{existence}}
Central symmetry reverses every strict ranking and pairs each type with
its reflection around \(\mu\).  Therefore, for every self-consistent
candidate \(x=(x_A,x_B)\in\mathcal C\),
\begin{equation}\label{candidatecenteringdirect}
x_B-\mu=-(x_A-\mu).
\end{equation}
Substituting the candidate first-order conditions into
\eqref{directfinitepayoffA}--\eqref{directfinitepayoffB} yields
\begin{equation}\label{candidatepayoffdirect}
V_A(x)=V_B(x)
=
c_0+\frac{\|x_A-x_B\|^2}{2\phi}.
\end{equation}
Using \eqref{potentialdefinition} and
\eqref{candidatecenteringdirect} also gives
\begin{equation}\label{candidatepotentialdirect}
\Phi(x)
=
c_0+\frac{\|x_A-x_B\|^2}{4\phi}.
\end{equation}

Let \(z\) be a global maximizer of \(\Phi\) on \(D^2\).
Proposition \ref{potentialgame} implies that \(z\) is a Nash equilibrium. Lemma
\ref{lemma:necandidates} therefore
implies \(z\in\mathcal C\), proving that
\(\mathcal C\ne\varnothing\).  Since \(\mathcal C\) is finite, choose
\[
x^\star\in
\arg\max_{x\in\mathcal C}\|x_A-x_B\|^2.
\]
Equation \eqref{candidatepotentialdirect} implies
\(\Phi(x^\star)\ge\Phi(z)\).  Since \(z\) is a global maximizer,
equality must hold.  Thus \(x^\star\) is itself a global potential
maximizer and, by Proposition \ref{potentialgame}, a Nash equilibrium.

Conversely, Lemma \ref{lemma:necandidates} places every Nash equilibrium
in \(\mathcal C\), and \eqref{candidatepayoffdirect} shows that both
parties rank equilibria by squared platform distance.  Every
distance-maximizing candidate is a global potential maximizer by the
preceding argument, while every other equilibrium gives both parties a
strictly lower payoff. Moreover, every global potential maximizer lies in
\(\mathcal C\) and must maximize distance there; otherwise
\eqref{candidatepotentialdirect} would give a strictly larger potential value.
Consequently,
\[
\mathcal E_A^\star
=\mathcal E_B^\star
=\arg\max_{(x_A,x_B)\in D^2}\Phi(x_A,x_B)
=
\arg\max_{(x_A,x_B)\in\mathcal C}
\|x_A-x_B\|^2
\ne\varnothing.
\qquad\blacksquare
\]

\subsection{Multidimensional voter polarization}

\paragraph{Proof of Proposition \ref{propspreadKD}}
Hold \(\nu\) and \(\phi\) fixed across the two electorates. Because the game is invariant
to a common translation of voter bliss points and platforms, center each electorate at
the origin. Let \(D^\cup\) be the smallest coordinate rectangle containing the centered
supports of both electorates, and use \((D^\cup)^2\) as the common strategy domain. The
maintained shock-support condition makes both potentials exact on this domain. For either
electorate, coordinatewise projection of a platform outside its own smaller rectangle
strictly reduces its distance from every voter type in at least one coordinate and weakly
reduces it in all others. Holding the rival's platform fixed, the projection therefore
strictly raises every voter's policy advantage for the deviating party. Under wide support
and the strict monotonicity of \(\nu\), it strictly raises that party's expected payoff.
Exactness then strictly raises the potential as well. Hence no equilibrium or global
potential maximizer on \((D^\cup)^2\) uses a platform outside the electorate's own rectangle,
so enlarging the strategy domain leaves both the global potential value and the
party-preferred equilibrium payoffs unchanged.

Let \(\hat x\in\mathcal E^\star(\mathbf x,\mathbf s)\), write
\[
d=\hat x_A-\hat x_B,
\]
and let \((\tilde{\mathbf x},\tilde{\mathbf s})\) denote the \(d\)-spread. Central
symmetry and the candidate identities in the proof of Proposition \ref{existence} imply
\[
\hat x_B=-\hat x_A,
\qquad
d=2\hat x_A.
\]

For the original electorate, let \(R\) be the ranking induced by the projections
\(x_i^\top d\), which is strict by Lemma \ref{lemma:notiebr}, and let
\(\bar w_{R_i}\) be the corresponding upper rank weights. For the
spread electorate, let \(\tilde R\) be any strict refinement of the weak order induced by
\(\tilde x_i^\top d\), with weights \(\bar w_{\tilde R_i}\). Permutations within a
projection tie do not affect the weighted sums because every member of the tie has the
same projected bliss point. Define
\[
z_{R_i}=x_{R_i}^\top\hat x_A,
\qquad
\tilde z_{\tilde R_i}=\tilde x_{\tilde R_i}^\top\hat x_A,
\]
and
\[
\bar z=\sum_i\bar w_{R_i}z_{R_i},
\qquad
\tilde{\bar z}=\sum_i\bar w_{\tilde R_i}\tilde z_{\tilde R_i}.
\]
The reflected lower rank weights generate \(\underline z=-\bar z\) and
\(\tilde{\underline z}=-\tilde{\bar z}\). Since \(d=2\hat x_A\), the maintained
\(d\)-spread is exactly a spread of these centered one-dimensional projected
distributions. The common-grid and tail-mass argument in the proof of Proposition
\ref{polarization} therefore gives
\[
\tilde{\bar z}-\tilde{\underline z}
>
\bar z-\underline z,
\]
and hence \(\tilde{\bar z}>\bar z\).

At the fixed symmetric profile \(\hat x\), direct integration over the uniform shock gives
\[
V_p(\hat x)
=
\frac{\nu(1)+\nu(0)}2+\frac{2}{\phi}\bar z,
\qquad
\tilde V_p(\hat x)
=
\frac{\nu(1)+\nu(0)}2+\frac{2}{\phi}\tilde{\bar z},
\quad p=A,B.
\]
Thus \(\tilde V_p(\hat x)>V_p(\hat x)\). Let \(\Phi\) and \(\tilde\Phi\) denote the two
potentials. The correction term in \eqref{potentialdefinition} depends only on the fixed
platforms, so
\[
\tilde\Phi(\hat x)>\Phi(\hat x).
\]

By Proposition \ref{existence}, \(\hat x\) is a global maximizer of \(\Phi\). Let
\(\tilde x^\star\) be a global maximizer of \(\tilde\Phi\). Then
\[
\tilde\Phi(\tilde x^\star)
\ge\tilde\Phi(\hat x)
>\Phi(\hat x).
\]
At every global potential maximizer, the direct candidate identities give
\[
V_p=2\Phi-\frac{\nu(1)+\nu(0)}2.
\]
The same identity holds for the spread electorate. It follows that
\[
\hat V_p(\tilde{\mathbf x},\tilde{\mathbf s})
>
\hat V_p(\mathbf x,\mathbf s),
\qquad p=A,B.
\]
\(\blacksquare\)

%% file: biblio.bib
@article{Alesina2020,
  author  = {Alesina, Alberto and Miano, Armando and Stantcheva, Stefanie},
  title   = {The Polarization of Reality},
  journal = {AEA Papers and Proceedings},
  year    = {2020},
  volume  = {110},
  pages   = {324--328}
}

@article{ali2024,
  author  = {Ali, S. Nageeb and Mihm, Maximilian and Siga, Lucas},
  title   = {The Political Economy of Zero-Sum Thinking},
  journal = {Econometrica},
  year    = {2025},
  volume  = {93},
  number  = {1},
  pages   = {41--70},
  doi     = {10.3982/ECTA22474}
}

@article{Aragones2015,
  author  = {Aragon{\`e}s, Enriqueta and Castanheira, Micael and Giani, Marco},
  title   = {Electoral Competition through Issue Selection},
  journal = {American Journal of Political Science},
  year    = {2015},
  volume  = {59},
  number  = {1},
  pages   = {71--90},
  doi     = {10.1111/ajps.12120}
}

@article{AshMorelliVanWeelden2017,
  author  = {Ash, Elliott and Morelli, Massimo and Van Weelden, Richard},
  title   = {Elections and Divisiveness: Theory and Evidence},
  journal = {The Journal of Politics},
  year    = {2017},
  volume  = {79},
  number  = {4},
  pages   = {1268--1285},
  doi     = {10.1086/692587}
}

@article{Autor2020,
  author  = {Autor, David and Dorn, David and Hanson, Gordon H. and Majlesi, Kaveh},
  title   = {Importing Political Polarization? The Electoral Consequences of Rising Trade Exposure},
  journal = {American Economic Review},
  year    = {2020},
  volume  = {110},
  number  = {10},
  pages   = {3139--3183}
}

@article{Buisseret,
  author  = {Bernhardt, Dan and Buisseret, Peter and Hidir, Sinem},
  title   = {The Race to the Base},
  journal = {American Economic Review},
  year    = {2020},
  volume  = {110},
  number  = {3},
  pages   = {922--942},
  doi     = {10.1257/aer.20181606}
}

@article{BuisseretFolkePratoRickne2022,
  author  = {Buisseret, Peter and Folke, Olle and Prato, Carlo and Rickne, Johanna},
  title   = {Party Nomination Strategies in List Proportional Representation Systems},
  journal = {American Journal of Political Science},
  year    = {2022},
  volume  = {66},
  number  = {3},
  pages   = {714--729},
  doi     = {10.1111/ajps.12691}
}

@article{ColonnelliPremTeso2020,
  author  = {Colonnelli, Emanuele and Prem, Mounu and Teso, Edoardo},
  title   = {Patronage and Selection in Public Sector Organizations},
  journal = {American Economic Review},
  year    = {2020},
  volume  = {110},
  number  = {10},
  pages   = {3071--3099},
  doi     = {10.1257/aer.20181491}
}

@article{Plott1967,
  author  = {Plott, Charles R.},
  title   = {A Notion of Equilibrium and Its Possibility under Majority Rule},
  journal = {American Economic Review},
  year    = {1967},
  volume  = {57},
  number  = {4},
  pages   = {787--806}
}

@article{SvitakovaSoltes2024,
  author  = {Svit{\'a}kov{\'a}, Kl{\'a}ra and {\v{S}}olt{\'e}s, Michal},
  title   = {Ranking of Candidates on Slates: Evidence from 20,000 Electoral Slates},
  journal = {Party Politics},
  year    = {2024},
  volume  = {30},
  number  = {3},
  pages   = {465--478},
  doi     = {10.1177/13540688231156481}
}

@article{BERNHARDT_DUGGAN_SQUINTANI_2009,
  author  = {Bernhardt, Daniel and Duggan, John and Squintani, Francesco},
  title   = {The Case for Responsible Parties},
  journal = {American Political Science Review},
  year    = {2009},
  volume  = {103},
  number  = {4},
  pages   = {570--587},
  doi     = {10.1017/S0003055409990232}
}

@article{BertrandKamenica2023,
  author  = {Bertrand, Marianne and Kamenica, Emir},
  title   = {Coming Apart? Cultural Distances in the United States over Time},
  journal = {American Economic Journal: Applied Economics},
  year    = {2023},
  volume  = {15},
  number  = {4},
  pages   = {100--141}
}

@unpublished{BonomiSpillovers,
  author = {Bonomi, Giampaolo},
  title  = {Disagreement Spillovers},
  note   = {Mimeo},
  year   = {2026}
}

@article{BGT21,
  author  = {Bonomi, Giampaolo and Gennaioli, Nicola and Tabellini, Guido},
  title   = {Identity, Beliefs, and Political Conflict},
  journal = {The Quarterly Journal of Economics},
  year    = {2021},
  volume  = {136},
  number  = {4},
  pages   = {2371--2411},
  doi     = {10.1093/qje/qjab034}
}

@article{Bowen2023,
  author  = {Bowen, T. Renee and Dmitriev, Dmitry and Galperti, Simone},
  title   = {Learning from Shared News: When Abundant Information Leads to Belief Polarization},
  journal = {The Quarterly Journal of Economics},
  year    = {2023},
  volume  = {138},
  number  = {2},
  pages   = {955--1000}
}

@article{Brady2007,
  author  = {Brady, David W. and Han, Hahrie and Pope, Jeremy C.},
  title   = {Primary Elections and Candidate Ideology: Out of Step with the Primary Electorate?},
  journal = {Legislative Studies Quarterly},
  year    = {2007},
  volume  = {32},
  number  = {1},
  pages   = {79--105}
}

@techreport{Dzuida23,
  author      = {Bueno de Mesquita, Ethan and Dziuda, Wioletta},
  title       = {Partisan Traps},
  institution = {National Bureau of Economic Research},
  type        = {Working Paper},
  number      = {31827},
  year        = {2023}
}

@article{ButticeStone2012,
  author  = {Buttice, Matthew K. and Stone, Walter J.},
  title   = {Candidates Matter: Policy and Quality Differences in Congressional Elections},
  journal = {The Journal of Politics},
  year    = {2012},
  volume  = {74},
  number  = {3},
  pages   = {870--887}
}

@techreport{Carvalho2023,
  author      = {Bergeron, Augustin and Carvalho, Jean-Paul and Henrich, Joseph and Nunn, Nathan and Weigel, Jonathan L.},
  title       = {Zero-Sum Environments, the Evolution of Effort-Suppressing Beliefs, and Economic Development},
  institution = {National Bureau of Economic Research},
  type        = {Working Paper},
  number      = {31663},
  year        = {2023},
  note        = {Revised August 2025},
  doi         = {10.3386/w31663}
}

@article{Chinoy,
  author  = {Chinoy, Sahil and Nunn, Nathan and Sequeira, Sandra and Stantcheva, Stefanie},
  title   = {Zero-Sum Thinking and the Roots of {U.S.} Political Differences},
  journal = {American Economic Review},
  year    = {2026},
  volume  = {116},
  number  = {3},
  pages   = {1052--1096},
  doi     = {10.1257/aer.20240692}
}

@article{DavisMason2016,
  author  = {Davis, Nicholas T. and Mason, Lilliana},
  title   = {Sorting and the Split-Ticket: Evidence from Presidential and Subpresidential Elections},
  journal = {Political Behavior},
  year    = {2016},
  volume  = {38},
  number  = {2},
  pages   = {337--354}
}

@article{DellaVigna2007,
  author  = {DellaVigna, Stefano and Kaplan, Ethan},
  title   = {The {Fox News} Effect: Media Bias and Voting},
  journal = {The Quarterly Journal of Economics},
  year    = {2007},
  volume  = {122},
  number  = {3},
  pages   = {1187--1234}
}

@book{Downs1957,
  author    = {Downs, Anthony},
  title     = {An Economic Theory of Democracy},
  publisher = {Harper and Row},
  address   = {New York},
  year      = {1957}
}

@article{Feddersen2022,
  author  = {Feddersen, Alexandra and Adams, James F.},
  title   = {Public Opinion Backlash in Response to Party Messages: A Case for Party Press Releases},
  journal = {Electoral Studies},
  year    = {2022},
  volume  = {80},
  pages   = {102532}
}

@book{fiorina2011,
  author    = {Fiorina, Morris P. and Abrams, Samuel J. and Pope, Jeremy C.},
  title     = {Culture War? The Myth of a Polarized America},
  series    = {Great Questions in Politics},
  publisher = {Longman},
  year      = {2011}
}

@article{GT23,
  author  = {Gennaioli, Nicola and Tabellini, Guido},
  title   = {Presidential Address: Identity Politics},
  journal = {Econometrica},
  year    = {2025},
  volume  = {93},
  number  = {6},
  pages   = {1937--1967},
  doi     = {10.3982/ECTA22269}
}

@unpublished{Gentzkow2016,
  author = {Gentzkow, Matthew},
  title  = {Polarization in 2016},
  note   = {Mimeo},
  year   = {2016}
}

@article{Gentzkow2019,
  author  = {Gentzkow, Matthew and Shapiro, Jesse M. and Taddy, Matt},
  title   = {Measuring Group Differences in High-Dimensional Choices: Method and Application to Congressional Speech},
  journal = {Econometrica},
  year    = {2019},
  volume  = {87},
  number  = {4},
  pages   = {1307--1340}
}

@article{Gleaser2005,
  author  = {Glaeser, Edward L. and Ponzetto, Giacomo A. M. and Shapiro, Jesse M.},
  title   = {Strategic Extremism: Why Republicans and Democrats Divide on Religious Values},
  journal = {The Quarterly Journal of Economics},
  year    = {2005},
  volume  = {120},
  number  = {4},
  pages   = {1283--1330}
}

@article{Greene99,
  author  = {Greene, Steven},
  title   = {Understanding Party Identification: A Social Identity Approach},
  journal = {Political Psychology},
  year    = {1999},
  volume  = {20},
  number  = {2},
  pages   = {393--403}
}

@article{Hirano2009,
  author  = {Hirano, Shigeo and Snyder, James M. and Ting, Michael M.},
  title   = {Distributive Politics with Primaries},
  journal = {The Journal of Politics},
  year    = {2009},
  volume  = {71},
  number  = {4},
  pages   = {1467--1480}
}

@book{HuberShipan2002DeliberateDiscretion,
  author    = {Huber, John D. and Shipan, Charles R.},
  title     = {Deliberate Discretion? The Institutional Foundations of Bureaucratic Autonomy},
  publisher = {Cambridge University Press},
  address   = {Cambridge},
  year      = {2002}
}

@article{KatzMair1995CartelParty,
  author  = {Katz, Richard S. and Mair, Peter},
  title   = {Changing Models of Party Organization and Party Democracy: The Emergence of the Cartel Party},
  journal = {Party Politics},
  year    = {1995},
  volume  = {1},
  number  = {1},
  pages   = {5--28}
}

@article{Eunji2022,
  author  = {Kim, Eunji and Lelkes, Yphtach and McCrain, Joshua},
  title   = {Measuring Dynamic Media Bias},
  journal = {Proceedings of the National Academy of Sciences},
  year    = {2022},
  volume  = {119},
  number  = {32},
  pages   = {e2202197119}
}

@book{Lewis2008PoliticsPresidentialAppointments,
  author    = {Lewis, David E.},
  title     = {The Politics of Presidential Appointments: Political Control and Bureaucratic Performance},
  publisher = {Princeton University Press},
  address   = {Princeton, NJ},
  year      = {2008}
}

@article{Lindbeck,
  author  = {Lindbeck, Assar and Weibull, J{\"o}rgen W.},
  title   = {A Model of Political Equilibrium in a Representative Democracy},
  journal = {Journal of Public Economics},
  year    = {1993},
  volume  = {51},
  number  = {2},
  pages   = {195--209}
}

@article{LW87,
  author  = {Lindbeck, Assar and Weibull, J{\"o}rgen W.},
  title   = {Balanced-Budget Redistribution as the Outcome of Political Competition},
  journal = {Public Choice},
  year    = {1987},
  volume  = {52},
  number  = {3},
  pages   = {273--297}
}

@article{MartinYurukoglu2017,
  author  = {Martin, Gregory J. and Yurukoglu, Ali},
  title   = {Bias in Cable News: Persuasion and Polarization},
  journal = {American Economic Review},
  year    = {2017},
  volume  = {107},
  number  = {9},
  pages   = {2565--2599}
}

@book{MartinVanberg2011ParliamentsCoalitions,
  author    = {Martin, Lanny W. and Vanberg, Georg},
  title     = {Parliaments and Coalitions: The Role of Legislative Institutions in Multiparty Governance},
  publisher = {Oxford University Press},
  address   = {Oxford},
  year      = {2011}
}

@article{Matakos2016,
  author  = {Matakos, Konstantinos and Troumpounis, Orestis and Xefteris, Dimitrios},
  title   = {Electoral Rule Disproportionality and Platform Polarization},
  journal = {American Journal of Political Science},
  year    = {2016},
  volume  = {60},
  number  = {4},
  pages   = {1026--1043},
  doi     = {10.1111/ajps.12235}
}

@article{Kendall2022,
  author  = {Kendall, Chad and Matsusaka, John G.},
  title   = {The Common Good and Voter Polarization},
  journal = {Journal of the European Economic Association},
  year    = {2025},
  doi     = {10.1093/jeea/jvaf059}
}

@article{PeregoYuksel2022,
  author  = {Perego, Jacopo and Yuksel, Sevgi},
  title   = {Media Competition and Social Disagreement},
  journal = {Econometrica},
  year    = {2022},
  volume  = {90},
  number  = {1},
  pages   = {223--265},
  doi     = {10.3982/ECTA16417}
}

@article{OrtunoOrtin1997,
  author  = {Ortu{\~n}o-Ort{\'i}n, Ignacio},
  title   = {A Spatial Model of Political Competition and Proportional Representation},
  journal = {Social Choice and Welfare},
  year    = {1997},
  volume  = {14},
  number  = {3},
  pages   = {427--438},
  doi     = {10.1007/s003550050076}
}

@article{Saporiti2014,
  author  = {Saporiti, Alejandro},
  title   = {Power Sharing and Electoral Equilibrium},
  journal = {Economic Theory},
  year    = {2014},
  volume  = {55},
  number  = {3},
  pages   = {705--729},
  doi     = {10.1007/s00199-013-0772-0}
}

@article{McCubbinsNollWeingast1987AdministrativeProcedures,
  author  = {McCubbins, Mathew D. and Noll, Roger G. and Weingast, Barry R.},
  title   = {Administrative Procedures as Instruments of Political Control},
  journal = {The Journal of Law, Economics, and Organization},
  year    = {1987},
  volume  = {3},
  number  = {2},
  pages   = {243--277}
}

@article{Moreira2025,
  author  = {Moreira, Thiago M. Q.},
  title   = {Is It Still the Economy? Economic Voting in Polarized Politics},
  journal = {British Journal of Political Science},
  year    = {2025},
  volume  = {55},
  pages   = {e45}
}

@incollection{Duggan2006,
  author    = {Duggan, John},
  title     = {Candidate Objectives and Electoral Equilibrium},
  booktitle = {The Oxford Handbook of Political Economy},
  editor    = {Weingast, Barry R. and Wittman, Donald A.},
  publisher = {Oxford University Press},
  year      = {2006},
  pages     = {64--83}
}

@article{CaillaudTirole2002,
  author  = {Caillaud, Bernard and Tirole, Jean},
  title   = {Parties as Political Intermediaries},
  journal = {Quarterly Journal of Economics},
  year    = {2002},
  volume  = {117},
  number  = {4},
  pages   = {1453--1489}
}

@article{SnyderTing2002,
  author  = {Snyder, Jr., James M. and Ting, Michael M.},
  title   = {An Informational Rationale for Political Parties},
  journal = {American Journal of Political Science},
  year    = {2002},
  volume  = {46},
  number  = {1},
  pages   = {90--110}
}

@book{MullerStrom1999PolicyOfficeVotes,
  editor    = {M{\"u}ller, Wolfgang C. and Str{\o}m, Kaare},
  title     = {Policy, Office, or Votes? How Political Parties in Western Europe Make Hard Decisions},
  publisher = {Cambridge University Press},
  address   = {Cambridge},
  year      = {1999}
}

@book{PT2000,
  author    = {Persson, Torsten and Tabellini, Guido},
  title     = {Political Economics: Explaining Economic Policy},
  publisher = {The MIT Press},
  series    = {MIT Press Books},
  year      = {2000}
}

@techreport{Pew2022,
  author      = {{Pew Research Center}},
  title       = {How Religion Intersects with Americans' Views on the Environment},
  institution = {Pew Research Center},
  year        = {2022}
}

@article{SHAYO09,
  author  = {Shayo, Moses},
  title   = {A Model of Social Identity with an Application to Political Economy: Nation, Class, and Redistribution},
  journal = {American Political Science Review},
  year    = {2009},
  volume  = {103},
  number  = {2},
  pages   = {147--174}
}

@article{Stantcheva2021,
  author  = {Stantcheva, Stefanie},
  title   = {Understanding Tax Policy: How Do People Reason?},
  journal = {The Quarterly Journal of Economics},
  year    = {2021},
  volume  = {136},
  number  = {4},
  pages   = {2309--2369}
}

@article{Strom1990BehavioralTheory,
  author  = {Str{\o}m, Kaare},
  title   = {A Behavioral Theory of Competitive Political Parties},
  journal = {American Journal of Political Science},
  year    = {1990},
  volume  = {34},
  number  = {2},
  pages   = {565--598}
}

@incollection{LipsetRokkan1967,
  author    = {Lipset, Seymour Martin and Rokkan, Stein},
  title     = {Cleavage Structures, Party Systems, and Voter Alignments: An Introduction},
  editor    = {Lipset, Seymour Martin and Rokkan, Stein},
  booktitle = {Party Systems and Voter Alignments: Cross-National Perspectives},
  publisher = {Free Press},
  address   = {New York},
  year      = {1967},
  pages     = {1--64}
}

@book{Riker1986,
  author    = {Riker, William H.},
  title     = {The Art of Political Manipulation},
  publisher = {Yale University Press},
  address   = {New Haven},
  year      = {1986}
}

@article{BaldassarriGelman2008,
  author  = {Baldassarri, Delia and Gelman, Andrew},
  title   = {Partisans without Constraint: Political Polarization and Trends in American Public Opinion},
  journal = {American Journal of Sociology},
  year    = {2008},
  volume  = {114},
  number  = {2},
  pages   = {408--446},
  doi     = {10.1086/590649}
}

@article{DellaPosta2020,
  author  = {DellaPosta, Daniel},
  title   = {Pluralistic Collapse: The ``Oil Spill'' Model of Mass Opinion Polarization},
  journal = {American Sociological Review},
  year    = {2020},
  volume  = {85},
  number  = {3},
  pages   = {507--536},
  doi     = {10.1177/0003122420922989}
}

@article{CallanderCarbajal2022,
  author  = {Callander, Steven and Carbajal, Juan Carlos},
  title   = {Cause and Effect in Political Polarization: A Dynamic Analysis},
  journal = {Journal of Political Economy},
  year    = {2022},
  volume  = {130},
  number  = {4},
  pages   = {825--880},
  doi     = {10.1086/718200}
}

@book{Hopkins2018,
  author    = {Hopkins, Daniel J.},
  title     = {The Increasingly United States: How and Why American Political Behavior Nationalized},
  publisher = {University of Chicago Press},
  address   = {Chicago},
  year      = {2018}
}

@article{ZakharovSorokin2014,
  author  = {Zakharov, Alexei V. and Sorokin, Constantine S.},
  title   = {Policy Convergence in a Two-Candidate Probabilistic Voting Model},
  journal = {Social Choice and Welfare},
  year    = {2014},
  volume  = {43},
  number  = {2},
  pages   = {429--446},
  doi     = {10.1007/s00355-013-0786-3}
}

@article{SorokinZakharov2018,
  author  = {Sorokin, Constantine S. and Zakharov, Alexei V.},
  title   = {Vote-Motivated Candidates},
  journal = {Journal of Economic Theory},
  year    = {2018},
  volume  = {176},
  pages   = {232--254},
  doi     = {10.1016/j.jet.2018.03.010}
}

@unpublished{AcharyaDuggan2026,
author = {Acharya, Avidit and Duggan, John},
title  = {Multidimensional Elections},
year   = {2026},
month  = aug,
note   = {Working paper}
}

@unpublished{Mack2022,
  author = {Mack, Andrew},
  title  = {Incentives for Learning in Repeated Elections},
  year   = {2022},
  note   = {Unpublished manuscript, conditionally accepted at the
            American Journal of Political Science}
}

@unpublished{Vaeth2026,
  author = {Vaeth, Martin},
  title  = {Rational Voter Learning, Issue Alignment, and Polarization},
  year   = {2026},
  note   = {Unpublished manuscript}
}

@article{KamadaKojima2014,
  author  = {Kamada, Yuichiro and Kojima, Fuhito},
  title   = {Voter Preferences, Polarization, and Electoral Policies},
  journal = {American Economic Journal: Microeconomics},
  year    = {2014},
  volume  = {6},
  number  = {4},
  pages   = {203--236},
  doi     = {10.1257/mic.6.4.203}
}

@article{PolbornSnyder2017,
  author  = {Polborn, Mattias K. and Snyder, Jr., James M.},
  title   = {Party Polarization in Legislatures with Office-Motivated Candidates},
  journal = {The Quarterly Journal of Economics},
  year    = {2017},
  volume  = {132},
  number  = {3},
  pages   = {1509--1550},
  doi     = {10.1093/qje/qjx012}
}

@unpublished{BoeriNikiforovaTabellini2026,
  author = {Boeri, Tito and Nikiforova, Nina and Tabellini, Guido},
  title  = {Do Elections Moderate or Polarize Political Rhetoric?},
  year   = {2026},
  month  = aug,
  note   = {Working paper}
}
